\documentclass{aa}  

\usepackage{graphicx}
\usepackage{booktabs}
\usepackage{threeparttable}
\usepackage{xcolor}

\usepackage{amsmath}	% Advanced maths commands
\usepackage{booktabs}
\usepackage{multirow}
\usepackage{float}
\usepackage{epsfig}
\usepackage{placeins}
\usepackage{xurl}
\usepackage{txfonts}
\usepackage{hyperref}
\defcitealias{Schrabback2018}{S18}
\defcitealias{Schrabback2021}{S21}
\defcitealias{Bocquet2019}{B19}
\defcitealias{Dietrich2019}{D19}
\defcitealias{Raihan2020}{R20}
\defcitealias{Rafelski2015}{R15}
\defcitealias{Skelton2014}{S14}
\defcitealias{Oesch2018}{Oe18}

\newcommand{\myvec}[1]{\ensuremath{\begin{matrix}#1\end{matrix}}}

\begin{document}

   \title{Extending empirical constraints on the SZ--mass scaling relation to higher redshifts via \textit{HST} weak lensing measurements of nine clusters from the SPT-SZ survey at $z\gtrsim1$}

   \author{
Hannah Zohren$^{1}$\thanks{E-mail: hzohren@astro.uni-bonn.de}, 
Tim Schrabback$^{1}$, 
Sebastian Bocquet$^{2,3}$, 
Martin Sommer$^{1}$, 
Fatimah Raihan$^{1}$, 
Beatriz Hern\'andez-Mart\'in$^{1}$, 
Ole Marggraf$^{1}$, 
Behzad Ansarinejad$^{4}$,
Matthew B. Bayliss$^{5}$,
Lindsey E. Bleem$^{6,7}$,
Thomas Erben$^{1}$,
Henk Hoekstra$^{8}$,
Benjamin Floyd$^{9}$,
Michael D. Gladders$^{7,10}$,
Florian Kleinebreil$^{1}$,
Michael A. McDonald$^{11}$,
Mischa Schirmer$^{12}$, 
Diana Scognamiglio$^{1}$,
Keren Sharon$^{13}$,
and Angus H. Wright$^{14}$}

   \institute{
  Argelander Institut f\"ur Astronomie, Rheinische Friedrich-Wilhelms-Universit\"at Bonn, Auf dem H\"ugel 71, D-53121 Bonn, Germany
  \and
Faculty of Physics, Ludwig-Maximilians University, Scheinerstr. 1, D-81679 M\"unchen, Germany
\and
Excellence Cluster ORIGINS, Boltzmannstr. 2, D-85748 Garching, Germany
\and
School of Physics, University of Melbourne, Parkville, VIC 3010, Australia
\and
Department of Physics, University of Cincinnati, Cincinnati, OH 45221, USA
\and
High Energy Physics Division, Argonne National Laboratory, Argonne, IL, USA 60439
\and
Kavli Institute for Cosmological Physics, University of Chicago, Chicago, IL, USA 60637
\and
Leiden Observatory, Leiden University, PO Box 9513, 2300 RA, Leiden, the Netherlands
\and
Department of Physics and Astronomy, University of Missouri--Kansas City, 5110 Rockhill Road, Kansas City, MO 64110, USA
\and
Department of Astronomy and Astrophysics, University of Chicago, 5640 South Ellis Avenue, Chicago, IL 60637, USA
\and
Kavli Institute for Astrophysics and Space Research, Massachusetts Institute of Technology, 77 Massachusetts Avenue, Cambridge, MA 02139, USA 
\and
Max-Planck-Institut f\"ur Astronomie (MPIA), K\"onigstuhl 17, 69117 Heidelberg, Germany
\and
Department of Astronomy, University of Michigan, 1085 S. University Ave, Ann Arbor, MI 48109, USA
\and
Ruhr University Bochum, Faculty of Physics and Astronomy, Astronomical Institute (AIRUB), German Centre for Cosmological Lensing, 44780 Bochum, Germany
             }

   \date{Received 23 December 2021; accepted 13 August 2022}

% \abstract{}{}{}{}{} 
% 5 {} token are mandatory
 
  \abstract{}{}{}{}
%   \begin{abstract}
{We present a \textit{Hubble Space Telescope} (\textit{HST}) weak gravitational lensing study of nine distant and massive galaxy clusters with redshifts \mbox{$1.0 \lesssim z \lesssim 1.7$} ($z_\mathrm{median} = 1.4$) and Sunyaev Zel'dovich (SZ) detection significance $\xi > 6.0$ from the South Pole Telescope Sunyaev Zel'dovich (SPT-SZ) survey. We measured weak lensing galaxy shapes in \textit{HST}/ACS F606W and F814W images and used additional observations from \textit{HST}/WFC3 in F110W and VLT/FORS2 in $U_\mathrm{HIGH}$ to preferentially select background galaxies at $z\gtrsim 1.8$, achieving a high purity. 
We combined recent redshift estimates from the CANDELS/3D-HST and HUDF fields to infer an improved estimate of the source redshift distribution. 
We measured weak lensing masses by fitting the tangential reduced shear profiles with spherical Navarro-Frenk-White (NFW) models. We obtained the largest lensing mass in our sample for the cluster SPT-CL{\thinspace}$J$2040$-$4451, thereby confirming earlier results that suggest a high lensing mass of this cluster compared to X-ray and SZ mass measurements.
Combining our weak lensing mass constraints with results obtained by previous studies for lower redshift clusters, we extended the calibration of the scaling relation between the unbiased SZ detection significance $\zeta$ and the cluster mass for the SPT-SZ survey out to higher redshifts. We found that the mass scale inferred from our highest redshift bin ($1.2 < z < 1.7$) is consistent with an extrapolation of constraints derived from lower redshifts, albeit with large statistical uncertainties. 
Thus, our results show a similar tendency as found in previous studies, where 
the cluster mass scale derived from the weak lensing data is lower than the mass scale expected in a \textit{Planck} $\nu\Lambda$CDM (i.e. $\nu$ $\Lambda$ Cold Dark Matter) cosmology given the SPT-SZ cluster number counts. }
%   \end{abstract}

    \titlerunning{\textit{HST} WL study of nine high-$z$ SPT clusters} 
    \authorrunning{H. Zohren et al.}
    
   \maketitle 
%
%-------------------------------------------------------------------

\section{Introduction}

Galaxy clusters trace the densest regions of the large-scale structure in the Universe. Studying their number density as a function of mass and redshift, therefore, provides insights into the cosmic expansion and structure formation histories, allowing for constraints of cosmological models \citep[e.g. ][]{Haiman2001,Allen2011}{}. The expected number of dark matter haloes at a given mass and redshift is predicted by the halo mass \mbox{function (HMF)}, which can be obtained from numerical simulations \citep[e.g. ][]{Tinker2008,McClintock2019,Bocquet2020}{}. A comparison of these predictions to observations of galaxy clusters as representatives of these haloes and their abundance serves as a probe, which is particularly sensitive to a combination of the cosmological parameters $\Omega_\mathrm{m}$, the matter energy density of the Universe, and $\sigma_8$, the standard deviation of fluctuations in the linear matter density field at scales of \mbox{8\,Mpc/$h$}. At the same time, cluster studies can constrain the dark energy equation of state parameter $w$. 

Such studies require samples of galaxy clusters with a well-defined selection function and covering a large redshift range. Common methods for the assembly of such samples include detection via the overdensity of galaxies in the optical/near-infrared (NIR) regime \citep[e.g. ][]{Rykoff2016}{}, via the X-ray flux \citep[e.g. ][]{Piffaretti2011,Pacaud2018,Liu2021}{}, or via the signal from the Sunyaev Zel'dovich (SZ) effect \citep[e.g. ][]{Bleem2015,Planck2016clustercounts,Hilton2021}{}.
 
The thermal SZ effect \citep[][]{Sunyaev1972}{} describes a distortion of the cosmic microwave background (CMB) blackbody spectrum towards higher energy, caused when CMB photons experience an inverse Compton scattering with the energetic electrons in the intracluster medium. Since the signal is independent of redshift, detecting clusters through the SZ effect enables the assembly of cluster catalogues, which are nearly mass-limited and extend out to very high redshifts. Additionally, the uncertainties in the selection function are relatively low because the SZ-observable provides a mass proxy with a comparably low intrinsic scatter \citep[$\sim$ 20 per cent, e.g. ][]{Angulo2012}{}. These are promising prerequisites for cosmological studies through the comparison of the observed cluster mass function and the predicted HMF.

However, accurate and precise calibration of the scaling relations between the observable mass proxy and the underlying unobservable halo mass as predicted by the HMF over a wide redshift range is needed to obtain meaningful cosmological constraints. Especially since the remaining uncertainties in the observable-mass scaling relations are the limiting factor hampering the progress to tighter constraints \citep[e.g. ][]{Dietrich2019}. It is, therefore, imperative to improve the cluster mass calibration out to the highest redshifts that are now accessible in cluster samples \citep{Bocquet2019,Schrabback2018,Schrabback2021}.
Mass measurements from weak gravitational lensing are frequently used as a method to obtain an absolute calibration of the normalisation of these scaling relations \citep[e.g. ][]{Okabe2010,Kettula2015,Dietrich2019,Herbonnet2020,Chiu2021,Schrabback2021}{}. Weak gravitational lensing causes a systematic distortion of the shapes of background galaxies when their light travels through the gravitational field of a foreground mass distribution. The weak lensing reduced shear quantifies the tangential distortion with respect to the centre of the mass distribution. The differential projected cluster mass distribution can be inferred from measurements of the reduced shear, without the need for assumptions about the dynamical state of the clusters. This is especially advantageous for high-redshift clusters because these objects are still dynamically young and may not have settled into hydrostatic equilibrium yet. The assumption of hydrostatic equilibrium is an important ingredient for measurements of the cluster mass with X-ray observations.

A complementary method to weak lensing studies with optical/NIR data is CMB cluster lensing, which measures the (stacked) weak lensing signal by galaxy clusters in maps of the temperature and polarisation of the CMB \citep[e.g. ][]{Raghunathan2019,Zubeldia2019,Madhavacheril2020}. Due to the high redshift of the CMB, the mass scale for high-redshift clusters is more easily accessible with this method, and constraints will become increasingly competitive with upcoming instruments such as SPT-3G \citep[][]{Benson2014} and CMB-S4 \citep[][]{CMBs4Collab2019}. 

In the low to intermediate redshift regime, wide-field ground-based surveys such as the Kilo Degree Survey \citep[KiDS, ][]{Kuijken2015}, the Dark Energy Survey \citep[DES, ][]{DES2005} and the Hyper-Suprime-Cam Survey \citep[HSC, ][]{Miyazaki2012} can calibrate cluster masses at the few per cent level via weak lensing, but they are not suited to obtain the critically required cluster masses at high redshifts. Their limited depth and ground-based resolution are not sufficient to resolve the shapes of the small and faint background galaxies behind high-redshift clusters. 

The aforementioned optical lensing studies have been limited to low to intermediate redshift regimes up to $z\sim 1$. It is important to extend the calibration of scaling relations to higher redshifts because cluster properties (e.g. thermodynamic properties such as density, temperature, pressure, and entropy) evolve over time.
Upcoming surveys conducted with \textit{Euclid} \citep{Laureijs2011}, the \textit{Nancy Grace Roman Space Telescope} \citep[formerly known as WFIRST, ][]{Spergel2015}, and the Vera C. Rubin Observatory \citep{LSST2009} will provide improved and critically required constraints on the cluster masses over a wide redshift range, where the exquisite depth of the \textit{Nancy Grace Roman Space Telescope} will be particularly valuable for the very high-redshift regime. 

However, until these surveys become available, pointed follow-up studies provide the best option to constrain the cluster mass scale out to high redshifts.
With this work, we present the first weak lensing constraints on the mass scale of SZ-selected clusters extending to redshifts above $z \gtrsim 1.2$, using galaxy shape measurements from \textit{HST} imaging. The median redshift of the sample with nine clusters studied here is $z = 1.4$. This study is an extension of the works by \citet[][henceforth \citetalias{Schrabback2018}]{Schrabback2018}, \citet[][henceforth \citetalias{Dietrich2019}]{Dietrich2019}, \citet[][henceforth \citetalias{Bocquet2019}]{Bocquet2019}, and \citet[][henceforth \citetalias{Schrabback2021}]{Schrabback2021} to constrain the redshift evolution of the SZ mass scaling relation based on clusters from the 2500\,deg$^2$ South Pole Telescope SZ survey \citep[SPT-SZ survey, ][]{Bleem2015}{}. With our high-redshift sample, we aim to tighten the constraints on the scaling relation parameter $C_\mathrm{SZ}$, describing its redshift evolution, which in particular helps to break the degeneracy of $C_\mathrm{SZ}$ with the dark energy equation of state parameter $w$.

The structure of this paper is as follows: we provide a summary of the studied cluster sample in Sect. \ref{Sec:SPT cluster sample}. We then present the data reduction of our optical observations and describe the photometric calibration steps in Sect. \ref{Sec:Data+Data reduction}. The selection of background galaxies based on four photometric bands and the estimation of the source redshift distribution from photometric redshift catalogues are detailed in Sect. \ref{Sec: full section, BG gal selection + redshift distrib}. We describe the weak lensing shape measurements in Sect. \ref{sec:shapes}. We present our weak lensing mass constraints including an estimation of the weak lensing mass bias in Sect. \ref{sec:wlconstraints}. We constrain the observable-mass scaling relation incorporating the new lensing results for our high-redshift SPT cluster sample in Sect. \ref{Sec:ScalingRelAnalysis}.
Finally, we discuss our results in Sect. \ref{Sec:Discussion} and summarise and conclude in Sect. \ref{Sec:Summary+Conclusions}. 

Unless indicated otherwise, we assume a standard flat $\Lambda$ Cold Dark Matter ($\Lambda$CDM) concordance cosmology throughout this paper with $\Omega_\mathrm{m} = 0.3$,
$\Omega_\Lambda = 0.7$, and $H_0 = 70$\, km\,s$^{-1}$\,Mpc$^{-1}$, as approximately consistent with CMB constraints \citep[e.g. ][]{Planck2020}{}.
We express masses in terms of $M_{\Delta \mathrm{c}}$ corresponding to a sphere within which the density is $\Delta$ times higher than the critical density at the given redshift.

All reported magnitudes in this work are AB-magnitudes. We correct all magnitude measurements for Galactic extinction with the extinction maps by \citet{2011ApJ...737...103S}.

%--------------------------------------------------------------------

\section{The high-$z$ SPT cluster sample and previous studies}
\label{Sec:SPT cluster sample}

\begin{table*}
	\centering
	\caption{Properties of the galaxy cluster sample.}
	\label{tab:Cluster sample properties}
% 	\begin{centre}

	\begin{threeparttable}
	\begin{tabular}{lcr rrrr c} 
		\hline
		\hline
         Cluster name & $z_\mathrm{l}$ & $\xi$ & \multicolumn{4}{c}{Coordinates centres (deg J2000)} & $M_{500\mathrm{c,SZ}}$ \\
            & &  &SZ $\alpha$ & SZ $\delta$ & X-ray $\alpha$  & X-ray $\delta$ & [$10^{14}\,\mathrm{M}_\odot / h_{70}$] \\
        \hline
        SPT-CL{\thinspace}$J$0156$-$5541 & 1.288 \tnote{$a$} & 6.98 & 29.04490 & $-55.69801$ & 29.0405 & $-55.6976$ & $3.96^{+0.57}_{-0.65} $\\
        SPT-CL{\thinspace}$J$0205$-$5829 & 1.322\tnote{$b$} & 10.40 & 31.44282 & $-58.48521$ & 31.4459 & $-58.4849$ & $5.06^{+0.55}_{-0.68}$\\
        SPT-CL{\thinspace}$J$0313$-$5334 & 1.474\tnote{$a$} & 6.09 & 48.48090 & $-53.57809$ & 48.4813 & $-53.5718$ & $3.31^{+0.55}_{-0.61}$\\
        SPT-CL{\thinspace}$J$0459$-$4947 & 1.710\tnote{$d$}  & 6.29 & 74.92693 & $-49.78724$ & 74.9240 & $-49.7823$ & $3.08^{+0.53}_{-0.53}$\\
        SPT-CL{\thinspace}$J$0607$-$4448 & 1.401\tnote{$a$}  & 6.44 & 91.89841 & $-44.80333$ & 91.8940 & $-44.8050$ & $3.60^{+0.57}_{-0.63}$\\
        SPT-CL{\thinspace}$J$0640$-$5113 & 1.316\tnote{$a$}  & 6.86 & 100.06452 & $-51.22045$ & 100.0720 & $-51.2176$ & $3.89^{+0.58}_{-0.65}$\\
        SPT-CL{\thinspace}$J$0646$-$6236 & 0.995\tnote{$e$} & 8.67 & 101.63906	& $-62.61360$ & -- & -- & $4.97^{+0.64}_{-0.76}$\tnote{$^f$}\\
        SPT-CL{\thinspace}$J$2040$-$4451 & 1.478\tnote{$c$} & 6.72 & 310.24832 & $-44.86023$ & 310.2417 & $-44.8620$ & $3.76^{+0.58}_{-0.63}$\\
        SPT-CL{\thinspace}$J$2341$-$5724 & 1.259\tnote{$a$} & 6.87 & 355.35683 & $-57.41580$ & 355.3533 & $-57.4166$ & $3.58^{+0.51}_{-0.59}$\\
		\hline

	\end{tabular}

	\textbf{Notes.}
	 We list cluster names, SZ significance $\xi$, SZ coordinates of the centre and SZ masses as presented in \citetalias{Bocquet2019}. The X-ray coordinates correspond to the centroid positions estimated by \citet{McDonald2017}.\newline
	$^a$ Spectroscopic redshifts by \citet{Khullar2019}.
	$^b$ Spectroscopic redshift from \citet{Stalder2013}. $^c$ Spectroscopic redshift from \citet{Bayliss2014}.  $^d$ Best redshift constraint currently available \citep[based on a spectral analysis of \textit{XMM-Newton} data, using the 6.7\,keV Fe emission line complex, ][]{Mantz2020}{}.
	$^e$ Observation design and data reduction followed the same procedures as described in \citet{Khullar2019}. More general results will be discussed in a future paper on high-$z$ spectroscopic measurements of SPT clusters.
	$^f$ We list the SZ mass recalculated at the updated redshift of the cluster.

	\end{threeparttable}
% 		\end{centre}
		
\end{table*}

We investigate nine massive and distant galaxy clusters at redshifts \mbox{$1.0 \lesssim z \lesssim 1.7$} detected by the SPT via their SZ signal. They were originally selected to have $z > 1.2$ according to the best redshift estimate available at the time. However, our analysis of more recent spectroscopic observations place the cluster SPT-CL{\thinspace}$J$0646$-$6236 at lower redshift, $z = 0.995$ (see also note $^e$ in Table \ref{tab:Cluster sample properties}). Therefore, only the remaining eight clusters constitute the complete sample of galaxy clusters at high redshifts \mbox{$z \geq 1.2$} with the strongest detection significance of \mbox{$\xi \geq 6$} from the 2500\,deg$^2$ SPT-SZ survey \citep[][see Table \ref{tab:Cluster sample properties} for cluster properties]{Bleem2015}{}. The sample has a median redshift of \mbox{$z_\mathrm{med} = 1.4$}. Our study represents the first homogeneous weak lensing study of a cluster sample of this size with a clean SZ-based selection function at this high-redshift regime. \citetalias{Bocquet2019} derive cosmological constraints with galaxy clusters from the 2500\,deg$^2$ SPT-SZ survey and provide updated redshift and SZ mass estimates for the SPT cluster sample, including the clusters studied here \citep[redshift updates for clusters relevant to this work are from ][]{Khullar2019,Mantz2020}. The SZ mass estimates incorporate a weak lensing mass calibration using data from \citetalias{Dietrich2019} and \citetalias{Schrabback2018}. 

The nine clusters in this work are also part of several previous studies. 
\citet{McDonald2017} examine \textit{Chandra} X-ray data for eight of these clusters and investigate the redshift dependency and compatibility with self-similar evolution of the ICM in a large sample of galaxy clusters. Their study includes an estimation of the positions of the cluster X-ray centres (see also Table \ref{tab:Cluster sample properties}) and the X-ray-based masses \citep[derived from the $M_\mathrm{gas}-M$ relation from ][]{Vikhlinin2009}, as well as density profiles and morphologies of the clusters. 
\citet{Ghirardini2021} investigate thermodynamic properties, for example, density, temperature, pressure, and entropy with combined \textit{Chandra} and \textit{XMM-Newton} X-ray observations of seven clusters in our sample and compare them with the corresponding properties of low-redshift clusters.
Additionally, \citet{Bulbul2019} include two of the clusters in their analysis of X-ray properties of SPT-selected galaxy clusters observed with \textit{XMM-Newton}. They constrain the scaling relations between the X-ray observables of the ICM (luminosity $L_\mathrm{X}$, ICM mass $M_\mathrm{ICM}$, emission-weighted mean temperature $T_\mathrm{X}$, and integrated pressure $Y_\mathrm{X}$), redshift, and halo mass. 
Further X-ray studies investigating astrophysical properties featuring one or more clusters from our sample include \citet{McDonald2013}, \citet{Sanders2018}, and \citet{Mantz2020}.
There have also been efforts to obtain precise spectroscopic redshifts for the majority of clusters in our sample \citep[][]{Stalder2013, Bayliss2014, Khullar2019, Mantz2020}{}, where some studies specifically investigate the galaxy kinematics and velocity distributions \citep[][]{Ruel2014,Capasso2019}{}.
Several multi-wavelength studies of cluster samples (including one or more clusters from our sample) with varying size investigate different cluster components such as the baryon content \citep[][]{Chiu2016,Chiu2018}{}, the properties, growth and star formation in brightest cluster galaxies \citep[BCGs, ][]{McDonald2016,DeMaio2020,Chu2021}{}, the mass-richness relation \citep[][]{Rettura2018}{}, environmental quenching of the galaxy populations in clusters \citep[][]{Strazzullo2019}{}, and AGN-feedback \citep[][]{Hlavacek-Larrondo2015,Birzan2017}{}.
The cluster SPT-CL{\thinspace}$J$2040$-$4451 was already studied in a weak lensing analysis by \citet{Jee2017}, using infrared imaging from the Wide Field Camera 3 (WFC3) on the \textit{HST} for shape measurements. We compare their analysis strategy and ours in detail in Sect. \ref{Sec:Discussion}.
    
%--------------------------------------------------------------------

\section{Data and data reduction}
\label{Sec:Data+Data reduction}

\subsection{HST ACS and WFC3 data}
\label{Sec:HSTData reduction}

We used high-resolution imaging from the \textit{HST} to measure galaxy shapes for the weak lensing analysis as detailed in \mbox{Sect. \ref{sec:shapes}}. The observational data analysed in our study were obtained during 
Cycles 19, 21, 23, and 24 as part of the SPT follow-up GO programmes 12477 (PI: F. High), 13412 (PI: T. Schrabback), 14252 (PI: V. Strazzullo), and 14677 (PI: T. Schrabback)
in the filters F606W and F814W with the ACS/WFC instrument and F110W with the WFC3/IR instrument.
We measured the shapes of galaxies for our weak lensing analysis in the F606W and F814W imaging, which have a field of view of $202'' \times 202''$ at a pixel scale of $0\farcs05/\mathrm{pixel}$. The ACS (Advanced Camera for Surveys) observations were obtained in a single pointing except for SPT-CL{\thinspace}$J$0205$-$5829 for which an additional larger $2\times 2$ mosaic was obtained in F606W as part of GO programme 12477.
The field of view of WFC3 is $136'' \times 123''$ with a pixel scale of roughly $0\farcs128/\mathrm{pixel}$ (the pixels are not exactly square shaped). We observed $2\times2$ mosaics in the F110W filter (with partial overlap between pointings), which roughly match the field of view of the ACS observations.  We used the observations in the F110W filter exclusively for the photometric selection of the background galaxies carrying the weak lensing signal. The integration times range between \mbox{2.3 and 5.5\,ks} (F606W), \mbox{3.3 and 4.9\,ks} (F814W), and \mbox{2.4\,ks} (F110W, spread out over a $2\times2$ mosaic to reach a minimum depth of \mbox{0.6\,ks} over the full ACS footprint) (see \mbox{Table \ref{tab:exposure times}}).

\begin{table*}
	\centering
	\caption{Summary of the integration times, image quality, and depth from our observations with \textit{HST}/ACS, \textit{HST}/WFC3, and VLT/FORS2. }
	\label{tab:exposure times}
	\begin{threeparttable}
	\begin{tabular}{lccc  ccc  ccc  ccc} 
		\hline
		\hline

		   & & F606W & & & F814W  & & & F110W  & &  & $U_\mathrm{HIGH}$  &  \\
		 \cmidrule(lr){2-4}\cmidrule(lr){5-7}\cmidrule(lr){8-10}\cmidrule(lr){11-13}
		Cluster name & $t_\mathrm{exp}$& IQ & depth & $t_\mathrm{exp}$ & IQ  & depth & $t_\mathrm{exp}$ $^b$  & IQ  & depth$^b$ & $t_\mathrm{exp}$  & IQ  & depth  \\
		 & [ks] & [$''$] & [mag] & [ks] & [$''$] & [mag] & [ks] & [$''$] & [mag] & [ks]  & [$''$]  & [mag]  \\
		\hline
		SPT-CL{\thinspace}$J$0156$-$5541 & 5.5  & 0.10 & 27.0 & 4.9   & 0.10  & 26.6 & 0.6  & 0.29 & 26.3 & 4.8  & 0.73 & 26.9\\
		SPT-CL{\thinspace}$J$0205$-$5829 & 3.7$^a$  & 0.10 & 27.1$^a$ & 3.7  & 0.08  & 26.5 &  0.6  & 0.29 & 26.3 & 4.8  & 0.85 & 26.8\\
		SPT-CL{\thinspace}$J$0313$-$5334 &  3.7  & 0.10 & 26.9 &  3.7   & 0.09  & 26.1 & 0.6  & 0.29 & 26.3 &  4.8  & 0.80 & 27.1\\
		SPT-CL{\thinspace}$J$0459$-$4947 &  2.3  & 0.11 & 26.7 &  4.8   & 0.10  & 26.5 &  0.6  & 0.28 & 26.3 &  6.0  & 0.81 & 26.9\\
		SPT-CL{\thinspace}$J$0607$-$4448 &  2.3  & 0.10 & 26.7 &  4.8   &  0.10 & 26.3 &  0.6  & 0.28 & 26.4 &  4.8  & 0.97 & 26.4\\
		SPT-CL{\thinspace}$J$0640$-$5113 &  5.6  & 0.10 & 26.7 &  3.3   &  0.10 & 26.2 &  0.6  & 0.26 & 26.1 &  2.4  & 0.97 & 26.3\\
		SPT-CL{\thinspace}$J$0646$-$6236 &  4.0  & 0.10 & 26.8 &  4.0   & 0.10  & 26.1 &  0.6  & 0.27 & 26.1 &  4.8  & 1.07 & 26.3\\
		SPT-CL{\thinspace}$J$2040$-$4451 &  2.1  & 0.10 & 26.6 &  4.8   &  0.10 & 26.1 &  0.6  & 0.28 & 26.1 &  4.8  & 0.88 & 26.5\\
		SPT-CL{\thinspace}$J$2341$-$5724 &  5.3  & 0.10 & 26.5 & 4.8   &  0.10 & 26.2 &  0.6  & 0.29 & 26.1 &  4.8  & 0.92 & 26.9\\
        HUDF &  $-$ & $-$ & $-$ & $-$ & $-$ & $-$ & $-$ & $-$ & $-$ & 6.6  & 1.03 & 26.6\\
		\hline

	\end{tabular}

	\textbf{Notes.} 
	For the image quality (IQ), we report the full width at half maximum of the PSF, based on measurements with \texttt{Source Extractor}. The depth corresponds to $5\sigma$ limiting magnitudes, computed from the standard deviation of 1000 non-overlapping apertures without flux from detected sources. We used apertures with diameters of $0\farcs7$ for \textit{HST} bands and $1\farcs2$ for $U_\mathrm{HIGH}$.\newline
	$^a$ For the cluster SPT-CL{\thinspace}$J$0205$-$5829 a $2\times2$ ACS mosaic from GO programme 12477 and one single ACS pointing from GO programme 14677 are available in the F606W band. We have eight overlapping exposures in the region with the biggest overlap with our observations in the other bands. We report the integration time and depth based on this region. \\
	$^b$ The F110W stacks are mosaics of eight exposures. The highest/intermediate/lowest depth is achieved, where eight/four/two exposures overlap, respectively. Since regions with only two overlapping exposures make up the most area in the stacks, we report integration times and depths equivalent to two exposures. 

	\end{threeparttable}
\end{table*}

We performed the basic image reduction steps for the \textit{HST}/ACS imaging data with the ACS calibration pipeline \texttt{CALACS}\footnote{\url{https://hst-docs.stsci.edu/acsdhb}, Chapter 3}. However, we deviated from the standard processing steps regarding the correction for charge transfer inefficiency (CTI). As in previous studies by \citetalias{Schrabback2018} and \citetalias{Schrabback2021}, we performed the CTI correction with the algorithm by \citet{Massey2014} and applied it to both the \textit{HST}/ACS imaging data and the respective dark frames. Furthermore, we performed a quadrant-based sky background subtraction, improved the bad pixel masks by manually flagging satellite trails and cosmic ray clusters, and computed accurately normalised RMS maps following the prescription by \citet{Schrabback2010}.\\
The \textit{HST}/WFC3 imaging data reduction was performed similarly to the \textit{HST}/ACS imaging data reduction. We downloaded the pre-reduced \texttt{flt} frames, which had already undergone basic image processing via the WFC3 calibration pipeline \texttt{calwf3}\footnote{\url{https://hst-docs.stsci.edu/wfc3dhb}, Chapter 3}, 
but we did not perform a quadrant-based sky background subtraction because it is not suitable for the parallel read-out mechanism of WFC3. Instead, we used \texttt{Source Extractor} \citep[version 2.23.1, ][]{BertinArnouts1996} to obtain a background model, which we subtracted. This allowed us to account properly for gradients in the background level. These occasionally occur in particular due to a variable airglow line of He I in the lower \mbox{exosphere} at 10830\,\AA, which mostly affects the bands F105W and F110W (see Chapter 7.9.5 of the WFC3 instrument handbook\footnote{\url{https://hst-docs.stsci.edu/wfc3ihb}} and WFC3 ISR 2014-03). We did not perform a CTI correction for the WFC3 data, as they are not affected by this.

Subsequently to the initial data reduction, we employed the software \texttt{DrizzlePac}\footnote{\url{https://www.stsci.edu/scientific-community/software/drizzlepac.html}} for aligning and combining \textit{HST} images in particular using the tasks \texttt{TweakReg} and \texttt{AstroDrizzle}. This typically involved combining 4 to 10 exposures for \textit{HST}/ACS imaging or 8 exposures in a $2\times2$ mosaic for WFC3 imaging. 
For the stacking with \texttt{AstroDrizzle}, we used the \texttt{lanczos3} kernel at the native pixel scale of $0\farcs05$ ($0\farcs128$) of the ACS (WFC3) images to distribute the flux onto the output image. Additionally, we employed the RMS image as the weighting image. We produced the stack for the imaging in the F606W band first and subsequently used this stack as the astrometric reference image for the stacks in the F814W and F110W bands to ensure optimal astrometric alignment between the stacks.

\subsection{Very Large Telescope (VLT) FORS2 data}
\label{Sec:VLTData reduction}

We used additional observations from VLT/FORS2 in the $U_\mathrm{HIGH}$ passband obtained as part of the ESO programme 0100.A-0204(A) (PI: Schrabback) between November 18 and November 20, 2017. Together with the \textit{HST} imaging, these observations facilitated a robust photometric selection of background galaxies. The images were taken with the two blue-sensitive $2\mathrm{k}\times4\mathrm{k}$ E2V CCDs in standard resolution with $2\times2$ binning, providing observations over a field of view of $6\farcm8 \times 6\farcm8$ at a pixel scale of $0\farcs25/\mathrm{pixel}$. We observed the nine galaxy clusters in our sample and additionally one pointing centred on the \textit{Hubble Ultra Deep Field} \citep[HUDF, ][]{Beckwith2006}, which we used to assess the photometric calibration of the $U_\mathrm{HIGH}$ band. The integration times per cluster ranged between \mbox{2.4\,ks} and \mbox{6.6\,ks} (see Table \ref{tab:exposure times}).

We reduced the data with the software \texttt{THELI}\footnote{\url{https://www.astro.uni-bonn.de/theli/}} \citep{Erben2005,Schirmer2013}. We performed a bias subtraction, flat-field correction, and a subtraction of a background model. The latter was obtained by taking advantage of the dither pattern applied between exposures. The images were median combined, resulting in the background model. This enabled us to distinguish features at a fixed detector position from sky-related signals. The background model was rescaled to the illumination level of the individual exposures and then subtracted from them.  We applied a sky background subtraction using \texttt{Source Extractor} \citep{BertinArnouts1996}. We obtained the astrometric calibration based on the Gaia DR1 catalogue \citep{GaiaCollab2016a, GaiaCOllab2016b} as reference. Finally, the images were co-added. We did not match the astrometry of the VLT observations to the one of the \textit{HST} data. Checking for offsets between \textit{HST} and VLT astrometrty with \texttt{Source Extractor} detected sources, we found small offsets of the order of 0.1\,arcsec, which we corrected for in the photometric analysis.

\subsection{Photometry}
\subsubsection{Photometric measurements with LAMBDAR}

We performed photometric measurements on our fully reduced images with the Lambda Adaptive Multi-Band Deblending Algorithm in R \citep[\texttt{LAMBDAR}\footnote{\url{https://github.com/AngusWright/LAMBDAR}}, ][]{Wright2016}. This algorithm can perform consistent and matched aperture photometry across images with varying pixel scales and resolutions. Therefore, it is ideally suited for our analysis, which requires accurate and precise colour measurements between the \textit{HST} and VLT imaging with very different resolutions. In the following, we give a brief summary of the \texttt{LAMBDAR} algorithm. We refer the reader to \citet{Wright2016} for a more in-depth description.

\texttt{LAMBDAR} requires at least two inputs: a FITS image and a catalogue of object locations and aperture parameters. Additionally, we provide a point-spread function (PSF) model for the FITS image. These files are read in as the first step, then the aperture priors from the catalogue are transferred onto the same pixel grid as the input FITS image, using the image's astrometric solution (stored in the FITS header). Subsequently, the aperture priors are convolved with the input PSF, and object deblending is executed based on the convolved aperture priors. Images are deblended via multiplication with a deblending function.
For this, it is assumed that the total flux in a pixel equals the sum of the fluxes from sources with aperture models overlapping that pixel. The flux per source is distinguished with the help of the deblending function. This function is calculated using the second assumption that the PSF convolved aperture model is a tracer of the emission profile of each source.
Taking into account the estimation of the local sky-backgrounds and noise correlation using random/blank apertures, \texttt{LAMBDAR} calculates the object fluxes with the help of the deblended convolved aperture priors. Here, the code accounts for aperture weighting and/or missed flux through an appropriate normalisation of fluxes. Finally, flux uncertainties in relation to all of the above steps are determined.

For our purposes, using \texttt{LAMBDAR} has two main advantages. Firstly, we can comfortably perform matched aperture photometry across our images with varying PSF sizes between $0\farcs08$ and $1\farcs07$. Secondly, the prior aperture definitions derived from high-resolution optical imaging allow for deblending of sources leading to more accurate flux measurements, in particular in comparison to conventional fixed aperture photometry. 

For each cluster, we ran \texttt{Source Extractor} on the F606W image to obtain the input catalogue with object locations and aperture parameters. We set the detection and analysis thresholds to $1.4\sigma$. We required a minimum of 8 pixels for a source detection and set the deblending threshold to 32 with a minimum contrast parameter of 0.005. Before the detection, the images were smoothed with a Gaussian filter of 5 pixels with a full width at half maximum (FWHM) of 2.5 pixels. We checked for residual shifts in the astrometry of our images with respect to the F606W detection image and corrected for them with a linear shift if necessary to avoid biases in the flux measurements with \texttt{LAMBDAR}. For the \textit{HST} images, we used \texttt{TinyTim} \citep{Krist2011} to obtain a PSF model for the photometric analysis. For the ACS images (i.e. in F606W and F814W), we looked up the average focus from the duration of the observation at the \textit{HST} Focus Model tool\footnote{\url{http://focustool.stsci.edu/cgi-bin/control.py}}. Since this tool does not offer an estimate for WFC3/IR (i.e. for the images in F110W), we assumed a focus offset of 0.0 microns as default\footnote{To cross-check this assumption, we measured the photometry with an alternative PSF model with a very different focus offset of 4.0 microns. We found that both measurements differ by \mbox{0.001\,mag} (median), which is negligible for our purposes.}. We used the central chip position as the position of reference for the estimation of the PSF model. In the case of the ACS instrument with two chips, we took the central pixel of chip 1 as a reference.
For our VLT/FORS2 images, we obtained a PSF model with the help of the software \texttt{PSFEx} \citep{Bertin2011}. 

Some of our fully reduced images exhibited slight residual gradients in the background level. Therefore, we performed an initial run with \texttt{Source Extractor} to obtain a background-subtracted image. We used these as the FITS input images to be analysed with \texttt{LAMBDAR}. 

\subsubsection{Photometric zeropoints}
\label{Sec: Photometric zeropoints}

\begin{figure*}
\centering

	\includegraphics[width=1.10\columnwidth]{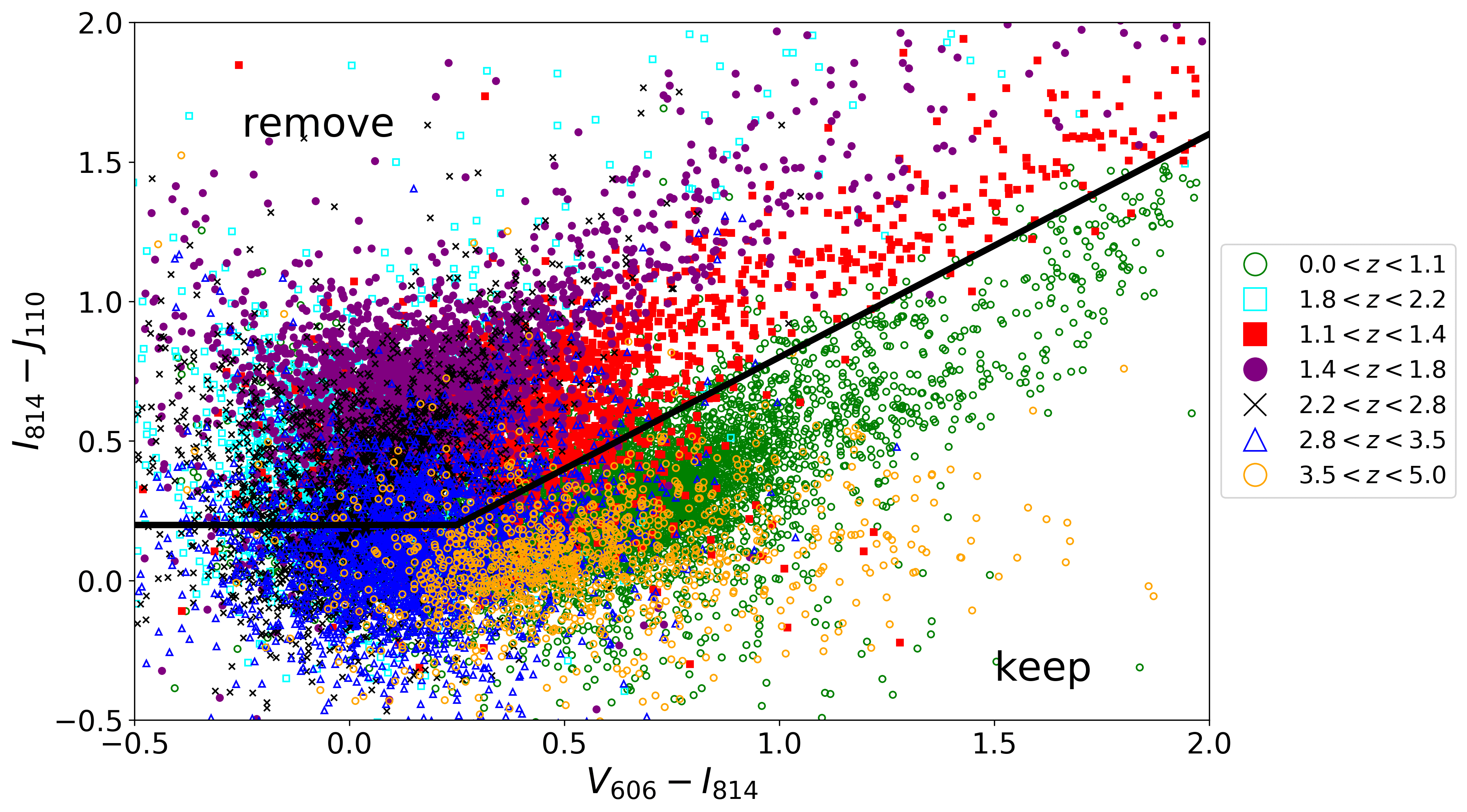}
	\includegraphics[width=0.90\columnwidth]{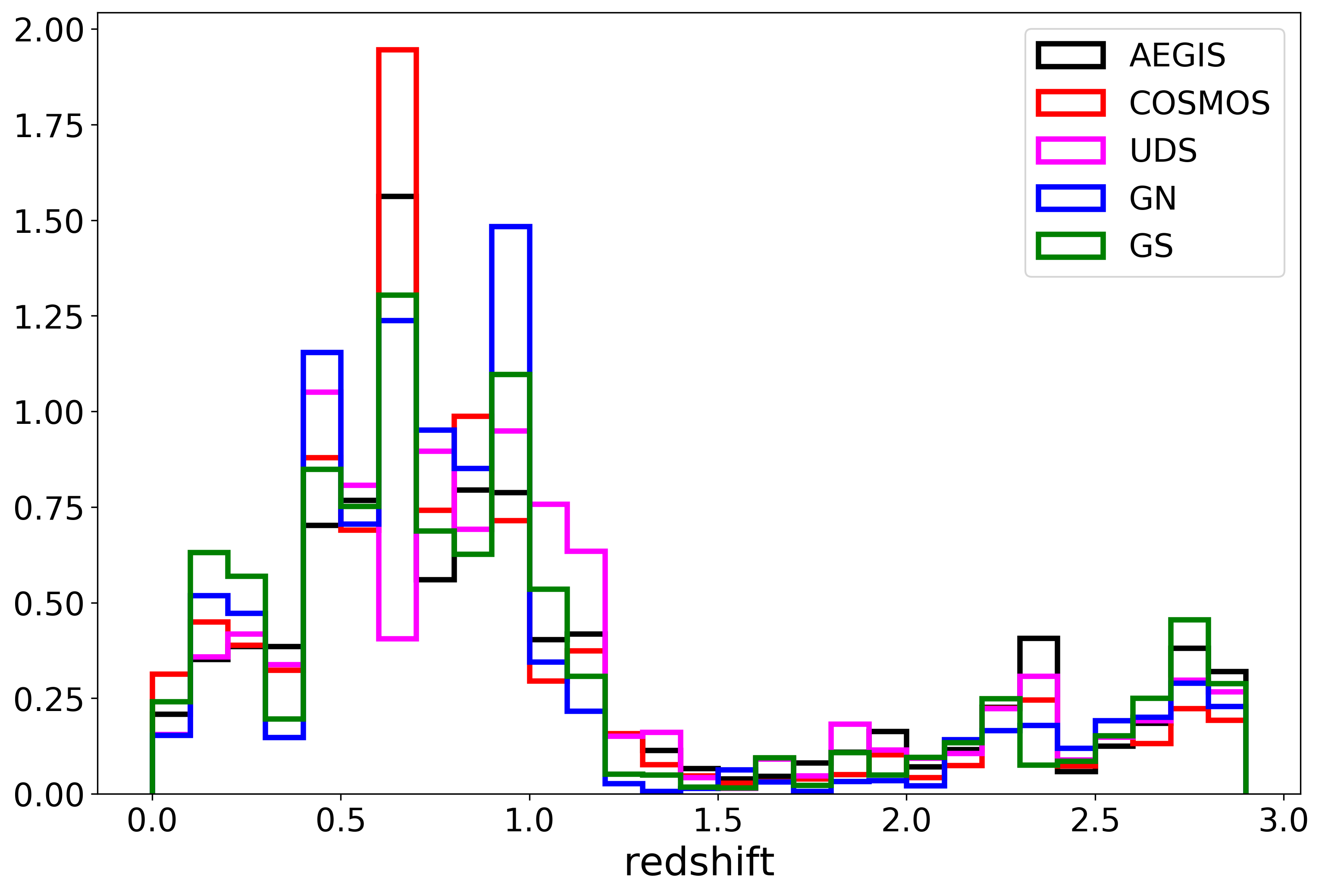}

    \caption{Removal of galaxies in the cluster redshift regime from the galaxy locus at magnitudes of  \mbox{$24.2 < V_{606} < 27.0$}. \textit{Left:} Cut in the $VIJ$ plane to remove galaxies with photometric redshifts $1.2 \lesssim z \lesssim 1.7$ according to the catalogues by \citet{Skelton2014} (that is galaxies in the regime of cluster redshifts of the sample studied here), illustrated for galaxies from the GOODS-South field with photometry from \citet{Skelton2014}. Red and purple symbols roughly correspond to galaxies in the cluster redshift regime. \textit{Right:} Redshift distribution of galaxies in our chosen galaxy locus from the five CANDELS/3D-HST fields. }
    \label{Fig:Cuts to remove only cluster gals}
\end{figure*}

The photometric calibration for the \textit{HST} bands is straightforward. We obtained a photometric zeropoint (ZP) for each coadd from the header keywords \texttt{PHOTFLAM} and \texttt{PHOTPLAM}:
\begin{equation}
    \begin{split}
       \mathrm{ZP}_\mathrm{AB} = & -2.5 \log_{10}(\mathrm{PHOTFLAM}) \\ 
              & - 5.0 \log_{10}(\mathrm{PHOTPLAM}) -2.408\,. 
    \end{split}
\end{equation}
\texttt{PHOTFLAM} is the inverse sensitivity, which facilitates the transformation from an instrumental flux in units of electrons per second to a physical flux density and \texttt{PHOTPLAM} denotes the pivot wavelength in units of \AA\footnote{\url{https://www.stsci.edu/hst/instrumentation/acs/data-analysis/zeropoints}}. 
Afterwards, we accounted for Galactic extinction with the extinction maps by \citet{2011ApJ...737...103S}\footnote{obtained from the website \url{https://irsa.ipac.caltech.edu/applications/DUST/}}. 

The challenge in the photometric calibration of the $U_{\mathrm{HIGH}}$ band is the lack of an adequate reference catalogue with well-calibrated $U$ band magnitudes for our cluster fields. In such a case, a common method for calibration is to use a stellar locus \citep{High2009}. It is based on the assumption that stars occupy a well-defined region, the stellar locus, in colour-colour space, independent of the line of sight. Then, the photometry can be calibrated by matching the photometry of stars in an observation to the universal stellar locus. However, we found that direct use of a stellar locus does not work well for our analysis due to the limited number of stars in the small fields of view. Additional large scatter resulted in substantial uncertainties of the stellar locus approach.
We, therefore, developed a calibration strategy based on a galaxy locus,
where we made use of the fact that galaxies have a characteristic distribution in colour-colour space, similar to stars occupying the stellar locus.
We identified a reference galaxy locus from the 3D-HST photometric catalogues as presented in \citet{Skelton2014}. They summarised photometric measurements in the five CANDELS/3D-HST fields (AEGIS, COSMOS, GOODS-North [abbreviated GN], GOODS-South [abbreviated GS], and UDS) over a total area of $\sim 900$\,arcmin$^2$. 
Among others, this includes the following bands relevant for our reference galaxy locus: the \textit{HST} bands F606W and F814W and $U$ bands from various instruments such as CFHT/MegaCam (AEGIS, COSMOS, and UDS), KPNO 4\,m/Mosaic (GOODS-North), and VLT/VIMOS (GOODS-South). We describe in \mbox{Sect. \ref{Sec:common photometric system}} how we accounted for the differences in these effective band-passes.
Compared to the CANDELS/3D-HST fields our cluster fields are overdense at the cluster redshift, changing the local galaxy distribution in colour-colour space. To account for this, we applied a preselection, which used the well-calibrated \textit{HST}-only colours to remove galaxies at the cluster redshift (see \mbox{Fig. \ref{Fig:Cuts to remove only cluster gals}}). In addition, the galaxy distribution varies locally due to line of sight variations. We reduced the impact of these by limiting the calibration with the galaxy locus to relatively faint galaxies in the regime \mbox{$24.2 < V_{606} < 27.0$}. We note that we optimised the calibration to be most accurate in this magnitude regime since it is the regime we used for the selection of background source galaxies (see Sect. \ref{Sec: full section, BG gal selection + redshift distrib}).
Together, this allowed us to calibrate $U-V_{606}$ colour estimates in the cluster fields by matching the galaxy distribution of the $VIJ$-selected galaxies in the \mbox{$V_{606} - I_{814}$} versus \mbox{$U-V_{606}$} colour-colour space.

For the calibration, we first accounted for Galactic extinction with the extinction maps by \citet{2011ApJ...737...103S}.
We then smoothed the distribution of the galaxies in the $UVI$ colour-colour space with a Gaussian kernel\footnote{using scipy.stats.gaussian\_kde in python} both for the galaxies of the reference galaxy locus and the galaxies in our observation. We identified the peak position of the highest density and applied a shift to the $U_{\mathrm{HIGH}}$ magnitudes according to the difference in $U - V_{606}$ of the peak positions. We quantified and propagated the statistical uncertainty of 0.08\,mag of our colour calibration scheme (see Appendix \ref{Appendix:ZP robustness with gal locus} for a robustness test of the $U_{\mathrm{HIGH}}$ band zeropoint calibration with the help of the reference galaxy locus; see Table \ref{Tab:Errorbudget of photometry,beta} for the effect of this statistical uncertainty on the average geometric lensing efficiency.)

\subsubsection{Defining a common photometric system}
\label{Sec:common photometric system}

\begin{figure}
	\includegraphics[width=\columnwidth]{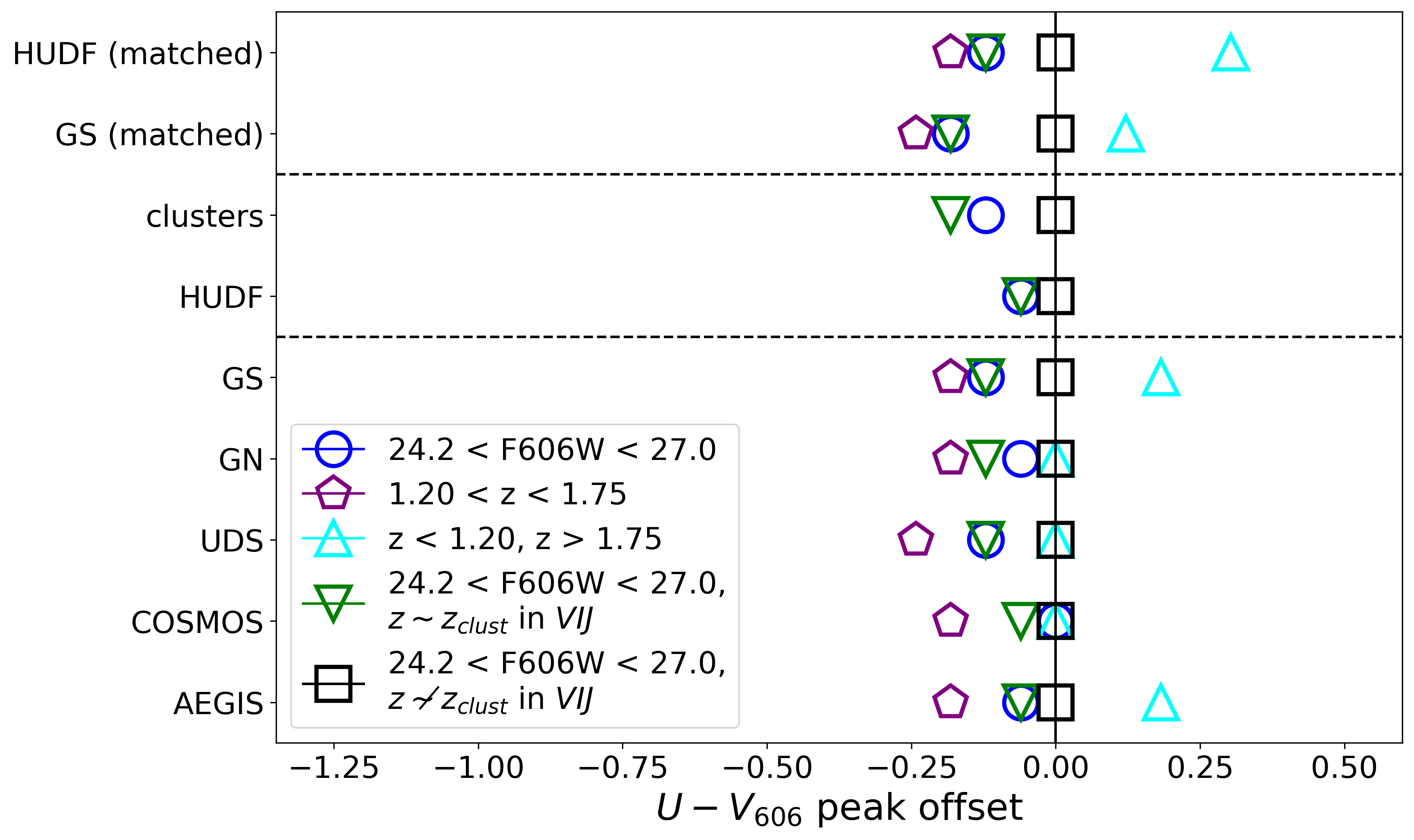}

    \caption{Offsets between different populations of galaxies and the reference galaxy locus (galaxies with \mbox{$24.2 < V_{606} < 27.0$}, where galaxies at the cluster redshifts $1.2 \lesssim z \lesssim 1.7$ are excluded according to a cut in the $VIJ$ colour plane; represented by black squares). Overall the populations exhibit quite similar offsets in \mbox{$U-V_{606}$} colour despite relying on different $U$ bands.  \textit{Top section:} Comparison of directly matched galaxies in the HUDF region based on our measurements and the catalogue in GOODS-South by \citet{Skelton2014}.  \textit{Mid section:} Comparison of cluster fields (measurements from all nine cluster fields combined) and our measurements in the HUDF area, where we have $U_\mathrm{HIGH}$ imaging. Since we do not have photometric redshifts available, the populations relying on these are missing (purple pentagons and cyan triangles). \textit{Bottom section:} Comparisons for five CANDELS/3D-HST fields.  }
    \label{Fig:UV-offsets in galaxy populations}
\end{figure}

When we investigate colour cuts for a suitable selection of background galaxies, we need to make sure to work in a consistent photometric framework. Regarding the $U$ bands, we have measurements from four different instruments at hand: $U_\mathrm{HIGH}$ from VLT/FORS2 (our observations), $U_\mathrm{MEGACAM}$ from CFHT/MegaCam, $U_\mathrm{KPNO}$ from KPNO 4\,m/Mosaic, and $U_\mathrm{VIMOS}$ from VLT/VIMOS \citep[the latter three filters are employed in different CANDELS/3D-HST fields in ][]{Skelton2014}. All of these have different effective filter curves.
We, therefore, had to make sure that we employed these different bands to select consistent source populations, in particular regarding the \mbox{$U-V_{606}$} colour. Comparing the \mbox{$U-V_{606}$} colour of these populations, we found that there are small offsets among the CANDELS/3D-HST fields. We quantified these by identifying the peak position of the galaxy loci after smoothing the distribution with a Gaussian kernel (galaxies with \mbox{$24.2 < V_{606} < 27.0$}, where galaxies in the cluster redshift regime $1.2 \lesssim z \lesssim 1.7$ are excluded according to a cut in the $VIJ$ colour plane; see Sect. \ref{Sec: Photometric zeropoints}). We applied a shift to the $U$ bands to make the peak positions coincide with the peak position of the galaxy locus in GOODS-South as an anchor. We used this field as an anchor because we have observations of our own in $U_\mathrm{HIGH}$ in the HUDF situated within GOODS-South. We list the applied shifts in Table \ref{Tab:Galloc_comparisons} in the Appendix. 
As a cross-check, we compared the peak positions in the \mbox{$U-V_{606}$} colour distribution for differently selected galaxy subsamples in \mbox{Fig. \ref{Fig:UV-offsets in galaxy populations}}. Here, we generally found good agreement. For example, for the full population of galaxies with \mbox{$24.2 < V_{606} < 27.0$}, we measured a standard deviation of the density peak positions between the five CANDELS/3D-HST fields of 0.045\,mag.
We conclude that the photometry is sufficiently comparable as a basis for the selection of background galaxies (we summarise systematic and statistical uncertainties connected to the photometry at the end of Sect. \ref{Sec: Background galaxy selection}). 
In addition to these considerations for the $U$ bands, we used \textit{HST} bands for which we have available observations for our cluster fields, that is, F606W, F814W, and F110W. Since not all reference catalogues have magnitude information on the galaxies in all of these bands, we needed to apply a few interpolations to estimate the fluxes and magnitudes of galaxies in our photometric system of filters. In this case, we performed an interpolation based on the closest available filters in effective wavelength, where one filter is redder (R) and one is bluer (B) than the missing filter (X):
\begin{equation}
    \begin{split} 
    	F_\mathrm{X} &= s  (\lambda_\mathrm{eff,X} - \lambda_\mathrm{eff,B}) + F_\mathrm{B}\,,\\
    	m_\mathrm{X} &= -2.5\log_{10}(F_\mathrm{X}) + \mathrm{ZP}\,,\\
    	\mathrm{with} \quad s &= \frac{(F_\mathrm{R} - F_\mathrm{B})}{(\lambda_\mathrm{eff,R} -\lambda_\mathrm{eff,B})}\,,
    \end{split}
\end{equation}
where $F$ denotes the flux, $m$ denotes the magnitude, $\mathrm{ZP}$ is the zeropoint (it is fixed to \mbox{$\mathrm{ZP} = 25.0$} for all bands in the \citet{Skelton2014} CANDELS/3D-HST photometric catalogues), and $\lambda_\mathrm{eff}$ is the effective wavelength of the respective filter. In a catalogue that covers the sources in all bands, we can gauge how well the interpolation typically represents the measured magnitude. 
Overall, there is a good match between the interpolated and the measured magnitudes. We do, however, see that the interpolation becomes increasingly noisy and asymmetric for fainter magnitudes. This is likely related to the (potentially different) depths of the available bands.

None of the available reference catalogues provides measurements in the band F110W. Options for interpolation are to use a combination of either F105W and F125W, or F850LP and F125W, or F814W and F125W. 
Depending on the method used, we found that a small median offset of the order of $0.04$\,mag with a standard deviation of $0.07$\,mag can be introduced. We did not attempt to correct for such differences but we investigated the impact of systematic photometric offsets on the estimate of the average lensing efficiency in Appendix \ref{Appendix:Impact of syst. shifts in photom}, finding that the impact of such a systematic offset can well be neglected given our current statistical uncertainties. We also checked how well our photometry compares to measurements from \citet{Skelton2014} in Appendix \ref{Appendix:Comparison of S14 and LAMBDAR photometry}. From this, we concluded that slight offsets in photometry can occur, and we included the expected uncertainties in the overall error budget of our analysis (summarised at the end of Sect. \ref{Sec: Background galaxy selection}).

%--------------------------------------------------------------------

\section{Photometric selection of source galaxies and estimation of the source redshift distribution}
\label{Sec: full section, BG gal selection + redshift distrib}

For a robust weak lensing analysis, it is important to preferentially select the galaxies at redshifts higher than the cluster redshifts. Only these galaxies carry the weak lensing signal that we are interested in. We need to estimate the expected source redshift distribution of the selected galaxies to quantify the average geometric lensing efficiency $\langle \beta \rangle$  defined as 
\begin{equation}
    \langle \beta \rangle = \frac{\sum \beta(z_i)w_i}{\sum w_i}  \,,
\end{equation}
with the shape weights $w_i$ \citep[see ][ and \mbox{Sect. \ref{sec:shapes}}]{Schrabback2018} and 
\begin{equation}
    \beta = \frac{D_\mathrm{ls}}{D_\mathrm{s}} H(z_\mathrm{s} - z_\mathrm{l}) \,,
\end{equation}
where $D_\mathrm{l}$, $D_\mathrm{s}$, and $D_\mathrm{ls}$ denote the angular diameter distances to the lens at redshift $z_\mathrm{l}$, the source at redshift $z_\mathrm{s}$, and between lens and source, respectively. The Heavyside step function is defined as $H(x) = 1$ if $x>0$ and  $H(x) = 0$ if $x \leq 0$. 
It is sufficient to estimate the averages $\langle \beta \rangle$ and $\langle \beta^2 \rangle$ to tie the measured weak lensing shear signal to the cluster mass \citep[e.g. ][]{Bartelmann2001}.

A straightforward but prohibitively observationally expensive way to identify the background galaxies is based on their spectroscopic redshifts. High-quality photometric redshifts can also be helpful if examined carefully for systematic outliers. Such redshift information is, however, not available for the galaxies in our observed cluster fields. Instead, we aim to use only the photometry from our observations to identify background galaxies. For this, we need reference catalogues of galaxies providing redshift and magnitude information in different bands. This allows us to understand how to distinguish background galaxies from contaminating foreground and cluster galaxies solely based on their colours. This is a commonly used strategy for weak lensing studies covering various redshift regimes \citep[e.g. ][]{Klein2019-APEX-WL,Schrabback2018,Schrabback2021}. In the following section, we first describe the reference catalogues used in this work. After that, we present suitable cuts in colour space to preferentially select background galaxies for the weak lensing analyses. These cuts identified in photometric redshift reference catalogues can be safely applied to the cluster fields, because gravitational lensing is a colour-indifferent effect.

\subsection{Redshift catalogues}
\label{Sec:Referenct Redshift cats}

\subsubsection{UVUDF}

The HUDF is a region of the sky that has been studied extensively both spectroscopically and in various photometric filters by the \textit{HST}. \citet[][henceforth \citetalias{Rafelski2015}]{Rafelski2015} conducted a joint analysis of imaging ranging from near-ultraviolet (NUV) bands F225W, F275W, and F336W \citep[UVUDF,][]{Teplitz2013}, over optical bands F435W, F606W, F775W, and F850LP \citep{Beckwith2006}, to near-infrared (NIR) bands F105W, F125W, F140W, and F160W \citep[UDF09 and UDF12, ][]{Oesch2010a,Oesch2010b,Bouwens2011,Koekemoer2013, Ellis2013}. These data sets cover an area of  $12.8$\,arcmin$^2$, but only $4.6$\,arcmin$^2$ have full NIR coverage. \citetalias{Rafelski2015} provide photometric redshifts obtained with the code \texttt{BPZ} \citep[][]{Benitez2000}, which are highly robust due to the exquisite depth and high wavelength coverage of the data sets \citep[e.g. demonstrated in][who found  a median of  $|(z_\mathrm{MUSE}-z_\mathrm{p}) / (1 + z_\mathrm{MUSE})| < 0.05$ from a comparison of photometric redshifts to high quality redshifts from the MUSE integral field spectrograph]{Brinchmann2017}. 
Given their accuracy, the \citetalias{Rafelski2015} photo-$z$s provide an important benchmark for our computation of the average lensing efficiency. However, the small area covered in the sky leads to a substantial impact of sampling variance. Consequently, we also need to incorporate other data sets, which are shallower but cover a larger footprint in the sky (see Sect. \ref{Sec:3D-HST cat description}).

\subsubsection{3D-HST}
\label{Sec:3D-HST cat description}     
    
\citet[][henceforth \citetalias{Skelton2014}]{Skelton2014} present catalogues with photometric measurements in filters covering a wide wavelength range and photometric redshifts for galaxies from the CANDELS/3D-HST fields over a total area of $\sim 900$\,arcmin$^2$. Their aim is to homogeneously combine various data sets available for these fields. Firstly, this includes the Cosmic Assembly Near-infrared Deep Extragalactic Legacy Survey \citep[CANDELS,][]{Grogin2011,Koekemoer2011}. It is an imaging survey conducted with \textit{HST}/WFC3 and \textit{HST}/ACS in five fields of the sky, namely AEGIS, COSMOS, GOODS-North, GOODS-South, and UDS. Secondly, the 3D-HST program \citep{Brammer2012} provides slitless spectroscopy obtained with the WFC3 G141 grism for galaxies across nearly 75 per cent of the CANDELS area and thus includes redshifts and spatially resolved spectral lines. Additionally, the WFC3 G141 grism spectroscopy data products are presented in \citet{Momcheva2016}, who also developed software to optimally extract spectra for the objects from the \citetalias{Skelton2014} photometric catalogues. 
\citetalias{Skelton2014} combined the photometric data sets from the CANDELS and 3D-HST programmes with available ancillary data sets in the five CANDELS/3D-HST fields by using a common WFC3 detection image, conducting consistent PSF-homogenised aperture photometry, and estimating photometric redshifts and redshift probability distributions with the code \texttt{EAZY} \citep[][]{Brammer2008}. The  \citetalias{Skelton2014} photometric redshift catalogues form an excellent basis to estimate the redshift distribution for our weak lensing study. They cover a large area on the sky distributed over five independent lines-of-sight. This helps to combat sampling variance when estimating the average lensing efficiency. Additionally, the wide wavelength coverage, especially including deep NIR observations, is particularly valuable for robust redshift measurements out to high redshifts, as required for this study. 

However, \citetalias{Schrabback2018} and \citet[][henceforth \citetalias{Raihan2020}]{Raihan2020} show that the photometric redshifts by \citetalias{Skelton2014} suffer from catastrophic outliers, which can significantly bias weak lensing mass measurements. Through the comparison of photometric redshift measurements from \citetalias{Skelton2014} and \citetalias{Rafelski2015}, \citetalias{Raihan2020} found that these outliers led to a systematic underestimation of the mean geometric lensing efficiency by $-13.2$ per cent (for clusters at a redshift of 0.9) with a catastrophic outlier fraction of 5\,per cent. \citetalias{Raihan2020} were able to mitigate this by recomputing the photometric redshifts using the code \texttt{BPZ} instead of \texttt{EAZY}. In particular, the interpolation of the implemented spectral energy distribution (SED) template set helped reduce the bias\footnote{When recomputing the photo-$z$s, \citetalias{Raihan2020}
employed an approximately homogeneous subset of broad-band filters (between $U$ and $H$ band), which are available for all five CANDELS fields. Since they dropped additional bands, this may increase the scatter in some of the photo-$z$ estimates compared to the \citetalias{Skelton2014} catalogue. However, for our analysis it is more important to have accurate estimates of the overall redshift distribution of colour-selected high-$z$ lensing source galaxies, as provided by the \citetalias{Raihan2020} catalogues.}. 
For our weak lensing study, we used the updated \citetalias{Raihan2020} photometric redshift catalogues in the five CANDELS/3D-HST fields to estimate the average redshift distribution and lensing efficiency of our samples of selected background galaxies.

Additionally, \citetalias{Schrabback2018} found some systematic deviations between the \citetalias{Rafelski2015} photometric redshifts and the grism redshifts \citep{Brammer2012,Momcheva2016}. Upon revisiting this comparison, now including MUSE spectroscopic redshifts \citep[][see Sect. \ref{Sec:MUSE} below for details]{Inami2017}, \citetalias{Raihan2020} identified the affected redshift regimes and corrected the respective bias by subtracting the median offset. This bias amounts to 0.081 (0.162) for the photo-$z$ regime \mbox{$1.0 < z < 1.7$} \mbox{($2.6 < z < 3.2$)}. The resulting `fixed' redshift catalogues do not suffer from issues with catastropic redshift outliers and are denoted as R15\_fix catalogues.

\subsubsection{HDUV}

The Hubble Deep UV Legacy Survey \citep[HDUV,][henceforth \citetalias{Oesch2018}]{Oesch2018} is an imaging programme, which expands on the \citetalias{Skelton2014} catalogues with deeper UV observations in the WFC3/UVIS bands F275W and F336W. It targets $\sim 100$\,arcmin$^2$ within the GOODS-North and GOODS-South fields. \citetalias{Oesch2018} conducted photometry consistent with \citetalias{Skelton2014} regarding the detection image and flux measurements and recalculated photometric redshifts with the \texttt{EAZY} code including their deeper UV images.

\subsubsection{MUSE}
\label{Sec:MUSE}

The MUSE Hubble Ultra Deep Field Survey \citep{Bacon2015,Inami2017,Brinchmann2017} comprises spectroscopic redshift measurements of almost 1400 sources in the HUDF region. This increases the number of available spectroscopic redshifts in this region by a factor of eight. It was conducted with the Multi Unit Spectroscopic Explorer (MUSE) at the Very Large Telescope. \citet{Inami2017} provide spectroscopic redshifts for sources with a completeness of 50 per cent at 26.5\,mag in F775W. The redshift distribution includes sources beyond \mbox{$z > 3$} and up to a F775W magnitude of $\sim 30$\,mag. This spectroscopic redshift catalogue is an excellent reference to judge the reliability of the photometric redshift catalogues used for the colour selection of background galaxies.

\subsection{Selection of background galaxies through colour cuts}
\label{Sec: Background galaxy selection}

\begin{figure*}
\centering
	
	\includegraphics[width=\columnwidth]{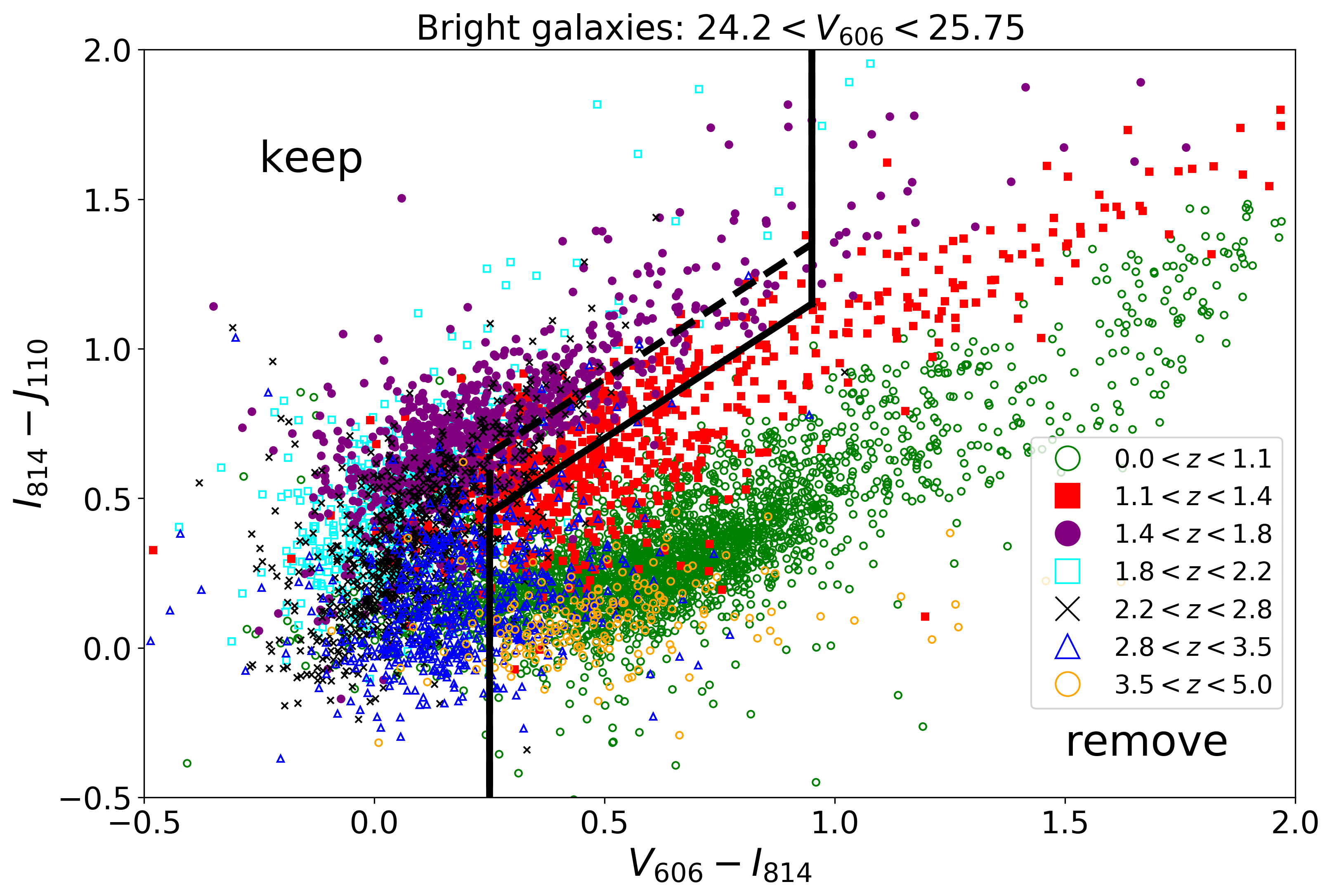}
	\includegraphics[width=\columnwidth]{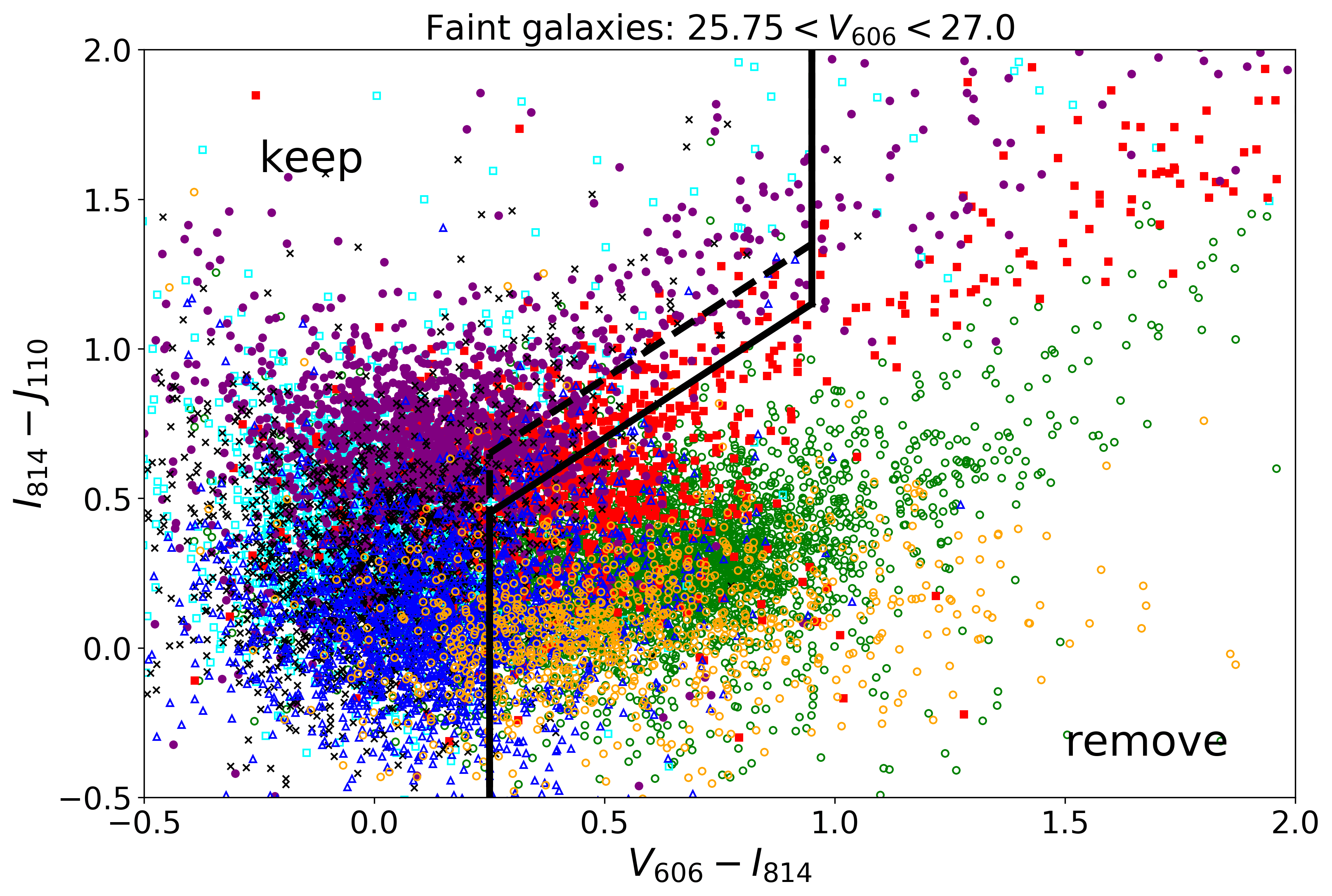}
	\includegraphics[width=\columnwidth]{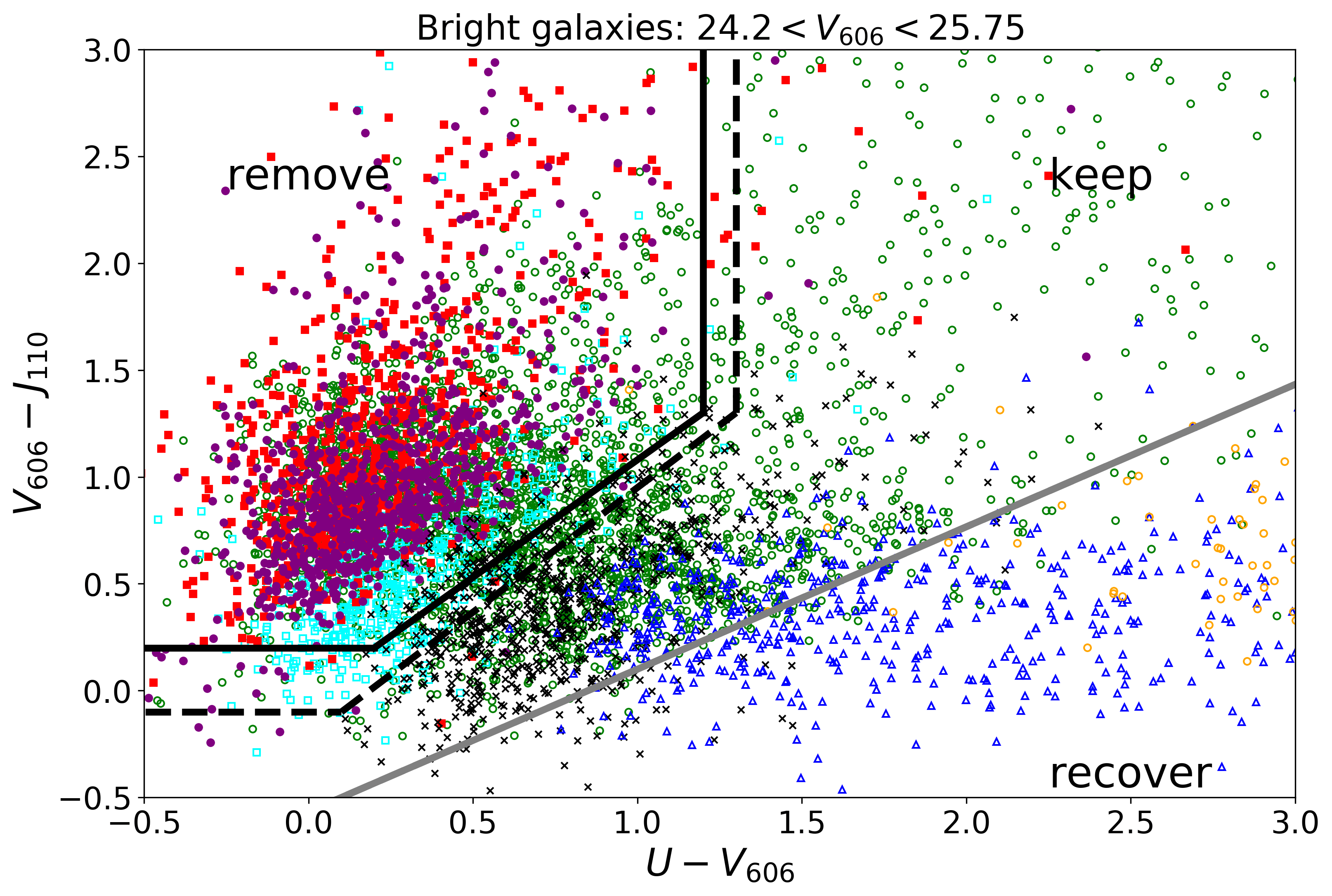}
	\includegraphics[width=\columnwidth]{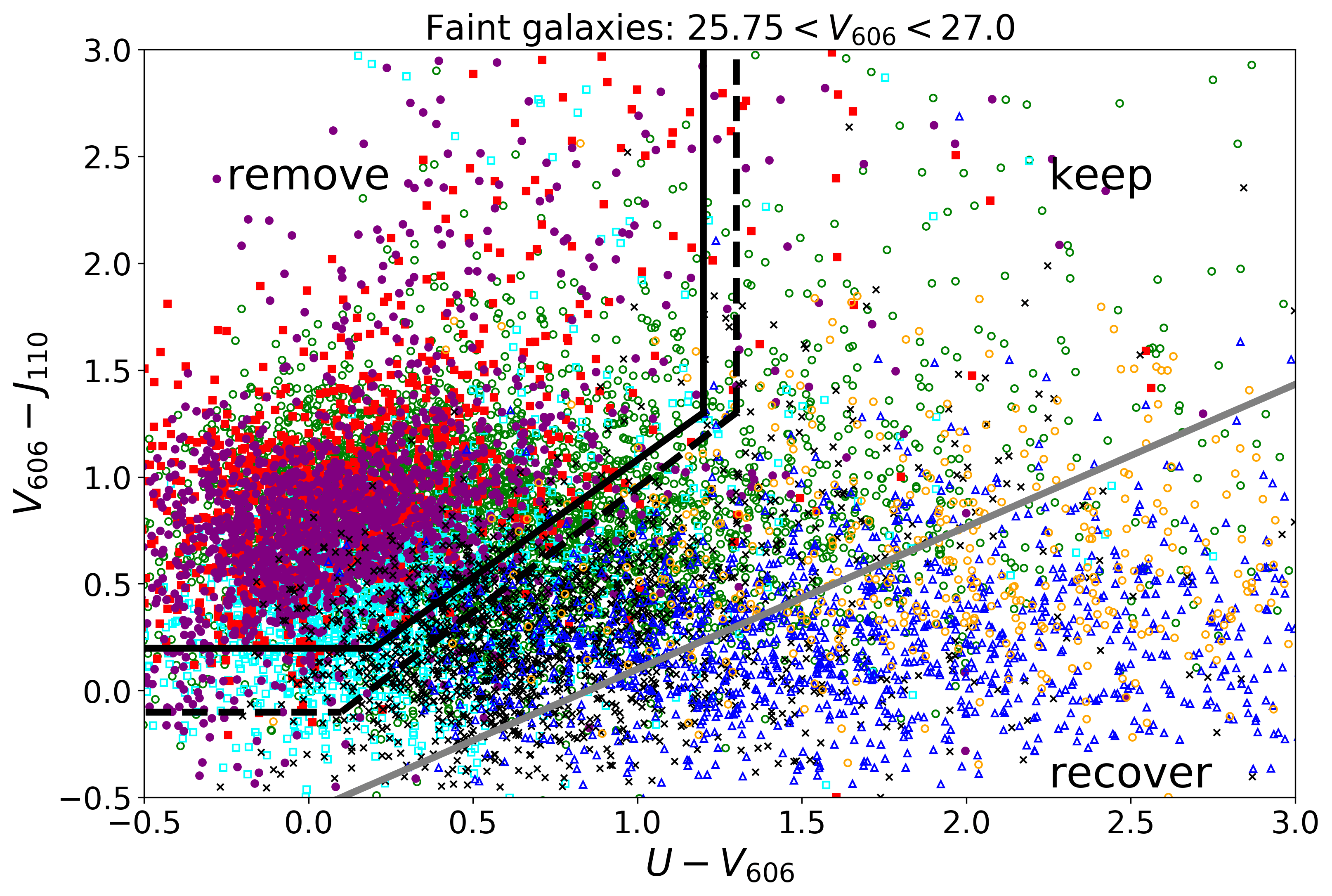}

    \caption{Colour selection for galaxy clusters at redshift \mbox{$1.2 \lesssim z \lesssim 1.7$}. The selected source galaxies are at redshift \mbox{$z \gtrsim 1.7$}. We display galaxies based on their photometry from \citetalias{Skelton2014} in the GOODS-South field. \textit{Top:} First selection step in the $VIJ$ plane for bright galaxies on the left and faint galaxies on the right. \textit{Bottom:}  Second selection step in the $UVJ$ plane for bright galaxies on the left and faint galaxies on the right. The solid black lines indicate cuts applied for bright galaxies, the dashed black lines show cuts for faint galaxies. Galaxies below the diagonal grey line are recovered in both the bright and the faint regime.}
    \label{Fig:Low-z colour selection cuts}
\end{figure*}

\begin{figure*}
\centering
	
	\includegraphics[width=\columnwidth]{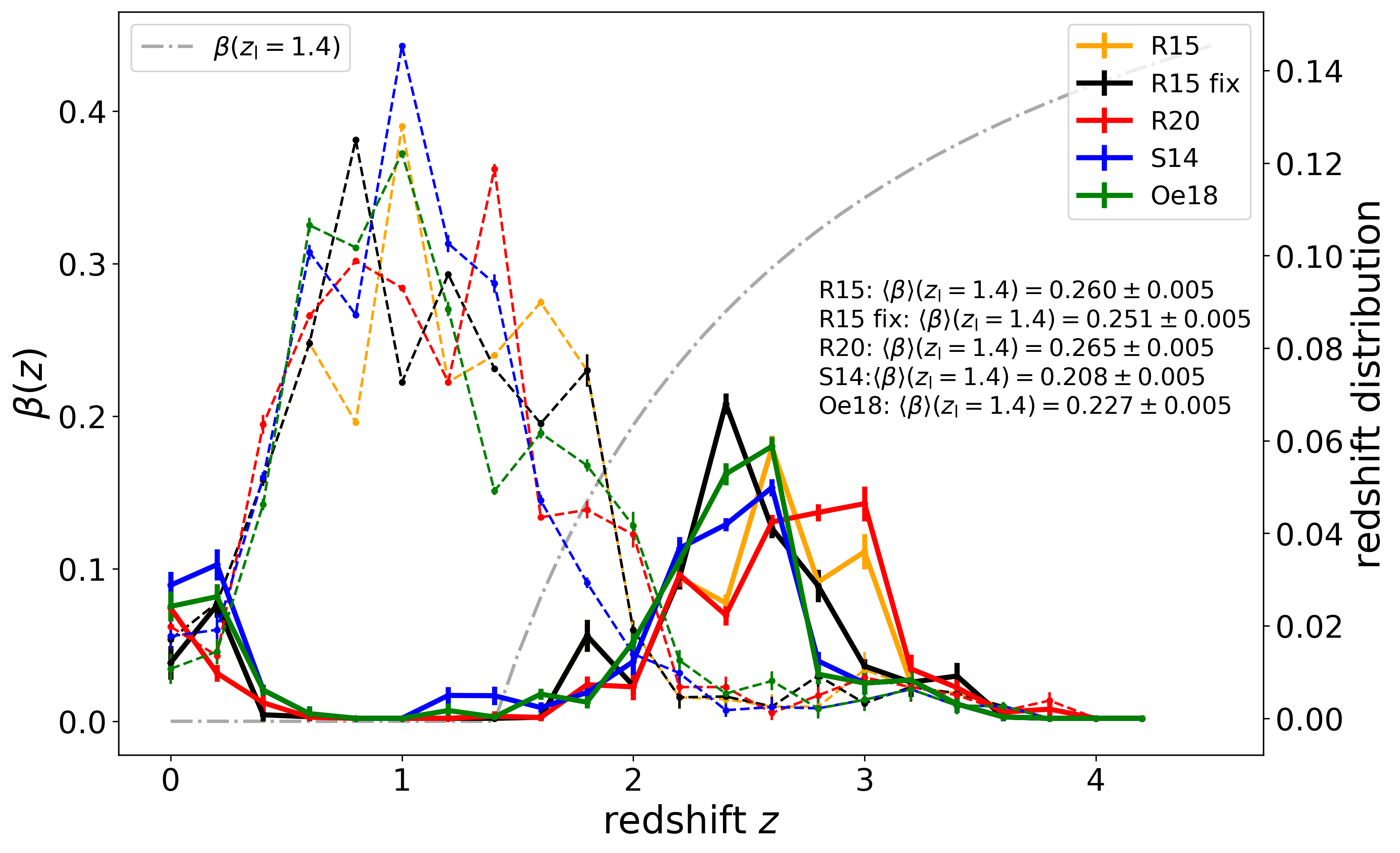}
	\includegraphics[width=\columnwidth]{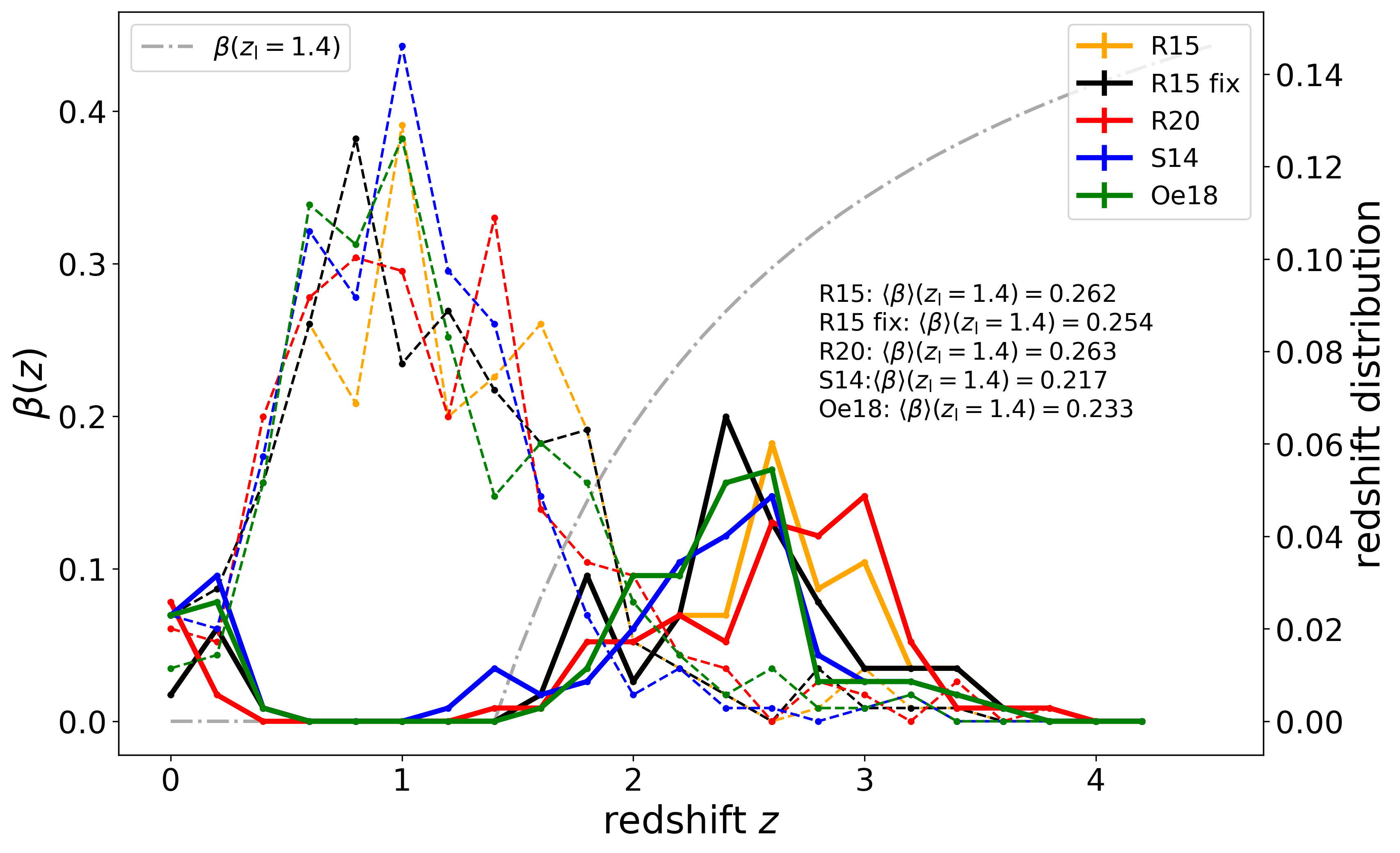}

    \caption{Redshift distributions resulting from the colour selection for galaxy clusters at redshift \mbox{$1.2 \lesssim z \lesssim 1.7$} in the HUDF region. The selected source galaxies (solid lines) are mostly at redshift \mbox{$z \gtrsim 1.7$}. Removed galaxies (dashed lines) are mostly at redshifts \mbox{$z \lesssim 1.7$}. The distributions only show galaxies matched between the five reference redshift catalogues (\citetalias{Rafelski2015}, R15\_fix, \citetalias{Raihan2020}, \citetalias{Skelton2014}, and \citetalias{Oesch2018}) and the photometric catalogue from this work. We additionally display the average lensing efficiency curve as a function of redshift (grey dash-dotted line) at the median lens redshift of the clusters at $z_\mathrm{l} = 1.4$. \textit{Left:} Redshift distributions for the five redshift catalogues and employing a colour selection based on the \citetalias{Skelton2014} photometry. The uncertainties represent the standard deviations from 50 noise realisations of the $U$ band in the \citetalias{Skelton2014} photometry. \textit{Right:} Redshift distributions for the five redshift catalogues and employing a colour selection based on the LAMBDAR photometry measured from our observations of HUDF in $U_\mathrm{HIGH}$ and from the \citetalias{Skelton2014} stacks in different \textit{HST}-bands (see text).}
    \label{Fig:Low-z redshift distribution}
\end{figure*}

\subsubsection{Defining the colour and magnitude cuts}
\label{Sec:Colour_selection, defining mag and colour cuts}
We aim to find criteria based on colours and magnitudes that help us distinguish the background galaxies of interest from the contaminating foreground and cluster galaxies. To this end, we made use of the \citetalias{Skelton2014}/\citetalias{Raihan2020} catalogues providing photometry and photometric redshifts for the largest number of galaxies.
First, we decided to focus on the magnitude regime \mbox{$24.2 < V_{606} < 27.0$} for the selection strategy. Inspecting the redshift distributions of galaxies in the CANDELS/3D-HST fields, we found that there is no significant amount of background galaxies at redshifts \mbox{$z \gtrsim 1.8$} present at magnitudes brighter than \mbox{$ V_{606} < 24.2$}. By focusing on galaxies fainter than this limit, we could exclude many bright foreground galaxies. Additionally, our cluster fields roughly reach limiting magnitudes of 27\,mag in the F606W band.

Second, we inspected the colour-colour plots of different combinations of colours to identify a suitable strategy. We found that a combination of the colour plane including $V_{606}$, $I_{814}$, and $J_{110}$ and the colour plane including $U$, $V_{606}$, and $J_{110}$ provided a useful basis for a selection of background galaxies, that is galaxies at redshifts higher than the cluster redshifts of \mbox{$1.2 \lesssim z \lesssim 1.7$}. We developed a selection consisting of two steps. For the first step, a strategic cut in the colour plane \mbox{$V_{606} - I_{814}$} and \mbox{$I_{814} - J_{110}$} (short $VIJ$ plane) allowed us to remove a significant fraction of foreground galaxies at \mbox{$0.0 < z < 1.1$}. We discarded all galaxies to the right of this cut (redder in  \mbox{$V_{606} - I_{814}$}, see the black line in upper panels of \mbox{Fig. \ref{Fig:Low-z colour selection cuts}}). With this cut, we did, however, still keep a lot of galaxies at the cluster redshift while discarding a substantial fraction of background galaxies at \mbox{$z > 2.2$}. The second step using the colour plane \mbox{$U - V_{606}$} and \mbox{$V_{606} - J_{110}$} (short $UVJ$ plane) helped us to refine the selection. Here, we could remove almost all galaxies at the cluster redshift (galaxies that are blue in \mbox{$U - V_{606}$} and red in \mbox{$V_{606} - J_{110}$}, occupying the upper left corner of the $UVJ$ plane in \mbox{Fig. \ref{Fig:Low-z colour selection cuts}}), and at the same time recover high-redshift sources we had discarded in the first selection step (galaxies that are red in \mbox{$U - V_{606}$}, occupying the lower right corner of the $UVJ$ plane in \mbox{Fig. \ref{Fig:Low-z colour selection cuts}}). Additionally, we slightly varied these cuts depending on whether the galaxies were bright \mbox{($24.2 < V_{606} < 25.75$)} or faint \mbox{($25.75 < V_{606} < 27.0$)}. Fainter galaxies typically exhibit a larger photometric scatter than brighter galaxies. We could, therefore, apply slightly tighter cuts for brighter galaxies without a high risk of contamination by cluster galaxies due to scatter. \mbox{Fig. \ref{Fig:Low-z colour selection cuts}} illustrates our cuts in the two colour planes and for the bright and faint magnitude regimes for clusters at redshift \mbox{$1.2 \lesssim z \lesssim 1.7$}.
For clarity, we summarise the selection strategy as follows: we selected all galaxies below the grey line in the $UVJ$ plane and all galaxies that are both to the left of the black line in the $VIJ$ plane \textit{and} to the right of the black line in the $UVJ$ plane.

We also investigated if it is possible to optimise the selection depending on the cluster redshift. For instance galaxies at redshift \mbox{$1.3 < z < 1.7$} could be used for a cluster at redshift \mbox{$z = 1.2$}, but have to be removed for a cluster at redshift \mbox{$z = 1.7$}. Unfortunately, such an optimisation was not possible with the available filters because all the galaxies in the redshift regime \mbox{$1.2 \lesssim z \lesssim 1.7$}  occupy a similar location in the $UVJ$ plane (see red and purple symbols in \mbox{Fig. \ref{Fig:Low-z colour selection cuts}}). We investigated two alternative selection strategies in Appendix \ref{Appendix:Coloursel_alternatives}, which did not improve the signal-to-noise ratio of the lensing analysis. We, therefore, decided to use common selection criteria for background galaxies, independent of the cluster redshift for the majority of our cluster sample. The only exception is the cluster SPT-CL{\thinspace}$J$0646$-$6236 at the lowest redshift of $z =0.995$. We used an optimised selection strategy for this particular cluster, which we describe in Appendix \ref{Appendix:Coloursel_alternatives0995}. 

Additionally, we investigated how beneficial the use of the $U$ band is for an efficient source selection since it is the band introducing the largest uncertainties. We found that it is possible to select sources with a similar average geometric lensing efficiency only based on the bands F606W, F814W, and F110W. However, the resulting source density of such a selection is significantly lower. In conclusion, the signal-to-noise ratio of the lensing measurement (proportional to the product of the average lensing efficiency and the square root of the source density) is about 1.4 times higher when the $U$ band is included for the source selection.

\subsubsection{Comparison of selections based on the S14 and the LAMBDAR photometry}
\label{Sec:Colour_selection, compare S14 + LAMBDAR}
We calculated the average lensing efficiency $\langle \beta \rangle$ for the selection based on the \citetalias{Skelton2014} photometry and for five catalogues with photometric redshift information, namely the original \citetalias{Skelton2014} redshifts, the updated \citetalias{Raihan2020} redshifts by \citetalias{Raihan2020}, the redshifts given in \citetalias{Rafelski2015}, a modified version of the \citetalias{Rafelski2015} redshifts from \citetalias{Raihan2020} called R15\_fix, and the redshifts from \citetalias{Oesch2018}. Throughout this section, we used the median lens redshift of our cluster sample of \mbox{$z_\mathrm{l} = 1.4$} for the calculation of $\langle \beta \rangle$. In addition to the selection as described in Sect. \ref{Sec:Colour_selection, defining mag and colour cuts}, we employed a signal-to-noise ($S/N$) threshold of $S/N_\mathrm{flux,606} > 10$ as applied for the shape measurements of galaxies (the signal-to-noise ratio is defined via the ratio of \texttt{FLUX\_AUTO} and \texttt{FLUXERR\_AUTO} from \texttt{Source Extractor}; see also Sect. \ref{sec:shapes}). We note that \citetalias{Raihan2020} optimised the redshifts for a source selection targeting background galaxies behind clusters of \mbox{$0.6 \lesssim z \lesssim 1.1$} (the cluster sample from \citetalias{Schrabback2018}). They apply a cut based on \mbox{$V-I$} colour at \mbox{$V-I < 0.3$} and a magnitude cut of \mbox{$V_{606} < 26.5$}. Even though these settings differ from ours, we found that the \citetalias{Raihan2020} catalogues are still applicable for our analysis because on average \mbox{84 per cent} of galaxies in our selection in the cluster fields also fulfil the condition \mbox{$V-I < 0.3$}. Additionally, we found that the average lensing efficiency calculated based on \citetalias{Raihan2020} photo-$z$s for our colour-selected galaxies in the HUDF was not significantly affected by a change of the magnitude limit from \mbox{$V_{606} < 27.0$} to \mbox{$V_{606} < 26.5$}.

The five redshift catalogues (denoted \citetalias{Rafelski2015}, R15\_fix, \citetalias{Raihan2020}, \citetalias{Skelton2014}, and \citetalias{Oesch2018}) overlap in the HUDF region. We matched the sources from our five reference catalogues based on their coordinates through the function \texttt{associate} from the \texttt{LDAC} tools\footnote{\url{https://marvinweb.astro.uni-bonn.de/data_products/THELIWWW/LDAC/}}. For a match, we required a distance smaller than $0\farcs3$. In \mbox{Fig. \ref{Fig:Low-z redshift distribution}}, we show the redshift distribution of the galaxies that we selected with our strategy. We note that the \citetalias{Skelton2014} $U$ band \mbox{($5\sigma\, \mathrm{depth} = 27.9$)} is considerably deeper than our observations in the $U_\mathrm{HIGH}$ band in the HUDF \mbox{($5\sigma\, \mathrm{depth} = 26.6$)}. To account for this difference, we added Gaussian noise to the \citetalias{Skelton2014} $U$ band photometry and show the average redshift distribution derived from 50 noise realisations of galaxies in the HUDF for a $U_\mathrm{HIGH}$ band depth of 26.6\,mag in Fig.\thinspace {\ref{Fig:Low-z redshift distribution}}. We note that, when we estimated the average lensing efficiency for the cluster fields, we added Gaussian noise to both the $U$ band and \textit{HST} photometry from the \citetalias{Skelton2014} catalogues to account for the difference between the depths in the respective cluster fields and in the CANDELS/3D-HST fields.
When we calculated the average lensing efficiency, we employed the shape weights from \citetalias{Schrabback2018} that depend on the signal-to-noise ratio (\texttt{FLUX\_AUTO}/\texttt{FLUXERR\_AUTO}) in $V_{606}$. Since the \citetalias{Skelton2014} catalogues do not provide measurements of \texttt{FLUX\_AUTO}, we used the listed total fluxes and respective errors instead\footnote{As a cross-check, we calculated the average lensing efficiency with the shape weights based on the total fluxes in the \citetalias{Skelton2014} catalogues and the \texttt{AUTO} fluxes in catalogues by \citetalias{Schrabback2018}. They have analysed shallower stacks in the CANDELS/3D-HST fields, including measurements of \texttt{FLUX\_AUTO}, which allowed us to draw a direct comparison. We found that the difference between both options is less than 1 per cent.}. The redshift distributions show that \citetalias{Skelton2014} and \citetalias{Oesch2018} have an excess of galaxies at the cluster redshifts and in the foreground at \mbox{$z < 0.4$} compared to the other catalogues. This is connected to the reported contamination by catastrophic redshift outliers (see \mbox{Sect. \ref{Sec:3D-HST cat description}}). We can see this effect as well in \mbox{Fig. \ref{Fig:Low-z redshift distribution}} where the \citetalias{Skelton2014} and \citetalias{Oesch2018} redshift catalogues yield lower values of the average lensing efficiency than the other redshift catalogues. In contrast to that, the average lensing efficiency results from the \citetalias{Raihan2020} redshift catalogues are in good agreement with the robust photometric redshift catalogues \citetalias{Rafelski2015} and R15\_fix. According to these catalogues, we expect nearly no contamination by cluster galaxies for our selection strategy (only $\sim1$ per cent of selected galaxies are within the cluster redshift range). Fig. \ref{Fig:Low-z redshift distribution} displays a small residual contribution of foreground galaxies in our source selection. This is, however, not a concern as long as the redshift distribution is modelled accurately. From a comparison of the average lensing efficiency based on \citetalias{Raihan2020} and R15\_fix we infer a systematic uncertainty of \mbox{$\Delta(\langle \beta \rangle)/ \langle \beta \rangle_\mathrm{R15\_fix} = 5.6\%$}. 

Since we measured fluxes in our observations with LAMBDAR, we additionally inspected the redshift distributions that we obtained when we used the LAMBDAR photometry measured from our observations of the HUDF in $U_\mathrm{HIGH}$ and from the \citetalias{Skelton2014} stacks in the \textit{HST} filters F606W, F814W, F850LP and F125W\footnote{\url{https://archive.stsci.edu/prepds/3d-hst/} ; (F606W + F850LP: GO programme 9425 with PI M. Giavalisco, F814W: GO programme 12062 with PI S. Faber, F125W: GO programme 13872 with PI G. Illingworth).} (we interpolated between the latter two filters to estimate the magnitude in the filter F110W). The resulting distribution is shown on the right-hand side of \mbox{Fig. \ref{Fig:Low-z redshift distribution}}. 
This corresponds to a systematic uncertainty of \mbox{$\Delta(\langle \beta \rangle)/ \langle \beta \rangle_\mathrm{R15\_fix} = 3.5\%$}.
Overall, the average lensing efficiency results based on \citetalias{Skelton2014} and LAMBDAR photometry agree within the uncertainties (see \mbox{Fig. \ref{Fig:Low-z redshift distribution}}).

\begin{figure}
	\includegraphics[width=\columnwidth]{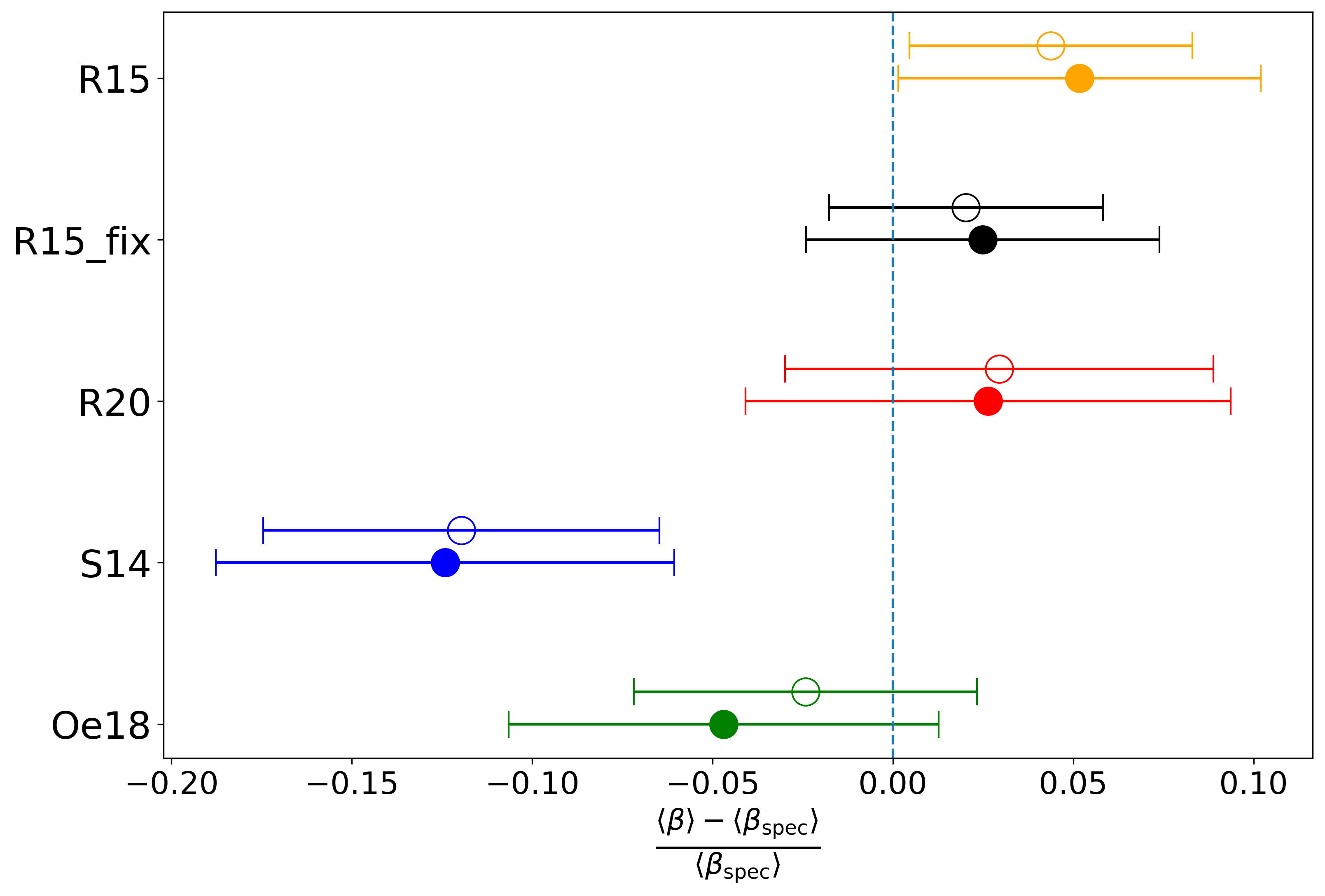}

    \caption{Relative bias in the average lensing efficiency normalised to the result based on spectroscopic/grism redshifts for the five reference redshift catalogues (\citetalias{Rafelski2015}, R15\_fix, \citetalias{Raihan2020}, \citetalias{Skelton2014}, and \citetalias{Oesch2018}). We performed the colour selection for all coordinate-matched galaxies in the HUDF with available spectroscopic/grism redshifts. The uncertainties represent the scatter from 1000 bootstrap resamples. Filled symbols represent source selections based on the \citetalias{Skelton2014} photometry, open symbols represent source selections based on the LAMBDAR photometry.}
    \label{Fig:Low-z MUSE/grism beta comparisons}
\end{figure}

\begin{figure}
	\includegraphics[width=\columnwidth]{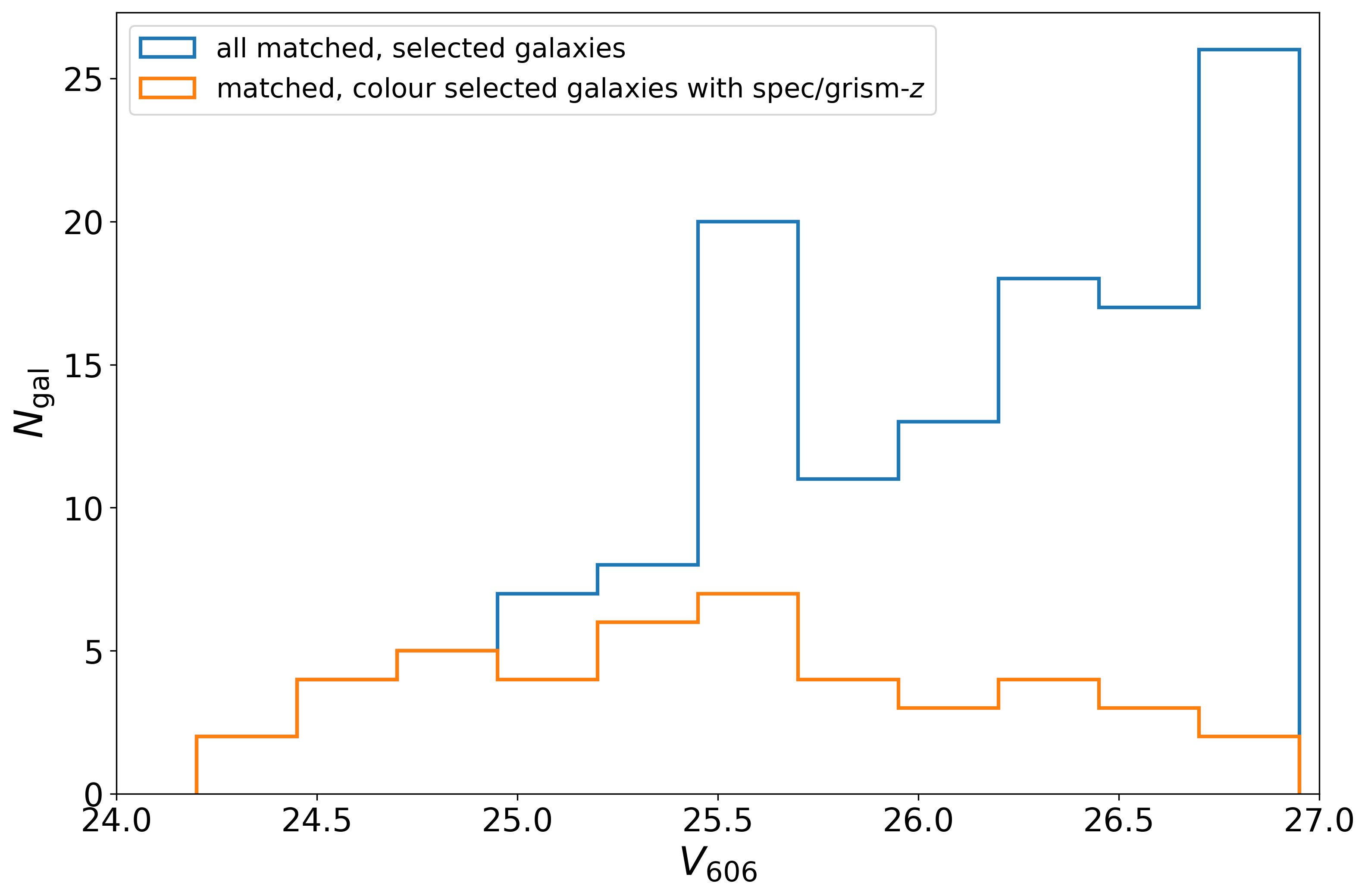}
	\includegraphics[width=\columnwidth]{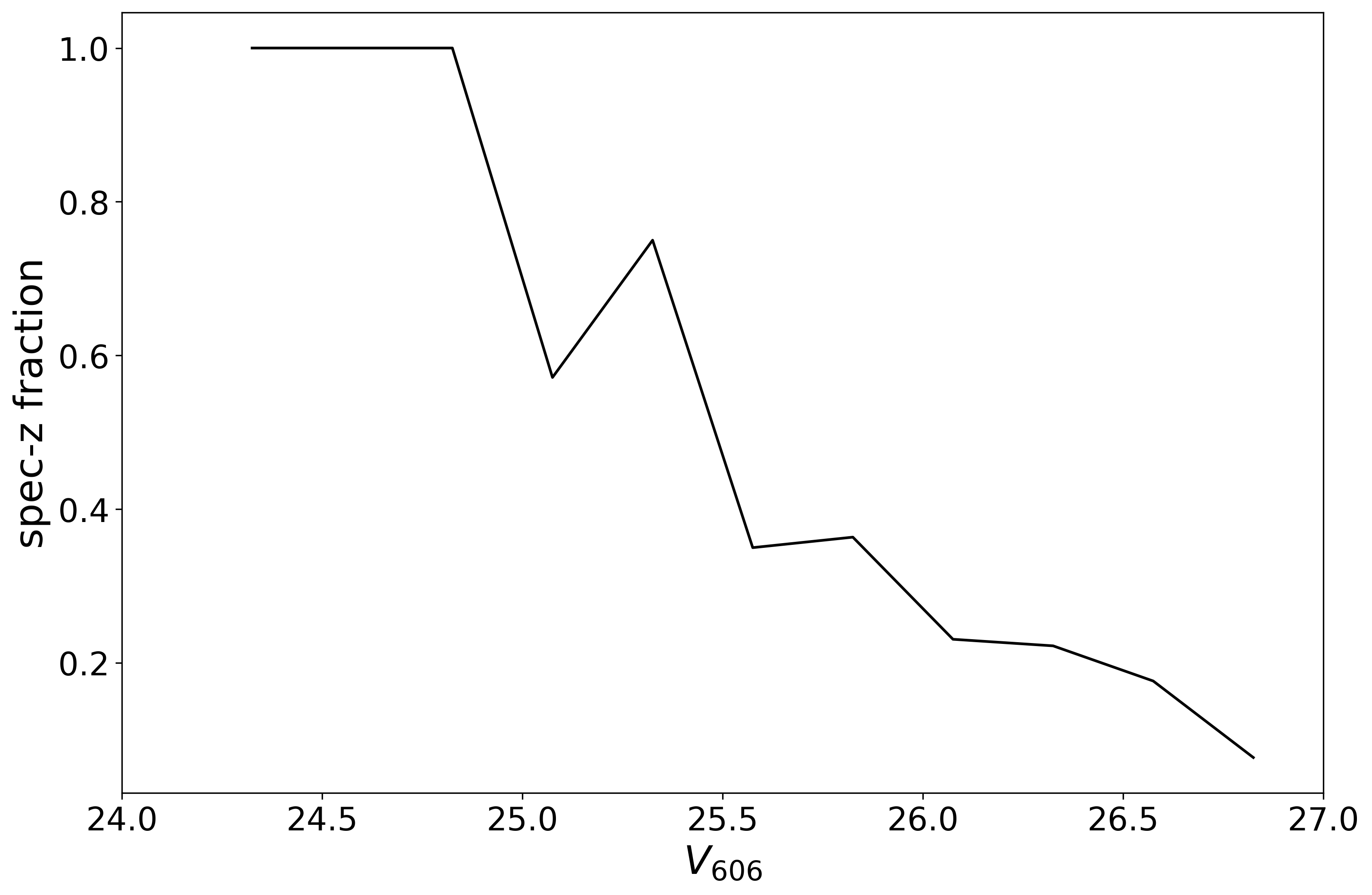}

    \caption{Overview about available spectroscopic redshift information as a function of magnitude. \textit{Top:} Histogram of all matched and colour-selected galaxies within the HUDF region (blue). The orange histogram shows how many of these have a robust spec-$z$ from MUSE or grism-$z$. \textit{Bottom:} Fraction of matched and colour-selected galaxies within the HUDF region with a robust spec-$z$ from MUSE or grism-$z$, corresponding to the ratio of the orange and blue histogram in the top panel.}
    \label{Fig:muse-z gals,completeness-histo+fraction}
\end{figure}

\begin{figure}

	\includegraphics[width=\columnwidth]{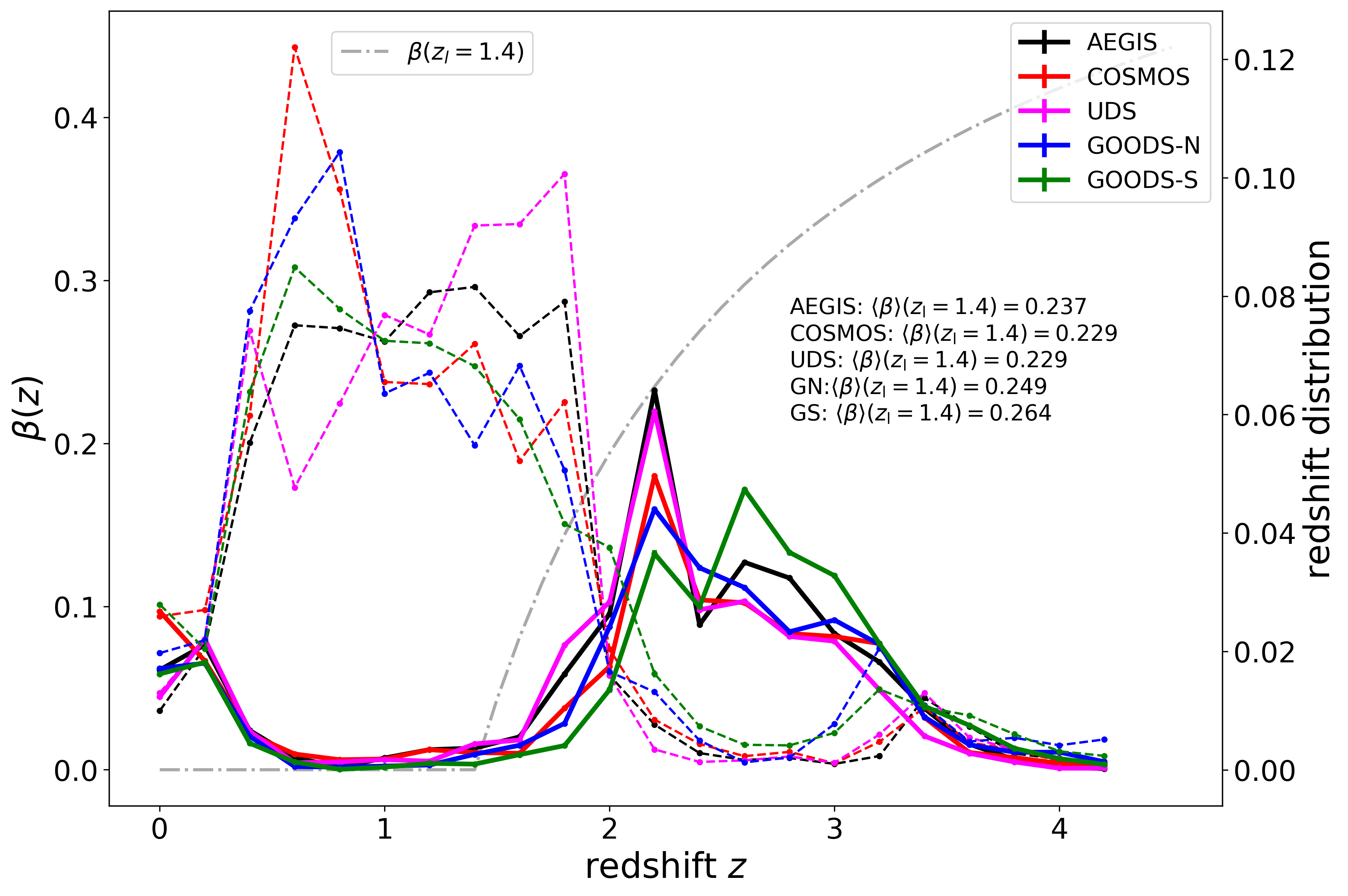}
    \caption{Redshift distribution of the galaxies in the CANDELS/3D-HST fields for the colour selection for clusters at \mbox{$1.2 \lesssim z \lesssim 1.7$}, employing the \citetalias{Raihan2020} photometric redshift catalogues. The selected source galaxies (solid lines) are mostly at redshift \mbox{$z \gtrsim 1.7$}. Removed galaxies (dashed lines) are mostly at redshifts \mbox{$z \lesssim 1.7$}. We additionally display the average lensing efficiency curve as a function of redshift (grey dash-dotted line) at the median lens redshift of the clusters at $z_\mathrm{l} = 1.4$.
    }
    \label{Fig:Low-z CANDELS redshift distribution}
\end{figure}

\subsubsection{Comparison of selections based on photo-$z$s and spec-$z$s}
As a cross-check for the photometric redshift catalogues,
we retrieved spectroscopic/grism redshifts from the MUSE and 3D-HST catalogues, respectively, for all galaxies matched by their coordinates in the HUDF field. As a reference, we then calculated the average lensing efficiency of the colour-selected sources based on the spectroscopic/grism redshifts. Here, we only used the MUSE spec-$z$s with the highest quality flags 3 (secure redshift, determined by multiple features) and 2 \citep[secure redshift, determined by a single feature, see][]{Inami2017}{}. In the case of galaxies with both spectroscopic redshifts from MUSE and grism redshifts from 3D-HST, we used the former for the calculation of $\langle \beta_ \mathrm{spec}\rangle$. To estimate the uncertainty, we bootstrapped the colour-selected galaxies and recalculated the average lensing efficiency 1000 times. \mbox{Fig. \ref{Fig:Low-z MUSE/grism beta comparisons}} shows how the average lensing efficiency calculated from the five photometric redshift catalogues compares to the one calculated based on spectroscopic/grism redshifts. We did not find a bias within the uncertainties, but we notice that the average lensing efficiency based on R15\_fix, \citetalias{Raihan2020} and \citetalias{Oesch2018} matches closest to the result based on the spectroscopic/grism redshifts. It also has to be noted that the spectroscopic/grism redshifts are only complete in comparison to the full sample of matched galaxies in the HUDF region up to a magnitude of \mbox{$V_{606} \lesssim 25.0$\,mag} (see \mbox{Fig. \ref{Fig:muse-z gals,completeness-histo+fraction}}). We still decided to correct our measurements of the average lensing efficiency by the roughly three per cent offset between the \citetalias{Raihan2020} redshift-based and the spectroscopic redshift-based lensing efficiency for all clusters except SPT-CL{\thinspace}$J$0646$-$6236. For the specific source selection used for this cluster, such an offset did not occur.

\subsubsection{Differences between the five CANDELS/3D-HST fields}
\label{Sect: Differences between CANDELS beta}

\begin{table}

                \caption{
		Summary of our systematic and statistical error budget. 
		}
	\begin{center}
    \begin{threeparttable}
		\begin{tabular}{ l c c c} 
			\hline\hline
			
            Source of \textbf{systematic}    & Rel. error &  Rel. error & Sect./ \\ 
            uncertainties &  signal    &  $M_{500c}$ & App. \\ 
            \hline
            \textbf{Redshift distribution:} &  &  & \\ 
            - \citetalias{Raihan2020} vs. R15\_fix comp.& 5.6\,\% & 8.4\,\% & \ref{Sec:Colour_selection, compare S14 + LAMBDAR} \\
            - Variations between & 5.7\,\% & 8.6\,\% & \ref{Sect: Differences between CANDELS beta} \\
            CANDELS/3D-HST fields &  &  \\
            - F110W band & 2.2\,\% & 3.3\,\% & \ref{Appendix:Comparison of S14 and LAMBDAR photometry}/\ref{Appendix:Impact of syst. shifts in photom} \\ 
            (LAMBDAR/\citetalias{Skelton2014}, interp.)  &  &  &  \\
            - $V - I$ colour  & 2.2\,\% & 3.3\,\% & \ref{Appendix:Comparison of S14 and LAMBDAR photometry}/\ref{Appendix:Impact of syst. shifts in photom} \\
            (LAMBDAR/\citetalias{Skelton2014})  &  &  &  \\
            
            \textbf{Shape measurements:} &  &  & \\ 
            - Shear calibration & 2.3\,\% & 3.4\,\% & \ref{sec:shapes} \\
            
            \textbf{Mass model:} &  &  & \\ 
            - $c(M)$ relation & & 4.0\,\% &  \\
            - Miscentring for &  &  & \\
                    \quad \quad X-ray centres & & 3.8\,\% /&  \ref{Sec:corr_for_mass_modelling_bias}\\
                    \quad \quad SZ centres & & 9.2\,\% &  \ref{Sec:corr_for_mass_modelling_bias}\\
   
            \hline
            total (added in quadrature) & & 14.4\,\% / & \\
                  & &  16.7\,\% & \\
			\hline\hline
            Source of \textbf{statistical}   & Rel. error &  Rel. error & Sect./ \\ 
            uncertainties &  signal    &  $M_{500c}$ & App. \\ 
            \hline
            \textbf{Redshift distribution:} &  &  &\\ 
            - Line of sight variations & 6.9\,\% & 10.4\,\% & \ref{Sect: Differences between CANDELS beta}\\
            - $U_\mathrm{HIGH}$ band calibration & 4.1\,\% & 6.2\,\% & \ref{Appendix:ZP robustness with gal locus}/\ref{Appendix:Impact of syst. shifts in photom} \\
            
            \hline
            total (added in quadrature) & & 12.1\,\% & \\
			\hline
			
		\end{tabular}
			\textbf{Notes.} 
            In the upper part of the table, we list all systematic uncertainties, which ultimately translate into an uncertainty in the weak lensing mass measurement, where we added the individual contributions in quadrature to obtain an estimate for the total uncertainty. We report the relative uncertainties in per cent in the second column, the resulting relative uncertainty on the mass in the third column, and refer the reader to the respective sections or appendices listed in the last column for more detailed information about the contributions to the error budget. In the lower table, we list statistical uncertainties in the redshift distribution, which affect the calculation of the average geometric lensing efficiency $\langle \beta \rangle$ for individual cluster fields. We note that the final statistical uncertainties reported in Tables \ref{tab:mass-Xray} and \ref{tab:mass-SZ} do include additional contributions from shape noise and uncorrelated large-scale structure projections.
	\end{threeparttable}	
	\end{center}
		
	\label{Tab:Errorbudget of photometry,beta}
\end{table}

Since we estimate the average lensing efficiency from all CANDELS fields, we want to evaluate the expected systematic uncertainties arising from differences in the depths, available filters, and calibrations in the five CANDELS/3D-HST fields.
Additionally, we expect statistical sampling variance due to line of sight variations.

We quantified the systematic uncertainties by measuring the average lensing efficiency for colour-selected galaxies independently in the five CANDELS/3D-HST fields (see \mbox{Fig. \ref{Fig:Low-z CANDELS redshift distribution}}). We obtained a mean of the average lensing efficiencies of \mbox{$\langle \beta \rangle_\mathrm{mean} = 0.242$} with a standard deviation of \mbox{$\sigma(\langle \beta \rangle) = 0.014$} between the \mbox{$N = 5$} fields (using the photometric redshifts from \citetalias{Raihan2020}). This translates into a systematic uncertainty of \mbox{$\sigma(\langle \beta \rangle)/\langle \beta \rangle_\mathrm{mean} = 5.7 \%$}. We calculated this more conservative systematic uncertainty without dividing by $\sqrt{N-1}$ because we noticed that the value of the GOODS-South field is notably higher, and thus, one field might not automatically be a good representation of the average of all. We added this uncertainty in quadrature to our systematic error budget (see Table \ref{Tab:Errorbudget of photometry,beta}). We note that this uncertainty also contains a statistical contribution as each CANDELS/3D-HST field represents a different line of sight. However, since the fields are each much larger than the small sub-patches studied in the paragraph below, we conservatively assume that the variations between the CANDELS/3D-HST fields are dominated by systematic uncertainties.

We gauged the expected statistical uncertainty from line of sight variations in the average lensing efficiency by placing non-overlapping apertures with the same area as the field of view of our observations (about 11 arcmin$^2$) in the CANDELS/3D-HST fields. We can fit exactly eight apertures in each of the fields. We calculated the average lensing efficiency independently for all of the apertures, where we obtained the mean \mbox{$\langle \beta \rangle_\mathrm{mean} = 0.243$} with a scatter of \mbox{$\sigma(\langle \beta \rangle) = 0.017$}. Hence, we added a statistical uncertainty of \mbox{$\sigma(\langle \beta \rangle)/\langle \beta \rangle_\mathrm{mean} = 6.9 \%$} to our statistical error budget (see Table \ref{Tab:Errorbudget of photometry,beta}).

Regarding uncertainties of the source redshift distribution, we estimated a total statistical uncertainty of \mbox{8.0 per cent} on the average lensing efficiency corresponding to 12.1 per cent on the mass scale. This includes uncertainties in the $U_\mathrm{HIGH}$ band calibration (see Appendix \ref{Appendix:ZP robustness with gal locus}) and line of sight variations (this section), which we added in quadrature. Furthermore, we estimated a total systematic uncertainty of \mbox{8.6 per cent} on the average geometric lensing efficiency. Here, we took into account systematics for the F110W band (interpolation versus direct measurement, aperture photometry versus LAMBDAR photometry, see Appendices \ref{Appendix:Comparison of S14 and LAMBDAR photometry} and \ref{Appendix:Impact of syst. shifts in photom}), uncertainties in the measurement of \mbox{$V-I$} colours (see Appendices \ref{Appendix:Comparison of S14 and LAMBDAR photometry} and \ref{Appendix:Impact of syst. shifts in photom}), uncertainties of the \citetalias{Raihan2020} redshift catalogues (see Sect. \ref{Sec:Colour_selection, compare S14 + LAMBDAR}), and variations between the CANDELS/3D-HST fields (differences of the filters, depths, availability of $U$ bands, and usage of different bands to interpolate the $J_{110}$ magnitudes, see this section). Again, we added these contributions in quadrature. All of these uncertainties are summarised in Table \ref{Tab:Errorbudget of photometry,beta}.

\subsection{Check for cluster member contamination}

\begin{figure}
\centering

	\includegraphics[width=\columnwidth]{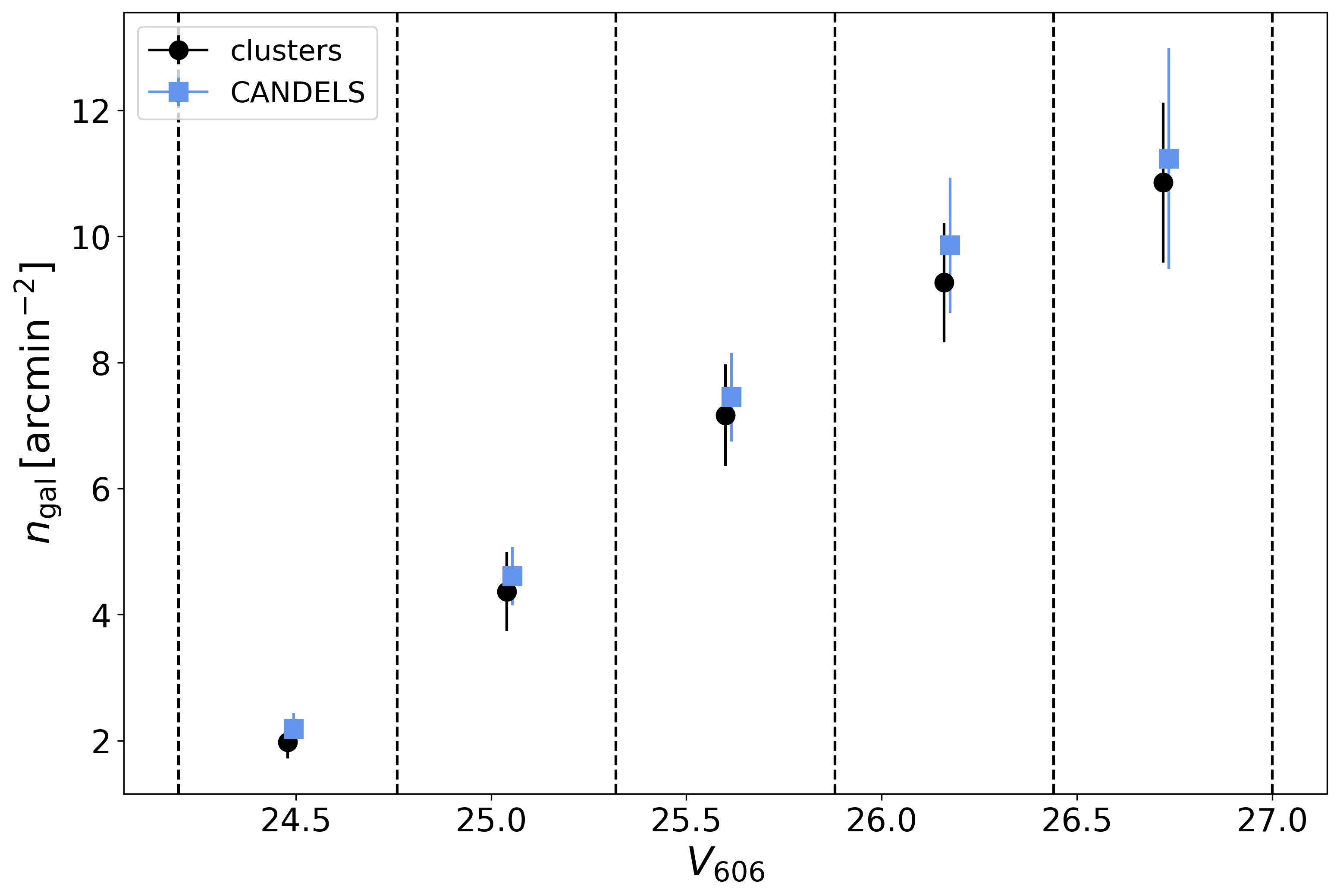}
	\includegraphics[width=\columnwidth]{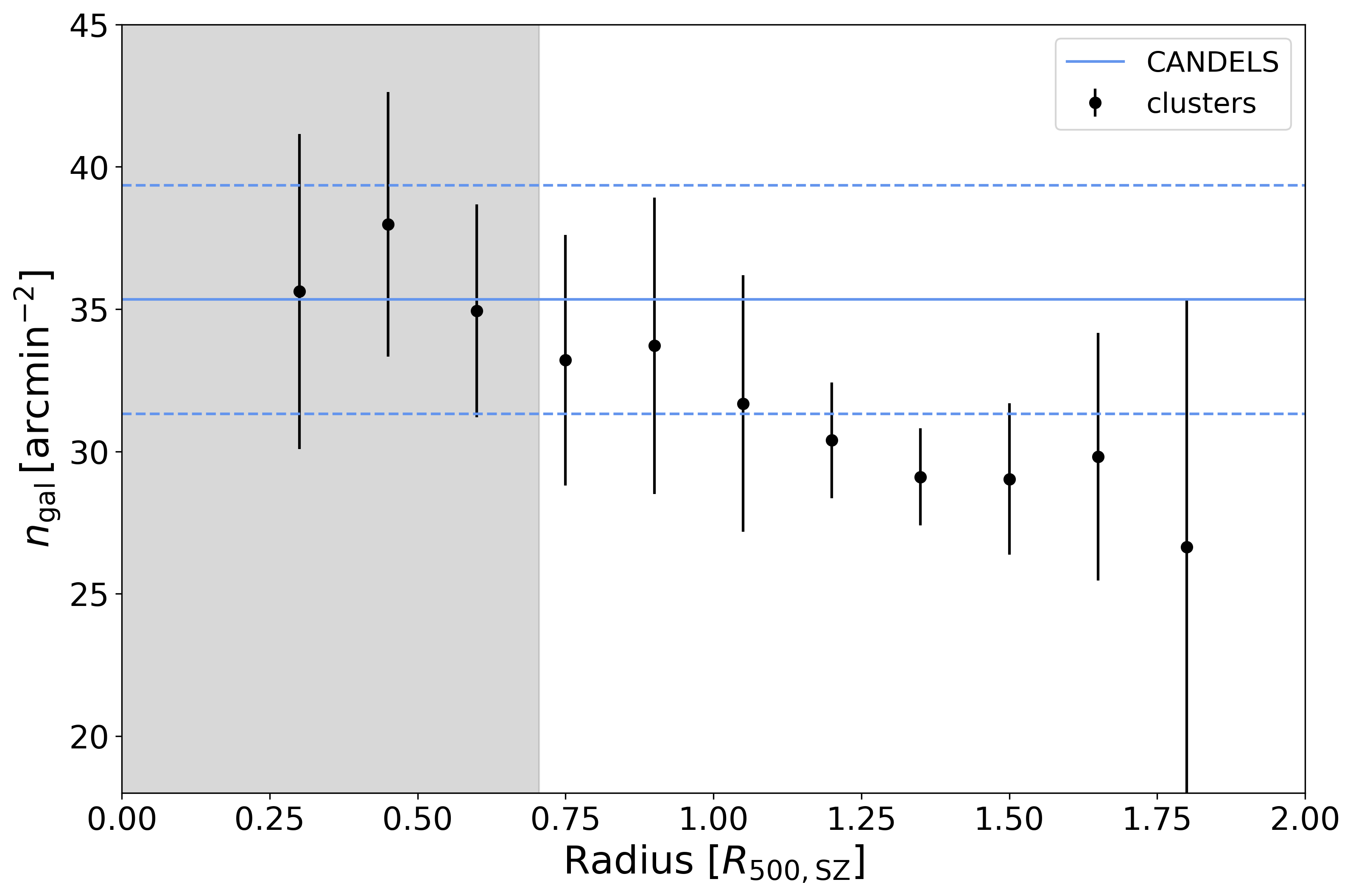}

    \caption{Number density profiles of selected source galaxies. \textit{Top:} We show the number density of selected galaxies $n_\mathrm{gal}$ averaged over the nine cluster fields (black symbols) and averaged over the five 3D-HST/CANDELS fields (blue symbols) as a function of magnitude. We took into account the masks, for example, from bright stars in the images, and we only considered photometrically selected galaxies, that is, no flags from shape measurements or signal-to-noise ratio cuts were considered here. The error bars correspond to the uncertainty of the mean from the variation between the contributing cluster fields or 3D-HST/CANDELS fields, respectively.  \textit{Bottom:} Average density of selected sources as a function of the distance to the X-ray cluster centre (except for the cluster SPT-CL{\thinspace}$J$0646$-$6236, where we used the SZ centre). These distances are given in units of the radius $R_\mathrm{500c,SZ}$ based on the SZ mass $M_\mathrm{500c,SZ}$.  Blue lines indicate the average density and $1\sigma$ uncertainties from the five 3D-HST/CANDELS fields. The error bars correspond to the uncertainty of the mean from the variation between the contributing cluster fields or 3D-HST/CANDELS fields, respectively. We excluded the grey-shaded region when we measured weak lensing masses. It corresponds to 500\,kpc or about $0.71\,R_{500}$ for a cluster with $R_{500} = 700$\,kpc. }
    \label{Fig:Radial+Magnitude number densitiy profiles}
\end{figure}

We aim to preferentially select background galaxies with our magnitude and colour cuts both in the cluster fields and the CANDELS/3D-HST fields. Investigating the source density of the selected galaxies and their radial dependence allows us to test if we have a substantial amount of contamination by cluster galaxies and if our method provides a consistent selection in the cluster fields and the CANDELS/3D-HST fields in the presence of noise \citepalias{Schrabback2018}.

To this end, we added Gaussian noise to the \citetalias{Skelton2014} photometric catalogues according to the difference between the depth of the cluster observations and the depth of the CANDELS/3D-HST fields. This may vary depending on the field and filter. We only added Gaussian noise if the CANDELS/3D-HST observation in a filter were deeper than the corresponding observation in the cluster field. Occasionally, the cluster observations were slightly deeper than some of the CANDELS/3D-HST observations, but only by $\sim0.2$ mag. We considered this negligible for the validity of this test.

We measured the source density of selected sources accounting for masks, for example, due to bright stars for the cluster fields and CANDELS/3D-HST fields. We only considered photometrically selected galaxies and did not consider potential flags from the shape measurement pipeline. We also did not apply the signal-to-noise ratio cut $S/N_\mathrm{flux,606} > 10$ as mentioned in Sect. \ref{Sec: Background galaxy selection} and \ref{sec:shapes} for this test, since the quantities \texttt{FLUX\_AUTO} and \texttt{FLUXERR\_AUTO} required to calculate the signal-to-noise ratio are not available in the CANDELS/3D-HST catalogues.
In \mbox{Fig. \ref{Fig:Radial+Magnitude number densitiy profiles}} (left panel), we show the average source density of selected galaxies as a function of the $V_{606}$ band magnitude. We found a good agreement over the full magnitude range of interest in this study. 

Additionally, we investigated the radial dependence of the source density of selected galaxies. In principle, an increase of the number density towards the cluster centre can indicate cluster member contamination. However, the profile can also be affected by blending and/or masking of background galaxies by cluster member galaxies, magnification, or selection effects. We accounted neither for blending and/or masking by cluster galaxies nor magnification in our analysis. The blending/masking by cluster galaxies should be less important than for clusters at lower redshifts since the cluster galaxies are more cosmologically dimmed. Additionally, we conservatively excluded the core region $r < 500\,\mathrm{kpc}$, when we measured the weak lensing masses so that this effect should not play a significant role.
Regarding magnification, for \citetalias{Schrabback2021} the application of a magnification correction had only a minor impact on the source density profile. Given the higher redshifts of our clusters, the lensing strength and, therefore, the expected impact of magnification is even lower, which is why we ignore it here.

\mbox{Fig. \ref{Fig:Radial+Magnitude number densitiy profiles}} (right panel) displays the radial distance from the X-ray centre (except for the cluster SPT-CL{\thinspace}$J$0646$-$6236, where we used the SZ centre) in units of the radius $R_\mathrm{500c,SZ}$, which we derived from the SZ mass $M_\mathrm{500c,SZ}$. We found a very slight trend of a higher source density towards the centres of the clusters. However, the profile is consistent with flat within the uncertainties. Together, both measurements provide an important confirmation for the success of the photometric background selection and cluster member removal.

%--------------------------------------------------------------------

\section{Shape measurements}
\label{sec:shapes}
The shape of a galaxy can be quantified by its ellipticity, as a complex number \mbox{$\epsilon = \epsilon_1 + \mathrm{i}\epsilon_2$}. The observed ellipticity $\epsilon_\mathrm{obs}$ of a background galaxy can be related to the intrinsic ellipticity $\epsilon_\mathrm{orig}$ and reduced shear $g$ via \citep{Bartelmann2001}
\begin{equation}
    \epsilon_\mathrm{obs} = \frac{\epsilon_\mathrm{orig} + g}{1 + g^*\epsilon_\mathrm{orig}}  \,.
\end{equation}
According to the cosmological principle, the intrinsic orientation of galaxies should have no preferred direction\footnote{Despite this principle, intrinsic alignments of galaxies due to various physical effects can pose a challenge for weak lensing analyses, especially for cosmic shear studies. See for example \citet{Troxel2015} for a review. These intrinsic alignments are, however, not a concern for this work.}. Therefore, the expectation value for an average over many galaxies \mbox{$\langle \epsilon_\mathrm{orig} \rangle = 0$} vanishes. 
In conclusion, we can estimate the reduced shear, that is, the main observable for weak lensing studies, from the ensemble-averaged PSF-corrected ellipticities of the background galaxies via
\begin{equation}
    \langle \epsilon_\mathrm{obs} \rangle = g\,.
\end{equation}

We measured galaxy shapes in the ACS F606W ($V$) and \mbox{F814W ($I$)} images
using the
KSB+ formalism \citep{Kaiser1995,Luppino1997,Hoekstra1998} as implemented by \citet{Erben2001} and \citet{Schrabback2007}.
We modelled the spatially and temporally varying ACS  point-spread function using an interpolation based on principal component analysis, as calibrated on dense stellar fields \citep{Schrabback2010,Schrabback2018}.
We corrected for shape measurement and selection biases
as a function of the KSB+ galaxy signal-to-noise ratio from \citet{Erben2001}.
This correction was derived by \citet{Hernandez-Martin2020}, who analysed custom \texttt{Galsim} \citep{Rowe2015} image simulations with ACS-like image characteristics.
Importantly,  \citet{Hernandez-Martin2020} tuned their simulated source samples such that the measured distributions in galaxy size, magnitude, signal-to-noise ratio, and ellipticity dispersion closely matched the corresponding measured distributions of the magnitude and colour-selected source samples from  \citetalias{Schrabback2018}, while also incorporating realistic levels of blending.
Varying the properties of the simulations, \citet{Hernandez-Martin2020}
estimated a (post-correction) multiplicative  shear calibration uncertainty 
of the employed KSB+ pipeline of  \mbox{$\sim 1.5\%$}.
Our data are very similar to those analysed by \citetalias{Schrabback2018}. Therefore, we expect the \citet{Hernandez-Martin2020} shear calibration to be directly applicable for our analysis.
However, our colour selection selects galaxies at slightly higher redshifts on average compared to the \mbox{$V-I$} selection from \citetalias{Schrabback2018}.
Some of our image stacks are also slightly deeper.
We, therefore, conservatively increased the shear calibration uncertainty in our systematic error budget by a factor $\times 1.5$ (see Table \ref{Tab:Errorbudget of photometry,beta}).
%\textbf{Do you agree?}).

Given their greater average depth (see Table \ref{tab:exposure times}), 
we based our shear catalogue primarily on the F606W stacks.
Here, we included galaxies with a measured flux signal-to-noise ratio \mbox{$S/N_\mathrm{flux,606}>10$}\footnote{With the aim to potentially reduce statistical uncertainties in our analysis, we also computed results using an alternative signal-to-noise ratio cut of \mbox{$S/N_\mathrm{flux}>7$}. While this did increase the source number density, we found that it only marginally changes the constraints of our SZ--mass scaling relation analysis, likely due to the low shape weights and the increased photometric scatter of the additional faint galaxies. In an interest to keep our study consistent with previous studies, for example, by \citetalias{Schrabback2021}, we chose to use the cut of \mbox{$S/N_\mathrm{flux}>10$}.} (defined as the ratio of the \texttt{FLUX\_AUTO} and \texttt{FLUXERR\_AUTO} parameters from \texttt{Source Extractor}). 
This single-band selection matches the one employed in Sect. \ref{Sec: Background galaxy selection} in the computation of the average geometric lensing efficiency.
For galaxies that additionally
%obey
have \mbox{$S/N_\mathrm{flux,814}>10$}, we combined the shape measurements from both filters to reduce the impact of measurement noise.

\begin{table}

        \caption{Number densities of selected source galaxies measured in the cluster fields.}
	\begin{center}
	
	\begin{threeparttable}

		\begin{tabular}{l c} 
			\hline\hline
			
			Cluster name & $n_\mathrm{gal}$  \\ 
			& [arcmin$^{-2}$]  \\ 
			
			\hline
			SPT-CL{\thinspace}$J$0156$-$5541 & 14.3  \\ 
			SPT-CL{\thinspace}$J$0205$-$5829 & 12.7  \\
			SPT-CL{\thinspace}$J$0313$-$5334 & 20.1  \\
			SPT-CL{\thinspace}$J$0459$-$4947 & 10.7  \\
			SPT-CL{\thinspace}$J$0607$-$4448 & 13.3  \\ 
			SPT-CL{\thinspace}$J$0640$-$5113 & 10.2  \\
			SPT-CL{\thinspace}$J$2040$-$4451 & 11.2  \\
			SPT-CL{\thinspace}$J$2341$-$5724 & 12.6  \\
			\hline
			average & 13.1 \\
			\hline
			SPT-CL{\thinspace}$J$0646$-$6236 & 26.9  \\
			
		\end{tabular}
		
	\end{threeparttable}	
	\end{center}
			\textbf{Notes.}
		 We apply the source selection as described in Sect. \ref{Sec:Colour_selection, defining mag and colour cuts} including only sources that pass the lensing selections and have a signal-to-noise ratio $S/N_\mathrm{flux,606} > 10$, leading to lower numbers compared to Fig. \ref{Fig:Radial+Magnitude number densitiy profiles}. The cluster SPT-CL{\thinspace}$J$0646$-$6236 is listed separately because we applied a different selection strategy for this cluster (see Appendix \ref{Appendix:Coloursel_alternatives0995}). 
	\label{Tab:Number densities}
\end{table}

In order to compute shape weights and filter-combined estimates of the reduced shear,
we made use of the
\mbox{$\log_{10}{S/N_\mathrm{flux}}$}-dependent
fits
computed by \citetalias[][see their appendix A]{Schrabback2018}
for the total ellipticity dispersion $\sigma_{\epsilon,V/I}$,
the intrinsic ellipticity dispersion 
$\sigma_{\mathrm{int},V/I}$,
and the ellipticity measurement noise
$\sigma_{\mathrm{m},V/I}$
of \mbox{$V-I$} colour selected galaxies in custom CANDELS \citep{Grogin2011}
$V$ (F606W) and $I$ (F814W) band
stacks of approximately single-orbit depth\footnote{We employ the
  \mbox{$\log_{10}{S/N_\mathrm{flux}}$}-dependent fits instead of the magnitude-dependent fits
  provided by \citetalias{Schrabback2018} in order to account for the slightly higher depth of some of our stacks and the
significant dependence of the measurement noise on \mbox{$\log_{10}{S/N_\mathrm{flux}}$}. For comparison, the dependence of $\sigma_{\mathrm{int},V/I}$ on  \mbox{$\log_{10}{S/N_\mathrm{flux}}$} is weak in the regime covered by most of our sources.}.
With the complex reduced shear estimates $\epsilon_{V/I}$ obtained in the $V$ band and the $I$ band, respectively, and the shape weights
\begin{equation}
w_{V/I}=\left( \sigma_{\epsilon,V/I} \right)^{-2}\,,
\end{equation}
we computed the filter-combined reduced shear estimate as
\begin{equation}
\epsilon_\mathrm{comb}=\frac{w_V \epsilon_V + w_I \epsilon_I}{w_V+w_I} \,.
\end{equation}
The measurement noise is independent between the stacks in the different  filters, which is why the combined ellipticity measurement variance reads
\begin{equation}
\sigma_\mathrm{m,comb}^2=\frac{(w_V \sigma_{\mathrm{m},V})^2 + (w_I \sigma_{\mathrm{m},I})^2}{(w_V+w_I)^2} \,.
\end{equation}
In the relevant $S/N$ or magnitude regime, differences are small between
$\sigma_{\mathrm{int},V}$ and $\sigma_{\mathrm{int},I}$ for the colour-selected source samples from \citetalias{Schrabback2018}.
In addition, \citet{Jarvis2008}
found that intrinsic shapes are highly correlated between \textit{HST} images of galaxies in different optical filters.
Therefore, as an approximation, we interpolated the intrinsic ellipticity
dispersion between the filters
\begin{equation}
\sigma_\mathrm{int,comb}=\frac{w_V \sigma_{\mathrm{int},V} + w_I \sigma_{\mathrm{int},I}}{w_V+w_I}\,,
\end{equation}
allowing us to compute shape weights for the combined shear estimate as
\begin{equation}
  w_\mathrm{comb}=\left( \sigma_\mathrm{int,comb}^2 + \sigma_\mathrm{m,comb}^2  \right)^{-1}\,.
\end{equation}
We reached an average final source density after all photometry and shape cuts of 13.1\,arcmin$^{-2}$ (see Table \ref{Tab:Number densities}) for the clusters with $1.2 \lesssim z \lesssim 1.7$. We note that this is substantially lower than the values shown in Fig. \ref{Fig:Radial+Magnitude number densitiy profiles} because we now included the signal-to-noise ratio and lensing cuts\footnote{While the number density is affected by a change of the signal-to-noise ratio cut, we found that the average geometric lensing efficiency is not sensitive to it. The change is smaller than $\sim 1$ per cent comparing the results with or without the cut at $S/N_\mathrm{flux,606}>10$.}.

%--------------------------------------------------------------------

\section{Weak lensing results}
\label{sec:wlconstraints}

Our pipeline used to obtain weak lensing constraints largely follows \citetalias{Schrabback2018} and \citetalias{Schrabback2021} to which we refer the reader for more detailed descriptions.

\subsection{Mass reconstructions}
\label{Sec:Mass-maps}
The weak lensing convergence $\kappa$ and shear $\gamma$ are both second-order derivatives of the lensing potential \citep[e.g.][]{Bartelmann2001}.
As a result, it is possible to reconstruct the convergence distribution from the
shear field up to a constant, which is also known as the mass-sheet degeneracy \citep{Kaiser1993,Schneider1995}.
Here, we employed the  Wiener-filtered reconstruction algorithm from \citet{McInnes2009} and \citet{Simon2009}, where we fixed the mass-sheet degeneracy
by setting the average convergence inside the observed fields to zero.
We computed $S/N$ maps of the reconstruction, where the noise map is computed as the root mean square (r.m.s.) image of the $\kappa$ field reconstructions of 500 noise shear fields, which were created by randomising the ellipticity phases in the real source catalogue.
Given the limited field of view and our choice to set the average convergence to zero, we expect to slightly underestimate the true $S/N$ levels \citepalias{Schrabback2021}.
 
 The obtained  $S/N$ reconstructions are shown as contours in the left panels of Figs. \ref{fi:wl_results_1} and \ref{fi:wl_results_2} -- \ref{fi:wl_results_4} in Appendix \ref{Appendix:WeakLensingResults}.
 SPT-CL{\thinspace}$J$0646$-$6236 and SPT-CL{\thinspace}$J$2040$-$4451 show clear peaks in the mass reconstruction signal-to-noise ratio maps with \mbox{$S/N_\mathrm{peak}>3$} (see Table \ref{tab:masspeaklocations} for details).
 We find tentative counterparts to the clusters with \mbox{$2<S/N_\mathrm{peak}<3$}
 for SPT-CL{\thinspace}$J$0156$-$5541, SPT-CL{\thinspace}$J$0459$-$4947,
 SPT-CL{\thinspace}$J$0640$-$5113, and SPT-CL{\thinspace}$J$2341$-$5724.
 The other clusters either show no significant peak in their corresponding mass reconstruction $S/N$ maps, or only a  peak close to the edge of the field of view, which is less reliable and likely spurious. While some of the clusters remained undetected in the reconstructed mass maps, we note that these maps are only for illustration purposes. We still took the tangential reduced shear profiles of all clusters in our sample into account for the likelihood analysis (see Sect. \ref{Sec:ScalingRelAnalysis}).

\begin{table*}  
\caption{Constraints on the peaks in the mass reconstruction signal-to-noise ratio maps including their locations (\mbox{$\alpha, \delta$}),
  positional uncertainties (\mbox{$\Delta\alpha, \Delta\delta$}) as estimated by bootstrapping the galaxy
  catalogue \citep[we note that this underestimates the true uncertainty as found by][]{Sommer2021}, and their peak signal-to-noise ratios \mbox{$(S/N)_\mathrm{peak}$}.  We excluded
unreliable peaks close to the edge of the field of view (compare Fig. \ref{fi:wl_results_1} and Figs. \ref{fi:wl_results_2}--\ref{fi:wl_results_4}).
\label{tab:masspeaklocations}}
\begin{center}
\begin{tabular}{lccccccc}
\hline\hline
Cluster & $\alpha$ & $\delta$ & $\Delta\alpha$ & $\Delta\delta$ & $\Delta\alpha$ & $\Delta\delta$ &\mbox{$(S/N)_\mathrm{peak}$}   \\
 &  [deg J2000] &  [deg J2000] & [arcsec] & [arcsec] & [kpc] & [kpc] &\\
\hline
SPT-CL{\thinspace}$J$0156$-$5541  &   29.04676  & $ -55.69426 $ & 9.1  &  4.8  &  76  &  41  &  2.0 \\
SPT-CL{\thinspace}$J$0459$-$4947  &   74.92771  & $ -49.77739 $ & 8.1  &  9.6  &  69  &  81  &  2.2 \\
SPT-CL{\thinspace}$J$0640$-$5113  &   100.08319  & $ -51.21488 $ & 6.4  &  5.5  &  53  &  46  &  2.6 \\
SPT-CL{\thinspace}$J$0646-6236  &  101.63130  & $ -62.62127 $ & 1.5  &  2.2  &  12  &  18  & 5.5 \\
SPT-CL{\thinspace}$J$2040$-$4451  &   310.24056  & $ -44.86349 $ & 4.6  &  3.7  &  39  &  31  &  3.4 \\
SPT-CL{\thinspace}$J$2341$-$5724  &   355.34768  & $ -57.41418 $ & 7.7  &  8.1  &  64  &  68  &  2.2 \\

\hline
\end{tabular}
\end{center}
{\flushleft
}
\end{table*}

\subsection{Fits to the tangential reduced shear profiles}
\label{Sec: fits_to_tangential_reduced_shear_profiles}

When measuring the reduced shear signal with respect to the centre of a mass concentration such as a cluster, it is helpful to distinguish the tangential component $g_\mathrm{t}$ and the cross component $g_\times$:
\begin{equation}
    \begin{split}
    g_\mathrm{t} & = -g_1\cos{2\phi} - g_2\sin{2\phi}  \,, \\
    g_\times & = +g_1\sin{2\phi} - g_2\cos{2\phi} \,,
    \end{split}
\end{equation}
where $\phi$ indicates the azimuthal angle with respect to the centre.
We computed the tangential component (`t') and the cross component (`$\times$') of the reduced shear  in linear bins of width \mbox{100\,kpc} (see the right panels of Fig. \ref{fi:wl_results_1} and Figs. \ref{fi:wl_results_2} -- \ref{fi:wl_results_4} in Appendix \ref{Appendix:WeakLensingResults}) around both the X-ray centroids (when available) and the SZ centres of the targeted clusters.
We fitted the tangential reduced shear profiles using spherical Navarro-Frenk-White \citep[NFW,][]{Navarro1997} models following  \citet{WrightBrainerd2000}, employing the concentration--mass
relation from \citet{Diemer2015} with updated parameters from \citet{Diemer2019}. 
When deriving mass constraints, we excluded the cluster cores \mbox{($r<500$\,kpc)}, since the inclusion of smaller scales would both increase the intrinsic scatter and systematic uncertainties related to the mass modelling \citep[see e.g. ][]{Sommer2021,Grandis2021}. We note that weak lensing mass constraints can also be derived this way for clusters, which were undetected in the reconstructed mass maps (see Sect. \ref{Sec:Mass-maps}).
We summarise the resulting fit constraints in Tables \ref{tab:mass-Xray} and \ref{tab:mass-SZ}.
For clusters with both X-ray and SZ centres, we regarded the X-ray-centred analysis as our primary result given the smaller expected mass modelling biases (see  Sect. \ref{Sec:corr_for_mass_modelling_bias}).

\begin{figure*}
\centering
    \includegraphics[clip, trim=0.25cm 0.32cm 0.3cm 0.0cm,width=0.98\columnwidth]{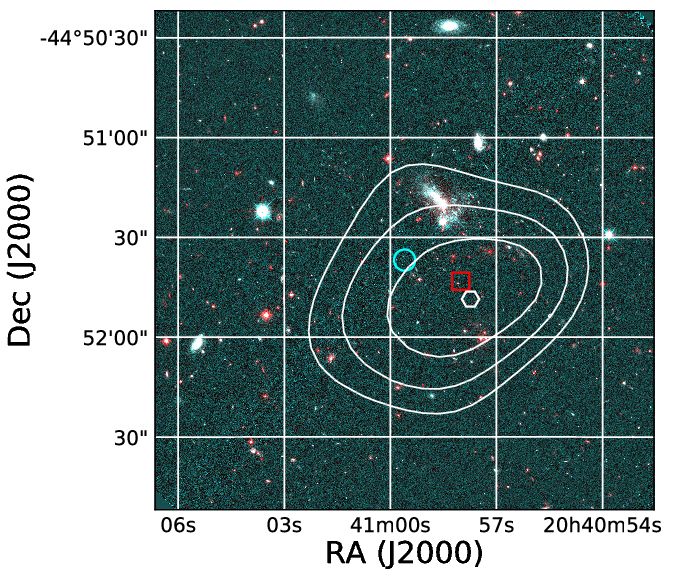}
  \includegraphics[clip, trim=0.5cm 5.5cm 0.5cm 3.5cm,width=0.83\columnwidth]{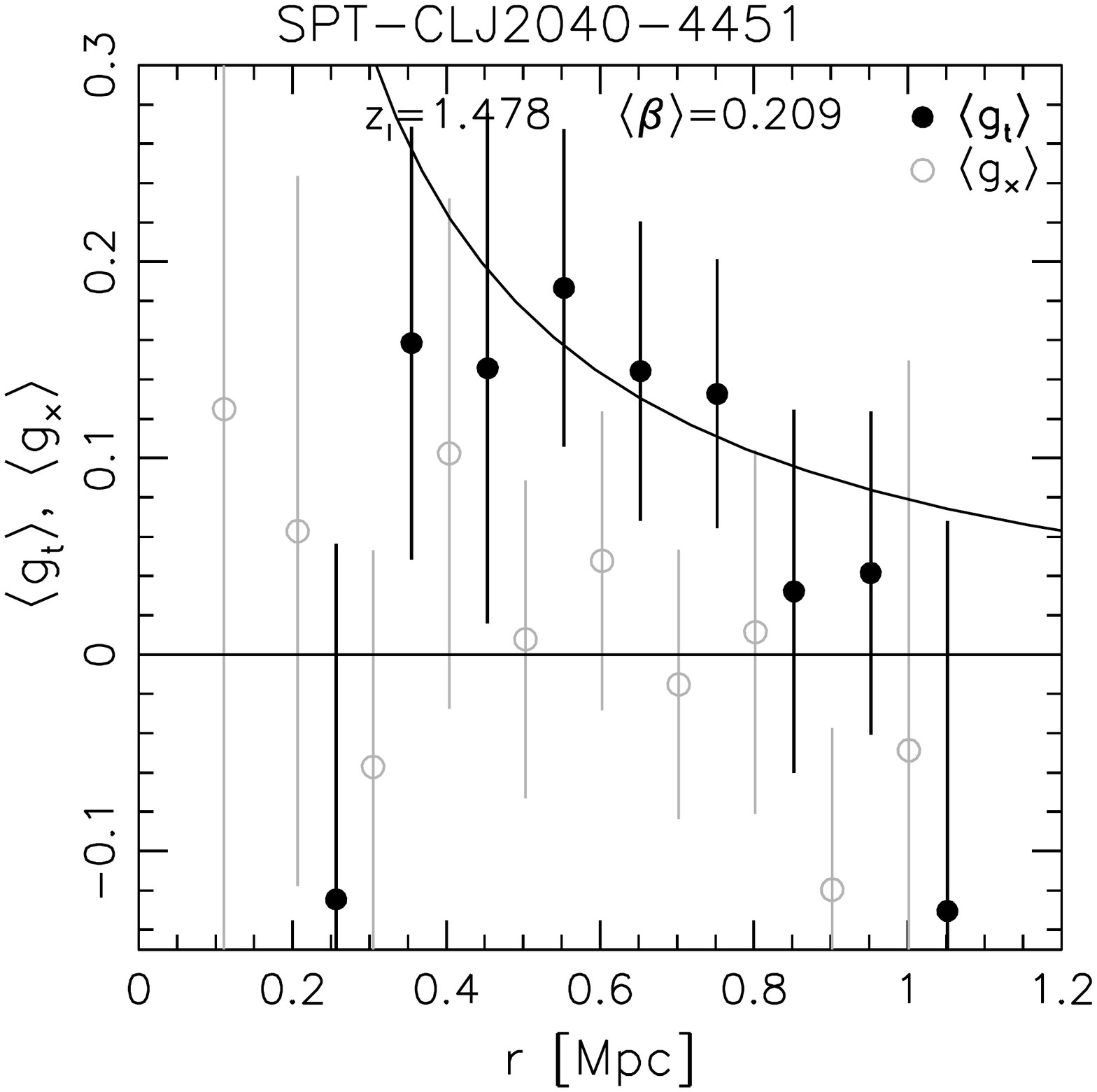}
 \caption{Weak lensing results for SPT-CL{\thinspace}$J$2040$-$4451.
     {\it Left:} Signal-to-noise ratio contours of the mass reconstruction, starting at $2\sigma$ in steps of $0.5\sigma$, overlaid on a F606W/F814W/F110W colour image (\mbox{$2\farcm5 \times 2\farcm5$} cutout).
     The peak in the $S/N$ map is indicated by the hexagon (excluding potential spurious secondary peaks near the edge of the field of view).
     The cyan circle and the red square show the locations of the SZ peak and the X-ray centroid,  respectively.
     {\it Right:} Reduced shear profile around the X-ray centre, including the tangential component (solid black circles including the best-fitting NFW model) and the cross component (open grey circles), which has been shifted along the $x$-axis for clarity. The results for the other clusters are shown in Appendix \ref{Appendix:WeakLensingResults}.
          \label{fi:wl_results_1}}
\end{figure*}

\begin{table*}  
	\caption{Weak lensing mass constraints derived from the fit of the  tangential reduced shear
		profiles around the X-ray centres using spherical NFW models assuming the $c(M)$ relation  from \citet{Diemer2015} with updated parameters from \citet{Diemer2019}
		for two different over-densities
		\mbox{$\Delta \in \{200\mathrm{c}, 500\mathrm{c}\}$}. 
		}
   
	\begin{center}
		
		\begin{tabular}{crccrcc}
			\hline
			\hline
			Cluster&  \multicolumn{1}{c}{$M_{200\mathrm{c}}^\mathrm{biased,ML}\,[10^{14}\mathrm{M}_\odot]$} &
			$\hat{b}_{200\mathrm{c,WL}}$&
			$\sigma(\mathrm{ln}\, b_\mathrm{\mathrm{200c,WL}})$   &                                                                   \multicolumn{1}{c}{$M_{500\mathrm{c}}^\mathrm{biased,ML}\,[10^{14}\mathrm{M}_\odot]$} & 
			$\hat{b}_{500\mathrm{c,WL}}$ &
			$\sigma(\mathrm{ln}\, b_\mathrm{\mathrm{500c,WL}})$ \\
			\hline
			SPT-CL{\thinspace}$J$0156$-$5541 & $4.5_{-2.9}^{+3.5}\pm 1.0\pm 0.5 $& $0.88\pm0.02$ &  $0.35\pm0.03$ & $3.1_{-2.1}^{+2.5} \pm 0.7\pm 0.3$ & $0.92\pm0.03$ &  $0.28\pm0.05$\\ 
			SPT-CL{\thinspace}$J$0205$-$5829 & $0.1_{-2.4}^{+2.8}\pm 0.5\pm 0.0 $& $0.76\pm0.03$ &  $0.41\pm0.05$ & $0.1_{-1.6}^{+1.9} \pm 0.3\pm 0.0$ & $0.79\pm0.03$ &  $0.41\pm0.04$\\ 
			SPT-CL{\thinspace}$J$0313$-$5334 & $2.8_{-2.4}^{+3.3}\pm 1.1\pm 0.3 $& $0.86\pm0.03$ &  $0.44\pm0.04$ & $1.9_{-1.7}^{+2.4} \pm 0.8\pm 0.2$ & $0.83\pm0.03$ &  $0.37\pm0.05$\\ 
			SPT-CL{\thinspace}$J$0459$-$4947 & $4.4_{-4.4}^{+6.8}\pm 1.5\pm 0.5 $& $0.85\pm0.05$ &  $0.51\pm0.08$ & $3.0_{-3.0}^{+5.0} \pm 1.1\pm 0.4$ & $0.79\pm0.05$ &  $0.43\pm0.10$\\ 
			SPT-CL{\thinspace}$J$0607$-$4448 & $0.6_{-2.2}^{+3.4}\pm 0.7\pm 0.1 $& $0.86\pm0.03$ &  $0.46\pm0.04$ & $0.4_{-1.5}^{+2.4} \pm 0.4\pm 0.0$ & $0.82\pm0.04$ &  $0.45\pm0.06$\\ 
			SPT-CL{\thinspace}$J$0640$-$5113 & $6.6_{-4.5}^{+5.1}\pm 1.1\pm 0.7 $& $0.93\pm0.03$ &  $0.27\pm0.08$ & $4.6_{-3.2}^{+3.8} \pm 0.8\pm 0.5$ & $0.85\pm0.04$ &  $0.37\pm0.05$\\ 
			SPT-CL{\thinspace}$J$2040$-$4451 & $16.4_{-5.7}^{+5.8}\pm 1.6\pm 1.9 $& $0.89\pm0.04$ &  $0.44\pm0.06$ & $12.0_{-4.4}^{+4.5} \pm 1.3\pm 1.4$ & $0.74\pm0.04$ &  $0.48\pm0.06$\\ 
			SPT-CL{\thinspace}$J$2341$-$5724 & $5.7_{-3.5}^{+3.9}\pm 1.1\pm 0.6 $& $0.88\pm0.03$ &  $0.35\pm0.04$ & $4.0_{-2.5}^{+2.9} \pm 0.8\pm 0.4$ & $0.87\pm0.03$ &  $0.25\pm0.05$\\ 
			\hline
		\end{tabular}
	\end{center}
	\textbf{Notes.}
	The  maximum likelihood mass estimates
		$M_{\Delta}^\mathrm{biased,ML}$ are given in $10^{14}\mathrm{M}_\odot$, where 
		errors correspond to statistical 68 per cent
		uncertainties from shape noise (asymmetric errors), followed by
		uncorrelated
		large-scale structure projections, the calibration of the $U_\mathrm{HIGH}$ band, and
		variations in the redshift
		distribution between different lines of sight (for systematic uncertainties see Table \ref{Tab:Errorbudget of photometry,beta}).
		Statistical corrections for mass modelling biases have not yet been applied for $M_{\Delta}^\mathrm{biased,ML}$.
		They are characterised by
		\mbox{$\hat{b}_\mathrm{\Delta,WL}=\text{exp}\left[\langle \text{ln}\,b_{\Delta,\text{WL}}\rangle\right]$}
		and $\sigma(\mathrm{ln}\,b_\mathrm{\Delta,WL})$, which relate to the mean and the width of the 
		estimated
		mass bias
		distribution \mbox{(see Sect. \ref{Sec:corr_for_mass_modelling_bias}).}
		\label{tab:mass-Xray}
\end{table*}

\begin{table*}  
	\caption{As Table \ref{tab:mass-Xray}, but for the analysis centring the shear profiles around the SZ centres.
		\label{tab:mass-SZ}}
   
	\begin{center}
		
		\begin{tabular}{crccrcc}
			\hline
			\hline
			Cluster&  \multicolumn{1}{c}{$M_{200\mathrm{c}}^\mathrm{biased,ML}\,[10^{14}\mathrm{M}_\odot]$} &
			$\hat{b}_{200\mathrm{c,WL}}$&
			$\sigma(\mathrm{ln}\, b_\mathrm{\mathrm{200c,WL}})$   &                             \multicolumn{1}{c}{$M_{500\mathrm{c}}^\mathrm{biased,ML}\,[10^{14}\mathrm{M}_\odot]$} & 
			$\hat{b}_{500\mathrm{c,WL}}$ &
			$\sigma(\mathrm{ln}\, b_\mathrm{\mathrm{500c,WL}})$ \\
			\hline
			SPT-CL{\thinspace}$J$0156$-$5541 & $3.9_{-2.8}^{+3.4}\pm 1.1\pm 0.4 $& $0.74\pm0.02$ &  $0.41\pm0.04$ & $2.7_{-1.9}^{+2.5} \pm 0.8\pm 0.3$ & $0.73\pm0.02$ &  $0.36\pm0.04$\\ 
			SPT-CL{\thinspace}$J$0205$-$5829 & $0.3_{-2.3}^{+3.1}\pm 0.5\pm 0.0 $& $0.76\pm0.03$ &  $0.38\pm0.05$ & $0.2_{-1.6}^{+2.2} \pm 0.4\pm 0.0$ & $0.72\pm0.03$ &  $0.40\pm0.05$\\ 
			SPT-CL{\thinspace}$J$0313$-$5334 & $4.3_{-3.1}^{+3.8}\pm 1.2\pm 0.4 $& $0.80\pm0.03$ &  $0.33\pm0.06$ & $3.0_{-2.2}^{+2.8} \pm 0.8\pm 0.3$ & $0.76\pm0.03$ &  $0.34\pm0.05$\\ 
			SPT-CL{\thinspace}$J$0459$-$4947 & $6.9_{-5.7}^{+7.0}\pm 1.7\pm 0.8 $& $0.83\pm0.07$ &  $0.49\pm0.12$ & $4.9_{-4.1}^{+5.3} \pm 1.2\pm 0.6$ & $0.67\pm0.06$ &  $0.65\pm0.09$\\ 
			SPT-CL{\thinspace}$J$0607$-$4448 & $2.4_{-2.5}^{+4.0}\pm 1.0\pm 0.3 $& $0.76\pm0.04$ &  $0.23\pm0.11$ & $1.7_{-1.7}^{+2.9} \pm 0.7\pm 0.2$ & $0.72\pm0.03$ &  $0.34\pm0.07$\\ 
			SPT-CL{\thinspace}$J$0640$-$5113 & $3.4_{-3.4}^{+5.1}\pm 1.0\pm 0.4 $& $0.66\pm0.03$ &  $0.56\pm0.05$ & $2.3_{-2.3}^{+3.7} \pm 0.7\pm 0.3$ & $0.70\pm0.03$ &  $0.36\pm0.07$\\ 
			SPT-CL{\thinspace}$J$0646$-$6236 & $12.1_{-3.3}^{+3.3}\pm 1.3\pm 1.1 $& $0.78\pm0.02$ &  $0.41\pm0.03$ & $8.6_{-2.5}^{+2.4} \pm 0.9\pm 0.8$ & $0.78\pm0.02$ &  $0.39\pm0.03$\\  
			SPT-CL{\thinspace}$J$2040$-$4451 & $15.7_{-5.8}^{+5.8}\pm 1.5\pm 1.8 $& $0.77\pm0.04$ &  $0.40\pm0.07$ & $11.5_{-4.4}^{+4.5} \pm 1.2\pm 1.3$ & $0.71\pm0.04$ &  $0.47\pm0.07$\\ 
			SPT-CL{\thinspace}$J$2341$-$5724 & $3.8_{-3.0}^{+3.8}\pm 1.0\pm 0.4 $& $0.71\pm0.03$ &  $0.46\pm0.04$ & $2.6_{-2.1}^{+2.7} \pm 0.7\pm 0.3$ & $0.70\pm0.03$ &  $0.41\pm0.05$\\ 
			\hline
		\end{tabular}
	\end{center}
\end{table*}

\subsection{Estimation of the weak lensing mass modelling bias}
\label{Sec:corr_for_mass_modelling_bias}

Weak lensing mass estimates can suffer from systematic biases caused by deviations of the cluster from an NFW profile, triaxial or complex mass distributions (e.g. due to mergers), both correlated and uncorrelated large-scale structure, and miscentring of the fitted shear profile. The measured weak lensing mass $M_{\Delta,\mathrm{WL}}$ at an overdensity $\Delta$ is typically smaller than the true mass of the halo $M_{\Delta,\mathrm{halo}}$ by a factor
\begin{equation}
    b_{\Delta,\mathrm{WL}} = \frac{M_{\Delta,\mathrm{WL}}}{M_{\Delta,\mathrm{halo}}} \,.
    \label{Eq:WLmass+bias+halomass}
\end{equation}
This bias also depends on the specific properties of the sample such as mass and redshift and the measurement setup regarding the employed concentration--mass relation and radial fitting range.

In this study, we obtained an estimate for the weak lensing mass bias distribution following the method described by \citet{Sommer2021}. They showed that the traditional, simplifying assumption of a log-normal bias distribution according to 
\begin{equation}
    \ln \left( \frac{M_{\Delta,\mathrm{WL}}}{M_{\Delta,\mathrm{halo}}} \right) \sim \mathcal{N}(\mu, \sigma^2)
\end{equation}
is a suitable choice in the absence of miscentring. Here, $\mathcal{N}(\mu, \sigma^2)$ is the log-normal distribution with expectation \mbox{value $\mu = \langle \ln b_{\Delta,\mathrm{WL}} \rangle$} and variance $\sigma^2$. The expectation value $\mu$ in log-space translates to a measure of the bias in linear space via the estimator 
\begin{equation}
    \hat{b}_{\Delta,\mathrm{WL}} = \exp [\langle \ln b_{\Delta,\mathrm{WL}} \rangle] \,.
\end{equation}
Following \citet{Sommer2021}, we used snapshots of the Millennium XXL simulations \citep[MXXL,][]{Angulo2012} at redshift $z = 1$ to estimate the weak lensing mass bias distribution. We obtained an estimate for each cluster individually by incorporating the given SZ mass and uncertainties of the measured radial tangential shear profile as input information. First, we used all haloes in the MXXL simulations with a halo mass within $2\sigma$ of the SZ mass of the respective cluster (see Table \ref{tab:Cluster sample properties}). Their mass distributions were projected along three mutually orthogonal axes increasing the effective sample size. We note that we did include a line of sight integration length of \mbox{$200\,h^{-1}$\,Mpc} and not the full line of sight. Consequently, this method takes into account only correlated but not uncorrelated large-scale structure. However, integration along a line of sight twice as long changes the mean results only marginally \citep[][]{Becker2011}. The projected mass distributions of the massive haloes served to calculate the shear and convergence fields on a grid with four arcsecond resolution. We converted the shear to the reduced shear using the same average lensing efficiency as in the respective cluster observations. This reduced shear field was azimuthally averaged in the same range and bins as in the cluster analysis to obtain a reduced shear profile. As the centre, we used either the 3D halo centre (most bound particle) or an offset centre drawn from an empirical miscentring distribution. We added noise to the reduced shear profile in each radial bin matching the corresponding uncertainties of the actual cluster tangential reduced shear estimates. We then obtained a weak lensing mass estimate by fitting the tangential reduced shear profile with an NFW profile, analogous to the analysis in our actual cluster observations. Subsequently, the comparison of the obtained weak lensing mass with the true halo mass provided the estimate for the weak lensing mass bias distribution for our specific setup. The full probability distribution  $P(M_{\Delta,\mathrm{WL}}|M_{\Delta,\mathrm{halo}})$ was modelled with the help of Bayesian statistics as described in \citet{Sommer2021}, where the SZ-derived mass estimates ($M_{200\mathrm{c},\mathrm{SZ}}$ and $M_{500c,\mathrm{SZ}}$) from \citetalias{Bocquet2019} served as a prior for the mass estimation. Thus, we did not take into account any mass dependence of the bias other than using the SPT-SZ masses as a prior. 

We incorporated miscentring into the estimation of the weak lensing mass bias distribution by applying an offset in a random direction before obtaining the reduced shear profile and subsequently fitting the masses. The offset was drawn from a miscentring distribution derived from the Magneticum Pathfinder Simulation \citep[][]{Dolag2016}{} measuring the offset between X-ray (or SZ) peaks from the simulation as a proxy for the centre and the position of the most bound particle \citepalias[see ][for a detailed description]{Schrabback2021}{}. We note that the log-normal assumption does not hold anymore for the weak lensing mass bias distribution in case of miscentring. However, the deviation is at the 3-5 per cent level regarding the true mass. Therefore, we could still obtain meaningful estimates of the mean bias and scatter from a log-normal fit.

We found that the weak lensing mass bias distribution is nearly independent of mass within the $2\sigma$ bounds of the given SZ-derived mass of the respective clusters. Thus, we averaged the bias and scatter over this mass range and report the results in Tables \ref{tab:mass-Xray} and \ref{tab:mass-SZ}. We found that the clusters exhibit a weak lensing mass bias $\hat{b}_{\Delta,\mathrm{WL}}$ between 0.74 and 0.92 in the presence of miscentring (using X-ray centres) with a scatter $\sigma$ between 0.25 and 0.48 regarding the weak lensing masses $M_{500c}$. On average, the masses computed with the X-ray centres are slightly less biased with a slightly smaller scatter when compared to the masses computed with the SZ centre (see Tables \ref{tab:mass-Xray} and \ref{tab:mass-SZ}). This is a result of the on average smaller offsets of the X-ray miscentring distribution compared to the offsets of the SZ miscentring distribution \citepalias{Schrabback2021}.

We note that we have derived these estimates from the MXXL snapshot at $z = 1$. \citetalias{Schrabback2021} report weak lensing mass bias estimates, which are interpolated between results at $z = 0.25$ and $z = 1$ according to the given cluster redshift. We found that the results using the $z = 0.25$ snapshot are very similar to those at $z = 1$. This suggests that there is no strong redshift evolution, and we decide to report the results from the $z = 1$ snapshot, closest to the redshift range of our sample.

%--------------------------------------------------------------------

\section{Constraints on the SPT observable-mass scaling relation}
\label{Sec:ScalingRelAnalysis}

In this section, we present how we combined the weak lensing mass measurements of our nine high-redshift SPT clusters with results for clusters at lower redshifts, namely weak lensing mass measurements of 19 SPT clusters with redshifts $0.29 \leq z\leq 0.61$ based on Magellan/Megacam observations \citepalias[][sample Megacam-19]{Dietrich2019} and of 30 SPT clusters with redshifts $0.58 \leq z \leq 1.13$ based on \textit{HST} observations \citepalias[][sample HST-30]{Schrabback2021}. 
We used this sample of in total 58 SPT clusters (we refer to it as HST-39 + Megacam-19) with weak lensing mass measurements to constrain the SPT observable-mass scaling relation.
Thereby, we extended the previous studies 
\citepalias{Schrabback2018,Dietrich2019,Bocquet2019,Schrabback2021}
out to redshifts of up to $z = 1.7$. 

\subsection{Likelihood formalism for the observable-mass scaling relation}
\label{Sec:Likelihood formalism}

In this section, we briefly summarise our likelihood formalism. It follows the definitions in \citetalias{Dietrich2019}, \citetalias{Bocquet2019}, and \citetalias{Schrabback2021}, which we refer the reader to for further details.

The SPT observable-mass scaling relation is based on the measured detection significance $\xi$ as a mass proxy. Its relation to the unbiased detection significance $\zeta$ can be quantified from simulations \citep{Vanderlinde2010} or analytically \citep{Zubeldia2021} and exhibits a scatter given by a Gaussian of unit width 
\begin{equation}
    P(\xi | \zeta ) = \mathcal{N} \left(\sqrt{\zeta^2 + 3},1\right) \,.
\end{equation}
Further following \citetalias{Bocquet2019} and \citetalias{Schrabback2021}, we define the scaling relation between the unbiased detection significance $\zeta$ and the mass $M_{500c}$ as a power-law in mass and the dimensionless Hubble parameter $E(z) \equiv H(z)/H_0$:
\begin{equation}
    \langle \ln \zeta \rangle = \ln \left[ \gamma_\mathrm{field} A_\mathrm{SZ} \left(\frac{M_{500c}}{3\times10^{14}\mathrm{M}_\odot /h}\right)^{B_\mathrm{SZ}} \left(\frac{E(z)}{E(0.6)}\right)^{C_\mathrm{SZ}} \right] \,,
\end{equation}
where $A_\mathrm{SZ}$, $B_\mathrm{SZ}$, and $C_\mathrm{SZ}$ parametrise the normalisation, mass slope, and redshift evolution, respectively, and $\gamma_\mathrm{field}$ characterises the effective depth of the individual SPT fields. Since we want to constrain this relation with the help of weak lensing mass measurements, we additionally need to consider the relation between lensing mass and true mass (see Eq. \ref{Eq:WLmass+bias+halomass}). We set $\Delta = 500c$ and omit this notation in this section for readability, so that the relation reads
\begin{equation}
    \ln \langle M_\mathrm{WL}\rangle = \ln b_\mathrm{WL} +  \ln M \,.
\end{equation}
Combining both relations, we therefore obtain the joint relation
\begin{equation}
    P\left(\left[ \myvec{\ln \zeta \\ \ln M_\mathrm{WL}}\right] | M,z \right) = \mathcal{N} \left( \left[\myvec{\langle \ln \zeta \rangle (M,z) \\ \langle \ln M_\mathrm{WL}\rangle (M,z) } \right] , \Sigma_{\zeta - M_\mathrm{WL}} \right) \,,
    \label{Eq:joint-scaling-rel}
\end{equation}
where the covariance matrix $\Sigma_{\zeta - M_\mathrm{WL}}$ summarises how the logarithms of the observables $\zeta$ and $M_\mathrm{WL}$ scatter. It is given by
\begin{equation}
    \Sigma_{\zeta - M_\mathrm{WL}} = \left( \myvec{ \sigma^2_{\ln \zeta} &  \rho_{\mathrm{SZ} - \mathrm{WL}}\sigma_{\ln \zeta} \sigma_{\ln M_\mathrm{WL}}\\ \rho_{\mathrm{SZ} - \mathrm{WL}}\sigma_{\ln \zeta} \sigma_{\ln M_\mathrm{WL}} & \sigma^2_{\ln M_\mathrm{WL}} } \right) \,.
\end{equation}
The quantities $\sigma_{\ln \zeta}$ and $\sigma_{\ln M_\mathrm{WL}}$ denote the widths of the normal distributions, which characterise the intrinsic scatter in $\ln \zeta$ and $\ln M_\mathrm{WL}$, respectively. They are assumed to be independent of redshift and mass. Correlated scatter between the SZ and the weak lensing observable is described by the correlation coefficient $\rho_{\mathrm{SZ} - \mathrm{WL}}$.

We note that the weak lensing observable is not the mass $M_\mathrm{WL}$, but rather the tangential reduced shear $g_\mathrm{t}$. Therefore, the likelihood for each cluster reads
\begin{equation}    
    \begin{split}
        P(g_\mathrm{t} | \xi, z, \boldsymbol{p}) &= \iiint \mathrm{d}M\, \mathrm{d}\zeta\, \mathrm{d}M_\mathrm{WL} \\
        & \times [ P(\xi | \zeta )  P(g_\mathrm{t}| M_\mathrm{WL}, N_\mathrm{source}(z), \boldsymbol{p}) \\
        & \times P(\zeta, M_\mathrm{WL} | M, z, \boldsymbol{p}) P(M | z, \boldsymbol{p}) ]\,.
    \end{split}
\end{equation}
Here, $P(\zeta, M_\mathrm{WL} | M, z, \boldsymbol{p})$ is the joint scaling relation introduced in Eq. (\ref{Eq:joint-scaling-rel}) and $P(M | z, \boldsymbol{p})$ denotes the halo mass function by \cite{Tinker2008}. It represents a weighting required to account for Eddington bias. The vector $\boldsymbol{p}$ summarises the astrophysical and cosmological modelling parameters. Furthermore, the source redshift distribution is given by  $N_\mathrm{source}(z)$ and the terms $P(\xi | \zeta )$ and $ P(g_\mathrm{t}| M_\mathrm{WL}, N_\mathrm{source}(z), \boldsymbol{p})$ contain information about the intrinsic scatter and observational uncertainties in the observables\footnote{We note that we already included the shape noise of the tangential reduced shear profiles when we quantified the mass modelling bias in Sect. \ref{Sec:corr_for_mass_modelling_bias}. However, the scatter $\sigma(\mathrm{ln}\, b_\mathrm{\mathrm{500c,WL}})$ of the weak lensing mass modelling bias changes only marginally for a noiseless estimation of the bias, so that our scaling relation results are not affected.}. 
Finally, the total log-likelihood corresponds to the sum of logarithms of the individual cluster likelihoods
\begin{equation}
    \ln \mathcal{L} = \sum_{i=1}^{N_\mathrm{cl}} \sum_{j=1}^{N_\mathrm{bin}}  \ln P(g_{\mathrm{t},ij} | \xi_{ij}, z_{ij}, \boldsymbol{p}) \,,
\end{equation}
where $N_\mathrm{cl} = 58$ is the total number of clusters considered to obtain constraints on the SPT observable-mass scaling relation and $N_\mathrm{bin}$ is the number of radial bins for the reduced shear profiles. We note that we naturally accounted for the selection function of the sample because we applied the established likelihood formalism only to the clusters from the SPT-SZ survey. Furthermore, the subsamples of clusters with weak lensing measurements were assembled randomly, independent of their lensing signal, so that the likelihood function is complete and does not suffer from biases due to weak lensing selections \citepalias{Dietrich2019,Bocquet2019}. In particular, this means that we also included the clusters that were not detected with a peak in the mass maps (see Sect. \ref{Sec:Mass-maps}), because we would otherwise have introduced unwanted selection effects.

We cannot constrain all parameters in this relation equally well with the current weak lensing mass measurements. In particular, our data set does not allow for meaningful constraints for $B_\mathrm{SZ}$ and $\sigma_{\ln \zeta}$ \citepalias{Schrabback2021}. Thus, we introduced the following priors. Regarding the slope parameter, we used a Gaussian prior  $B_\mathrm{SZ}\sim \mathcal{N}(1.53,0.1^2)$, which is motivated by the cosmological study in \citetalias{Bocquet2019}. We assumed $\sigma_{\ln \zeta} \sim \mathcal{N}(0.13,0.13^2)$ as used by \citet{deHaan2016} and derived based on mock observations of hydrodynamic simulations from \citet{LeBrun2014}. Additionally, we implemented the weak lensing mass modelling bias and corresponding scatter obtained in Sect. \ref{Sec:corr_for_mass_modelling_bias} and adopted a flat prior for the correlation coefficient, that is  $\rho_{\mathrm{SZ} - \mathrm{WL}} \in [-1,1]$.

We conducted the likelihood analysis with an updated version of the pipeline used in \citetalias{Bocquet2019} and \citetalias{Schrabback2021}, which is embedded in the \texttt{COSMOSIS} framework \citep{Zuntz2015} and where the likelihood is explored with the MULTINEST sampler \citep{Feroz2009}. The full, updated pipeline will be made available along with a future publication by Bocquet et al. (in prep.). 

We tested the likelihood machinery with mock cluster data. We simulated an SPT cluster catalogue with SZ detection significances and redshifts. We chose a number density and shape noise resembling the optical observations and implement an average source redshift distribution to simulate weak lensing cluster observations. These served as a basis to generate mock shear profiles, which we used as input for the likelihood analysis. Running the analysis on these mock data, we found that the resulting constraints on the scaling relation meet the expectation, thereby providing a valuable consistency check of our pipeline.

\begin{table*}
	\centering
	\caption{Fit results for the parameters of the $\zeta$--mass relation, analogously to table 12 in \citetalias{Schrabback2021}, now including the weak lensing measurements for the nine high-$z$ SPT clusters from this work. 
        }
	\label{tab:SZ mass-scaling-rel_results}
	\begin{threeparttable}
	\begin{tabular}{l cc ccc} % four columns, alignment for each
		\hline
            \hline
            Parameter & Prior & \multicolumn{2}{c}{HST-39 + Megacam-19} & SPTcl ($\nu\Lambda$CDM) & \textit{Planck} + SPTcl ($\nu\Lambda$CDM) \\
            & & fiducial & binned & \citepalias{Bocquet2019} & (no WL mass calibration) \\
            \hline

            $\ln A_\mathrm{SZ}$ & flat & $1.71\pm0.19$ & -- & $1.67\pm0.16$ & $1.27^{+0.08}_{-0.15}$\\
            $\ln A_\mathrm{SZ}(0.25<z<0.5)$ & flat & -- & $1.74\pm0.23$ & -- & -- \\
            $\ln A_\mathrm{SZ}(0.5<z<0.88)$ & flat & -- & $1.58\pm0.31$ & -- & -- \\
            $\ln A_\mathrm{SZ}(0.88<z<1.2)$ & flat & -- & $1.85\pm0.43$ & -- & -- \\
            $\ln A_\mathrm{SZ}(1.2<z<1.7)$ & flat & -- & $1.89\pm0.81$ & -- & -- \\
            $C_\mathrm{SZ}$ & flat/fixed & $1.34\pm1.00$ & $1.34$ & $0.63^{+0.48}_{-0.30}$ & $0.73^{+0.17}_{-0.19}$\\
            \hline
            \multicolumn{3}{l}{Prior-dominated parameters in our analysis:} &  & &  \\
            $B_\mathrm{SZ}$ & $\mathcal{N}(1.53, 0.1^2)$ & $1.56\pm0.09$ & $1.57\pm0.10$ & $1.53\pm0.09$ & $1.68\pm0.08$\\
            $\sigma_{\ln\zeta}$ & $\mathcal{N}(0.13, 0.13^2)$ & $0.16^{+0.06}_{-0.13}$ & $0.15^{+0.04}_{-0.13}$ & $0.17\pm0.08$ & $0.16^{+0.07}_{-0.12}$\\
            \hline
                    
	\end{tabular}
	
	\textbf{Notes.}
    SPTcl ($\nu\Lambda$CDM) denotes the results from the \citetalias{Bocquet2019} study, which combined SPT cluster counts with weak lensing and X-ray mass measurements. The results from the analysis denoted as \textit{Planck} + SPTcl ($\nu\Lambda$CDM) are based on a combination of measurements from the \textit{Planck} CMB anisotropies \citep[TT,TE,EE+low-E, ][]{Planck2020CMB} and SPT cluster counts.
	
	\end{threeparttable}
\end{table*}

\subsection{Redshift evolution of the $\zeta$--mass relation}

\begin{figure*}
\centering
	\includegraphics[width=1.6\columnwidth]{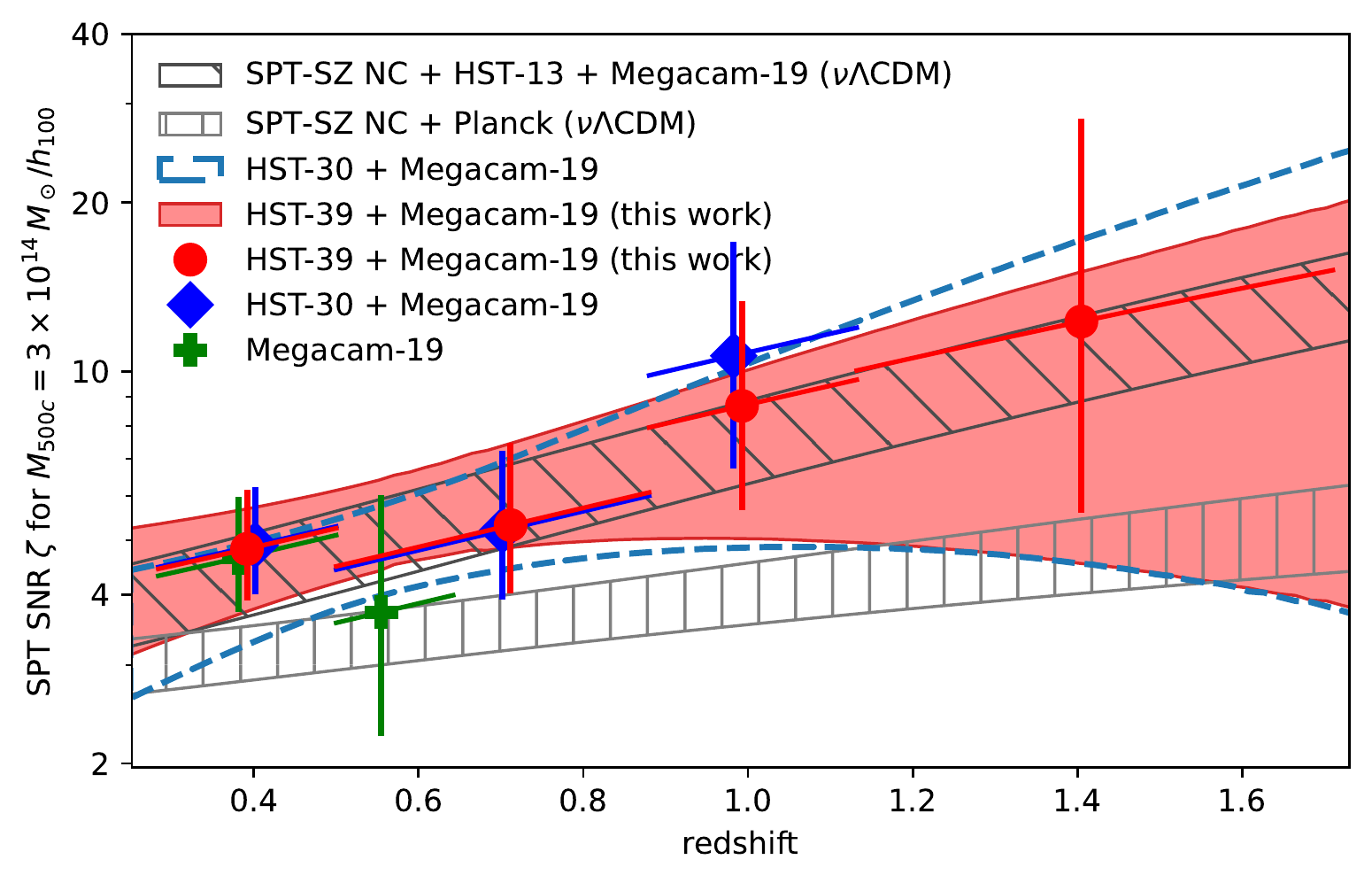}
	\caption{ Evolution of the unbiased SPT detection significance $\zeta$ at the pivot mass $3\times10^{14}\,\mathrm{M}_\odot/h_{100}$ as a function of redshift. The red band indicates the main result of this work. The blue dashed curves indicate the corresponding $1\sigma$ band from the \citetalias{Schrabback2021} analysis for comparison. The red and blue data points represent the corresponding binned analyses. They are placed in the centre of the bins. Horizontal error bars represent the bin widths. The redshift evolution parameter is fixed to $C_\mathrm{SZ} = 1.34$ for our binned analysis.
	The diagonally hatched and vertically hatched bands correspond to the relations from the \citetalias{Bocquet2019} study and the SPT cluster counts in combination with a flat \textit{Planck} $\nu\Lambda$CDM cosmology, respectively. 
	The displayed uncertainties correspond to the 68 per cent credible interval (bands for the full relation and error bars for the binned analysis). }
    \label{Fig:Redshift_evol_of_SZ mass-scaling-rel}
\end{figure*}

We applied the likelihood setup to our full cluster sample of 58 clusters with weak lensing mass measurements to constrain the $\zeta$--mass relation. We present our results in Table \ref{tab:SZ mass-scaling-rel_results}. With our analysis, we constrained the scaling relation parameters \mbox{$A_\mathrm{SZ} = 1.71\pm0.19$} and \mbox{$C_\mathrm{SZ} = 1.34\pm1.00$}, while the parameter $B_\mathrm{SZ}$ is dominated by the prior.
Fig. \ref{Fig:Redshift_evol_of_SZ mass-scaling-rel} displays the redshift evolution of the scaling relation, now for the first time extending out to redshifts up to $z\sim1.7$ (red band, result of the fiducial analysis). For comparison, we show the constraints from \citetalias{Schrabback2021} based on the HST-30 + Megacam-19 samples in blue, demonstrating that our findings in this study are fully consistent with these previous results. This was expected because we added only nine clusters to the previously used sample. In addition, our clusters are at the high-redshift end and therefore the statistical uncertainties are larger compared to clusters at lower and intermediate redshifts.
Furthermore, the diagonally hatched region represents the scaling relation constraints from \citetalias{Bocquet2019}, who analysed weak lensing measurements from the Megacam-19 sample and 13 clusters from \citetalias{Schrabback2018} in combination with X-ray measurements and cluster abundance information. They marginalised over cosmological parameters for a flat $\nu\Lambda$CDM cosmology. 
For comparison, we also show results computed for a joint analysis of \textit{Planck} primary CMB anisotropies \citep[TT,TE,EE+low-E, ][]{Planck2020CMB} and the SPT cluster abundance as the vertically hatched region. Again, this includes a marginalisation over cosmological parameters assuming a flat $\nu\Lambda$CDM cosmology. This analysis does not incorporate any weak lensing mass measurements. 

As also found in \citetalias{Schrabback2021}, we observe an offset between the red and vertically hatched regions implying that the mass scale preferred from our analysis with the weak lensing data sets is lower than the mass scale that would be consistent with the \textit{Planck} $\nu\Lambda$CDM cosmology by a factor of $0.72^{+0.09}_{-0.14}$ (at our pivot redshift of $z=0.6$).

Analogous to \citetalias{Schrabback2021}, we wanted to check if the simple scaling relation model is applicable over the full, wide redshift range investigated here by performing a binned analysis, where the amplitude $A_\mathrm{SZ}$ is allowed to vary individually for each bin. Therefore, we added a bin of $1.2 < z < 1.7$ to the bins that were already used before in \citetalias{Schrabback2021} (namely $0.25 < z < 0.5$, $0.5 < z < 0.88$, and $0.88 < z < 1.2$). We kept the redshift evolution parameter fixed to the value from the fiducial analysis at $C_\mathrm{SZ} = 1.34$. From \mbox{Fig. \ref{Fig:Redshift_evol_of_SZ mass-scaling-rel}}, we can see that the results in our new high-redshift bin are consistent with the scaling relation results from the full unbinned analysis. Additionally, we found that our results in the lower redshift bins are very similar to the results from the binned analysis in \citetalias{Schrabback2021}. This is also expected because the bins contain the same clusters except for SPT-CL{\thinspace}$J$0646$-$6236, which was added to the third redshift bin and causes a small shift towards a higher cluster mass scale due to its large cluster mass.

%--------------------------------------------------------------------

\section{Discussion}
\label{Sec:Discussion}

Weak lensing studies of galaxy clusters with ever higher redshifts face the increasingly difficult challenge to identify background galaxies carrying the lensing signal \citep[e.g. ][]{Mo2016,Jee2017,Finner2020}. In a simplified consideration, the signal-to-noise ratio of a lensing measurement scales with the product of the average geometric lensing efficient $\langle \beta \rangle$ and the square root of the source number density $\sqrt n$. For comparison purposes, we define the weak lensing sensitivity factor $\tau_\mathrm{WL}$ as the product of these two quantities:  $\tau_\mathrm{WL} = \langle \beta \rangle \sqrt{n}$\footnote{In principle, the signal-to-noise ratio of a lensing measurement also depends on other parameters such as cluster mass and fit range. However, the signal-to-noise ratio still scales with the weak lensing sensitivity factor $\tau_\mathrm{WL}$. We use it to represent how the source selection affects the lensing signal-to-noise ratio and compare this quantity for different studies.}. The average geometric lensing efficiency is tied to the purity of the source sample, that is, the fraction of true background source galaxies. A higher purity is desirable as it also increases the average geometric lensing efficiency. At the same time, cuts to identify true background source galaxies should not be too rigorous as this might reduce the overall source density potentially at the cost of also excluding true background galaxies. Additionally, a lower source density is more subject to shot noise, consequently reducing the lensing signal-to-noise ratio.

Some previous weak lensing studies were conducted with \textit{HST}/WFC3 in infrared bands to measure masses of clusters at redshifts $z \gtrsim 1.5$. They introduced varying techniques to select source galaxies for the lensing measurements. For their weak lensing analysis of cluster SpARCS1049$+$56 at redshift $z = 1.71$, \citet{Finner2020} selected sources via a magnitude cut of $H_\mathrm{F160W} > 25.0$\,mag and specific shape cuts aiming to remove galaxies with high uncertainty in the ellipticity measurement and objects that are too small or too elongated to be galaxies. Applying this method to their observations, they achieved a source density of 105\,arcmin$^{-2}$ and estimated an average geometric lensing efficiency of $\langle \beta \rangle = 0.107$. This translates into a signal-to-noise ratio of $\tau_\mathrm{WL} \sim 1.10$. Alternatively, \citet{Jee2017} performed a weak lensing study of clusters SPT-CL{\thinspace}$J$2040$-$4451 and IDCS{\thinspace}J1426$+$3508 at redshifts $z = 1.48$ and $z=1.75$, respectively.
They selected source galaxies requiring that they are bluer than the cluster red-sequence combined with a bright magnitude and shape measurement uncertainty cut. They obtained a source density of $\sim 240$\,arcmin$^{-2}$ with an average lensing efficiency of $\langle \beta \rangle = 0.086$ and $\langle \beta \rangle = 0.120$ for IDCS{\thinspace}J1426$+$3508 and SPT-CL{\thinspace}$J$2040$-$4451, respectively. This corresponds to $\tau_\mathrm{WL} \sim 1.33$ and $\tau_\mathrm{WL} \sim 1.86$, respectively.
 
\citet{Mo2016} conducted a weak lensing study of IDCS{\thinspace}J1426$+$3508 prior to \citet{Jee2017} using \textit{HST}/ACS and \textit{HST}/WFC3 data from the bands F606W, F814W, and F160W. They measured galaxy shapes with the F606W imaging selecting source galaxies with $24.0 < V_\mathrm{F606W} < 28.0$ (the latter is roughly the $10\sigma$ depth limit of their observations), $0\farcs27 < $ FWHM\footnote{measured with \texttt{Source Extractor} } $ < 0\farcs9$ (to exclude too large/small galaxies either because they are likely foreground galaxies or to avoid PSF problems, respectively), and $I_\mathrm{F814W} - H_\mathrm{F160W} < 3.0$ (to exclude cluster red-sequence galaxies). They achieved an average lensing efficiency of $\langle \beta \rangle = 0.086$ at a source density of 89\,arcmin$^{-2}$, resulting in $\tau_\mathrm{WL} \sim 0.81$.

In conclusion, both NIR studies \citep{Jee2017,Finner2020} achieved higher source densities, but lower average geometric lensing efficiencies than our study, which has an average source density of 13.1\,arcmin$^{-2}$ and an average geometric lensing efficiency of \mbox{$\langle \beta \rangle = 0.244$}, and thus \mbox{$\tau_\mathrm{WL} \sim 0.88$}. The studies by \citet{Jee2017} and  \citet{Finner2020} owe the high signal-to-noise ratios mainly to very deep observations enabling high source densities. 
In contrast, our study focuses on a high purity as visible in Figs. \ref{Fig:Low-z redshift distribution} and \ref{Fig:Low-z CANDELS redshift distribution}, which display that we selected almost only high-$z$ sources at $z\gtrsim 2$ with high lensing efficiency, while keeping the contamination of foreground, cluster, and near background galaxies low. This strategy resulted in an average lensing efficiency more than twice as high, and it helps to keep systematic uncertainties low for several reasons. First, excluding galaxies at the cluster redshift minimises uncertainties related to the correction for cluster member contamination. Second, galaxies in the near background are located in a regime where $\beta(z)$ is a steep function of $z$. Thus, systematic redshift uncertainties lead to larger systematic uncertainties in $\langle \beta\rangle$ than for the distant background galaxies selected in our approach. Finally, the efficient removal of foreground galaxies minimises the impact that catastrophic redshift outliers scattering between low and high redshifts have on the computation of $\langle\beta\rangle$ \citepalias[see ][]{Schrabback2018,Raihan2020}. 
While we found that the uncertainties in the redshift distribution (\citetalias{Raihan2020} versus R15\_fix comparison and variations between CANDELS/3D-HST fields) dominate the systematic error budget (see Table \ref{Tab:Errorbudget of photometry,beta}), our comparatively low number density introduced high statistical uncertainties, which (together with other statistical uncertainties) outweigh the systematic ones in our current analysis.
However, we stress that our approach, which aims to limit systematic uncertainties by using data of moderate depth and applying a stringent background selection, could directly be applied to similar data sets obtained for larger cluster samples.

In combination with the considerable measurement uncertainties and the substantial expected intrinsic scatter (see Sect. \ref{Sec:corr_for_mass_modelling_bias}), the best-fitting cluster mass estimates in our study are, therefore, expected to scatter significantly.
This likely explains the relatively low mass estimate of SPT-CL{\thinspace}$J$0205$-$5829, which remained undetected in the weak lensing data despite its high SZ-inferred mass, and the comparably high best-fitting mass estimate for SPT-CL{\thinspace}$J$2040$-$4451. Still, we emphasise that our study aims to provide mass constraints that are accurate on average for our sample of nine galaxy clusters. Indeed, the median ratio of lensing mass to SZ mass from SPT is close to unity. We found a median ratio of bias corrected weak lensing mass to SZ mass $M_\mathrm{500c,WL,corr}/M_\mathrm{500c,SZ}$ of $1.048\pm0.372$ or $1.064\pm0.462$ using the weak lensing masses with X-ray centres (8 clusters) or SZ centres (9 clusters), respectively. We estimated the uncertainties via bootstrapping of the cluster sample.

Deviations between the X-ray or SZ mass and the lensing mass for individual clusters can, for instance, be caused by their different sensitivities to large-scale structure projections, triaxiality, and variations in density profiles. For example, we measured the highest weak lensing mass for the cluster SPT-CL{\thinspace}$J$2040$-$4451, which is notably higher than the expectation from the SZ or X-ray mass estimates.
However, taking the statistical uncertainties of the weak lensing, SZ and X-ray mass estimates into account, as well as the mass modelling bias and scatter, we found that the bias-corrected weak lensing mass agrees with its SZ (X-ray) mass estimate at the $1.2\sigma$ ($1.2\sigma$) level. We used the SZ mass listed in Table \ref{tab:Cluster sample properties} and the X-ray mass \mbox{$M_\mathrm{500c,X-ray} = 3.10^{+0.79}_{-0.47}\times 10^{14}\,\mathrm{M}_\odot$} from \citet{McDonald2017} as reference. We quantified the expected discrepancy between the SZ or X-ray mass and the weak lensing mass further in Appendix \ref{Appendix:Consistency of WL with SZ + X-ray}.
For this particular cluster, \citet{Jee2017} found a weak lensing mass of $M_{200\mathrm{c}} = 8.6^{+ 1.7}_{-1.4}\,\times 10^{14}\,\mathrm{M}_\odot$ (not corrected for mass modelling bias), which is also higher than the X-ray and SZ mass estimates of the cluster. Our weak lensing mass constraint of $M_{200\mathrm{c}}^\mathrm{biased,ML} = 16.4_{-5.7}^{+5.8}\pm 1.6\pm 1.9 \,\times 10^{14}\,\mathrm{M}_\odot$ (for comparability with \citealt{Jee2017} not corrected for mass modelling bias) deviates only by $1.2\sigma$ from the result by \citet{Jee2017}, so that our results confirm the generally higher lensing mass for SPT-CL{\thinspace}$J$2040$-$4451 (albeit with larger statistical uncertainties), suggesting potential line of sight effects. This conclusion is additionally supported by a high dynamical mass measurement (albeit with large uncertainties) by \citet{Bayliss2014}.  

Several differences in the analyses especially regarding the source selection strategies and fit ranges may explain the difference between the lensing masses from \citet{Jee2017} and our study. \citet{Jee2017} obtained their weak lensing mass constraint from \textit{HST}/WFC3 imaging in F105W, F140W, and F160W.
They fitted a spherical NFW profile assuming the concentration--mass relation of \citet{Dutton2014} and centred at their measured X-ray peak position (from \textit{Chandra} data), including weak lensing sources outside of a minimum radius \mbox{$r_\mathrm{min} = 26$\,arcsec}, corresponding to 218\,kpc at the cluster redshift. The WFC3/IR observations by \citet{Jee2017} provide a full azimuthal coverage out to $r \lesssim 60$\,arcsec, while we have $r\lesssim 90$\,arcsec ($r\lesssim 72$\,arcsec) around the SZ (X-ray) centre in our observations. We note that our inner fit limit ($r_\mathrm{min} = 500$\,kpc) corresponds to an angular radius of 59\,arcsec. Accordingly, our analysis primarily employs reduced shear measurements at larger scales compared to the analysis of \citet{Jee2017}.

Additionally, we measured the weak lensing mass assuming the concentration--mass relation by \citet{Diemer2015} with updated parameters from \citet{Diemer2019}, we centred the fit around the X-ray centroid from \citet{McDonald2017}, which has a distance of 8.1\,arcsec to the X-ray peak employed by \citet{Jee2017}, and we used galaxies outside a minimum radius of $r_\mathrm{min} = 500$\,kpc. We excluded any scales smaller than this to minimise systematic mass modelling uncertainties and the impact of a potential residual cluster member contamination (below the detection limit). Since the X-ray peak and centroid positions are relatively close to each other, it is reasonable to compare the weak lensing mass results without applying the statistical mass modelling correction.

The largest difference between the \citet{Jee2017} study and ours is the source selection strategy. \citet{Jee2017} based their work on imaging that is significantly deeper (with a limiting magnitude of F140W $\sim 28$\,mag) than ours but limited to a smaller field of view. Their selection of background galaxies focussed on the exclusion of red-sequence galaxies (galaxies at $\mathrm{F105W} - \mathrm{F140W} < 0.5$ are selected) and resulted in a source number density of $\sim 240\,\mathrm{arcmin}^{-2}$ with a fraction of non-background sources (with $z \leq z_\mathrm{cluster}$) of approximately 45 per cent. Additionally, the inclusion of scales at \mbox{$218\,\mathrm{kpc}<r<500$\,kpc} likely shrinks statistical uncertainties since the lensing signal is high in the inner regions of the cluster. This allowed them, in turn, to achieve small statistical uncertainties of their weak lensing mass constraints. However, the inclusion of such core regions usually increases the intrinsic scatter and mass modelling uncertainties \citep[][see also Sect. \ref{Sec:corr_for_mass_modelling_bias}]{Sommer2021}. Our more strict selection strategy for the background galaxies based on magnitudes/colours from four bands is contaminated by 17 to 20 per cent of non-background galaxies. The shallower data finally resulted in a source number density of 11.2\,arcmin$^{-2}$ for SPT-CL{\thinspace}$J$2040$-$4451 so that our analysis exhibits substantially larger statistical uncertainties in the weak lensing mass constraints. 

\citet{Jee2017} reported the detection of the cluster in their weak lensing mass map at the location \mbox{$\alpha = 20^\mathrm{h}40^\mathrm{m}57\fs85$} and \mbox{$\delta = -44^\circ51^\prime42\farcs4$} with $6\sigma$ significance. In our mass map, we detected a peak at $3.4\sigma$, with a separation of 6.6\,arcsec from the location in \citet{Jee2017}.
While this offset is slightly larger than our estimate of the positional uncertainty derived using bootstrapping (see Table \ref{tab:masspeaklocations}), we note that \citet{Sommer2021} found that bootstrapping substantially underestimates the true uncertainty. The peaks from both studies are close to the X-ray centroid position from \cite{McDonald2017} so that they are overall in agreement. We also note that the peak in our weak lensing mass reconstruction for SPT-CL{\thinspace}$J$2040$-$4451 closely coincides with the X-ray centroid. Accordingly, the shear profile is approximately centred on the position that maximises the lensing signal. This likely scatters the mass result high, especially if the statistical correction for mass modelling bias is applied.

While several studies undoubtedly confirmed SPT-CL{\thinspace}$J$2040$-$4451 as one of the most massive high-redshift clusters known, our study shows that based on our weak lensing measurements, the SPT cluster population is less massive than what one would expect in a {\it Planck} $\Lambda$CDM cosmology, also at very high redshifts (see Sect. \ref{Sec:ScalingRelAnalysis}). 

With our cluster sample and analysis, we enabled constraints on the SZ--mass scaling relation and its redshift evolution for the first time out to the redshift regime of $z>1.2$. While lensing studies at lower redshifts can be calibrated more precisely and systematics are generally smaller, high-redshift clusters are particularly sensitive to probe, for example, models with massive neutrinos \citep{Ichiki2012}, or deviations from standard $\Lambda$CDM expectations, such as early dark energy \citep{Klypin2021}. Therefore, exploring the high-redshift regime is worthwhile to understand the cosmological $\Lambda$CDM model and its possible extensions. Our study provides a first step towards constraints from clusters at redshifts $z>1.2$.

%--------------------------------------------------------------------

\section{Summary and conclusions}
\label{Sec:Summary+Conclusions}

In this work, we studied the gravitational lensing signal of a  sample of nine clusters with high redshifts $z \gtrsim 1.0$ in the SPT-SZ survey. They all exhibit a strong SZ signal with a high SZ detection significance $\xi>6.0$. We obtained weak lensing mass constraints from shape measurements of galaxies with high-resolution \textit{HST}/ACS imaging in the F606W and F814W bands. With the help of additional \textit{HST} imaging using WFC3/IR in F110W and VLT/FORS2 imaging in $U_\mathrm{HIGH}$, we applied a strategy to photometrically select background galaxies, even for clusters at such challenging high redshifts. 

Using updated photometric redshift catalogues computed by \citetalias{Raihan2020} for the CANDELS/3D-HST fields as a reference, we estimated the source redshift distribution and calculated the average geometric lensing efficiency, applying the same selection criteria in the reference photometric redshift catalogues as in the cluster observations. We also added Gaussian noise to the reference catalogues if they were deeper than our cluster observations. We carefully investigated sources of systematic and statistical uncertainties for estimates of the average geometric lensing efficiency. We found consistent results in the HUDF field comparing our photometric measurements employing the algorithm \textsc{LAMBDAR} for adaptive aperture photometry and the \citetalias{Skelton2014} photometric measurements based on fixed aperture photometry. A comparison based on photometric and spectroscopic redshifts revealed a $\sim 3$ per cent difference in calculating the average geometric lensing efficiency, which we accounted for in the weak lensing analysis. 

We reconstructed the projected cluster mass distributions based on the shear measurements of the selected galaxies. In the resulting mass maps, we detected two of the clusters with a peak at $S/N > 3$, four clusters with $S/N > 2$, and three clusters were not detected. We obtained weak lensing mass constraints by fitting the tangential reduced shear profiles with spherical NFW models, employing a fixed concentration--mass relation by \citet{Diemer2015} with updated parameters from \citet{Diemer2019}. We reported statistical uncertainties from shape noise, uncorrelated large-scale structure projections, line of sight variations in the source redshift distribution, and uncertainties in the calibration of the $U_\mathrm{HIGH}$ band. We also estimated mass modelling biases using simulated clusters from the Millennium XXL simulations accounting for miscentring. Masses based on the X-ray centre were less biased ($\hat{b}_{\Delta\mathrm{c,WL}}$) and exhibited a slightly smaller scatter of the mass bias ($\sigma(\mathrm{ln}\, b_{\Delta\mathrm{c,WL}})$) than masses obtained using SZ centres. This is consistent with findings in previous studies \citep[e.g. ][\citetalias{Schrabback2021}]{Sommer2021}. 

We carefully investigated the sources of systematic uncertainties in our study. The total systematic uncertainty of our weak lensing mass estimates amounts to 14.4 per cent (16.7 per cent) for the analyses centring the reduced shear profiles around the X-ray (SZ) centres. Here, the largest contribution (12.9 per cent) comes from uncertainties related to the source selection and calibration of the source redshift distribution (see Table \ref{Tab:Errorbudget of photometry,beta}).

Our weak lensing mass constraints for SPT-CL{\thinspace}$J$2040$-$4451 are higher, but still consistent with the earlier results obtained by \citet{Jee2017}. 
Given the limited depth of our data and the high redshifts of the targeted clusters, our weak lensing mass estimates are relatively noisy. However, on average they are consistent with the SZ-inferred mass estimates from \citetalias{Bocquet2019}, which employ a weak lensing mass calibration based on data from \citet{Dietrich2019} and \citet{Schrabback2018}. We found a median ratio of $1.048\pm0.372$ or $1.064\pm0.462$ using the weak lensing masses with X-ray centres (8 clusters) or SZ centres (9 clusters), respectively.

Finally, we used the obtained weak lensing mass measurements in a joint analysis with measurements for clusters at lower \citepalias{Dietrich2019} and intermediate \citepalias{Schrabback2021} redshifts to constrain the scaling relation between the debiased SPT cluster detection significance $\zeta $ and cluster mass, thereby expanding the previous studies by \citetalias{Bocquet2019} and \citetalias{Schrabback2021} to higher redshifts $z > 1.2$. Our binned analysis of the redshift evolution of the $\zeta $--mass scaling relation revealed that the new highest redshift bin at $1.2 < z < 1.7$ is consistent with the scaling relation behaviour predicted from lower redshifts, albeit with large statistical uncertainties. Even with these large uncertainties at the high redshift end, our results for the full, unbinned analysis support previous findings where the mass scale preferred in an analysis including the weak lensing measurements is lower than the mass scale required for consistency with the \textit{Planck} $\nu\Lambda$CDM cosmology presented in \citet{Planck2020CMB}.

In our pilot study, we developed an approach for weak lensing mass measurements of high-$z$ clusters with well-controlled systematics,
thereby obtaining such measurements for a first significant sample of SZ-selected clusters at \mbox{$z\gtrsim 1.2$}.
However, the small sample size and limited depth of the data imply large statistical uncertainties, which can be addressed by applying the approach to new weak lensing data of additional high-redshift clusters. While statistical uncertainties dominate in our study, there also remain notable systematic uncertainties, which need to be reduced in the future.
Our study shows that the largest systematic uncertainty for lensing studies of high-redshift galaxy clusters arises from the calibration of the source redshift distribution. Here, surveys such as the planned \textit{James Webb Space Telescope} Advanced Deep Extragalactic Survey\footnote{\url{https://pweb.cfa.harvard.edu/research/james-webb-space-telescope-advanced-deep-extragalactic-survey-jades}} (JADES) will help to calibrate the redshift distributions especially for high-redshift clusters, which are observed with deep imaging data. This survey will provide imaging and spectroscopy to unprecedented depth and infer photometric and spectroscopic redshifts over an area of 236\,arcmin$^2$ in the GOODS-South and GOODS-North fields. Additionally, direct calibration methods and those utilizing the stacked redshift probability distribution functions of galaxies already show promising results and need to be further explored to help reduce systematic uncertainties in the redshift calibration \citep[e.g. ][]{Ilbert2021}. Furthermore, in-depth analyses of hydrodynamical simulations will help to better understand and reduce systematics due to the concentration--mass relation, the weak lensing mass modelling, and miscentring distribution uncertainties.

%--------------------------------------------------------------------

\begin{acknowledgements}
This research is based on observations made with the NASA/ESA Hubble Space Telescope obtained from the Space Telescope Science Institute, which is operated by the Association of Universities for Research in Astronomy, Inc., under NASA contract NAS 5-26555. These observations are associated with GO programmes 12477, 13412, 14252, and 14677 (observations of the nine clusters targeted for this study), as well as  
14043, 9425, 12062, and 13872 (archival data in the GOODS-South region).
This work is based on observations taken by the 3D-HST Treasury Program (HST-GO-12177 and HST-GO-12328) with the NASA/ESA Hubble Space Telescope.

This work made use of HDUV Data Release 1.0 data products \citep{Oesch2018}. The team members involved in the HDUV survey are: P. Oesch, M. Montes, N. Reddy, R. J. Bouwens, G. D. Illingworth, D. Magee, H. Atek, C. M. Carollo, A. Cibinel, M. Franx, B. Holden, I. Labbe, E. J. Nelson, C. C. Steidel, P. G. van Dokkum, L. Morselli, R. P. Naidu, S. Wilkins. 

This work is based on observations collected at the European Organisation for Astronomical Research in the Southern Hemisphere under ESO programme 0100.A-0204(A).

This work has made use of data from the European Space Agency (ESA) mission {\it Gaia} (\url{https://www.cosmos.esa.int/gaia}), processed by the {\it Gaia} Data Processing and Analysis Consortium (DPAC, \url{https://www.cosmos.esa.int/web/gaia/dpac/consortium}). Funding for the DPAC has been provided by national institutions, in particular, the institutions participating in the {\it Gaia} Multilateral Agreement.

The Bonn group acknowledges support from  the German Federal Ministry for
Economic Affairs and Climate Action (BMWK) provided through DLR under projects 50OR1803, 50OR2002, 50OR2106, and 50QE2002, as well as support provided  by the Deutsche Forschungsgemeinschaft (DFG, German Research Foundation) under grant 415537506.
HZ, FR, and DS are members of and received financial support from the International Max Planck Research School (IMPRS) for Astronomy and Astrophysics at the Universities of Bonn and Cologne. DS acknowledges support from the European Union's Horizon 2020 research and innovation programme under grant agreement No. 776247. HHo acknowledges support from Vici grant 639.043.512 financed by the Netherlands Organization for Scientic Research.
AHW is supported by an European Research Council Consolidator Grant (No. 770935). 
Argonne National Laboratory's work was supported by the U.S. Department of Energy, Office of High Energy Physics, under contract DE-AC02-06CH11357. 
This work was performed in the context of the South Pole Telescope scientific programme. SPT is supported by the National Science Foundation through grants  OPP-1852617. Partial support is also provided by the Kavli Institute of Cosmological Physics at the University of Chicago.
The authors would like to thank Peter Schneider and the anonymous referee for useful comments, which helped to improve this manuscript.\\
\\
Data availability: The full, updated pipeline, which we used for the likelihood analysis in this work, will be made available along together with an upcoming publication by Bocquet et al. (in prep.). The data underlying this article will be shared upon reasonable request to the corresponding author.

\end{acknowledgements}

  \bibliographystyle{aa} % style aa.bst
  \bibliography{HST-WL-SPT-high-z_A+A} % your references Yourfile.bib

\begin{appendix}

\section{Comparison of S14 and LAMBDAR photometry}
\label{Appendix:Comparison of S14 and LAMBDAR photometry}

While we measured fluxes in our observations with the LAMBDAR software, we only had the \citetalias{Skelton2014} photometry available when we estimated the redshift distribution from the CANDELS/3D-HST fields. Therefore, we checked how consistent we expect our measurements to be with the \citetalias{Skelton2014} photometry. We can perform this check in the central region of the GOODS-South field covering the HUDF, which we observed in the VLT FORS2 $U_\mathrm{HIGH}$ band. In addition to our stack in the $U_\mathrm{HIGH}$ band, we downloaded the stacks\footnote{\url{https://archive.stsci.edu/prepds/3d-hst/}} the \citetalias{Skelton2014} team used in the bands F606W, F814W, F850LP, and F125W (F606W + F850LP: GO programme 9425 with PI M. Giavalisco, F814W: GO programme 12062 with PI S. Faber, F125W: GO programme 13872 with PI G. Illingworth) and measured the photometry on these stacks with LAMBDAR. We used the PSF models provided on the 3D-HST website. We then matched the galaxies in our catalogue with the galaxies in the \citetalias{Skelton2014} photometric catalogue with the \texttt{associate} function from the \texttt{LDAC} tools, requiring a distance of not more than $0\farcs3$ for a match. We interpolated the magnitude $J_{110}$ from our measurements in the filters F850LP and F125W. 

\begin{figure*}
 \centering

	\includegraphics[width=0.95\columnwidth,trim={0.0 0.0 0.0 0.0},clip]{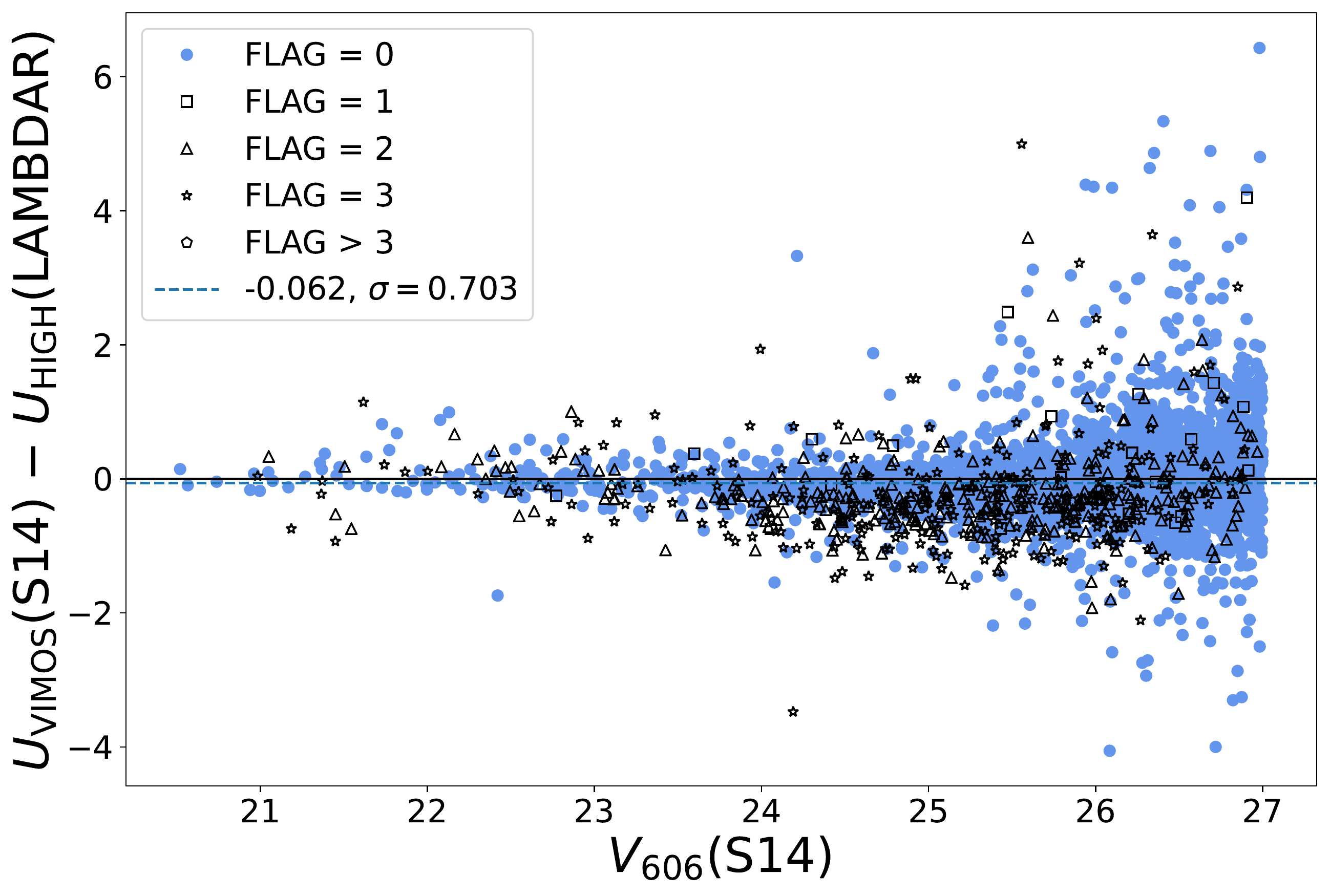}
	\includegraphics[width=0.95\columnwidth,trim={0.0 0.0 0.0 0.0},clip]{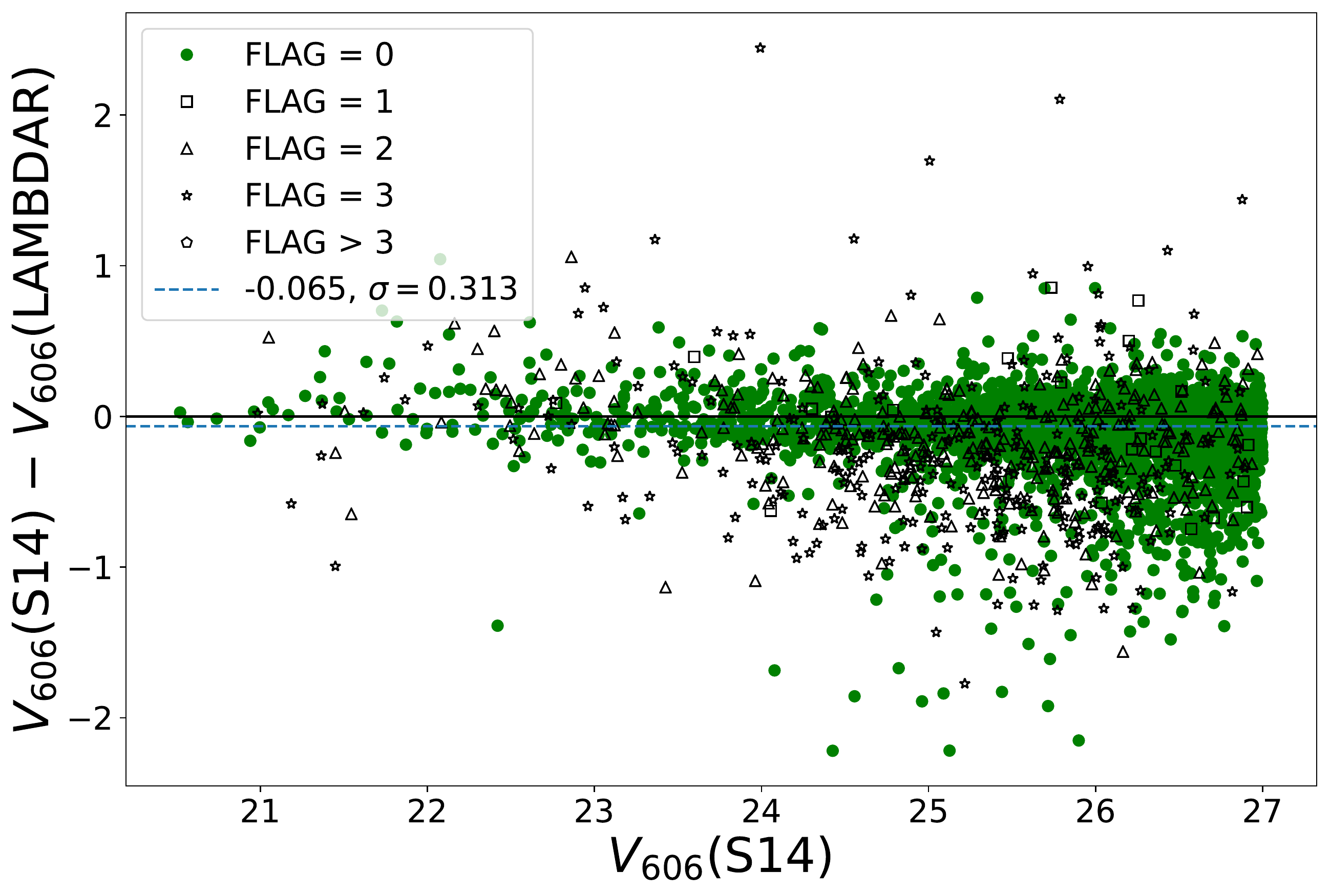}
	\includegraphics[width=0.95\columnwidth,trim={0.0 0.0 0.0 0.0},clip]{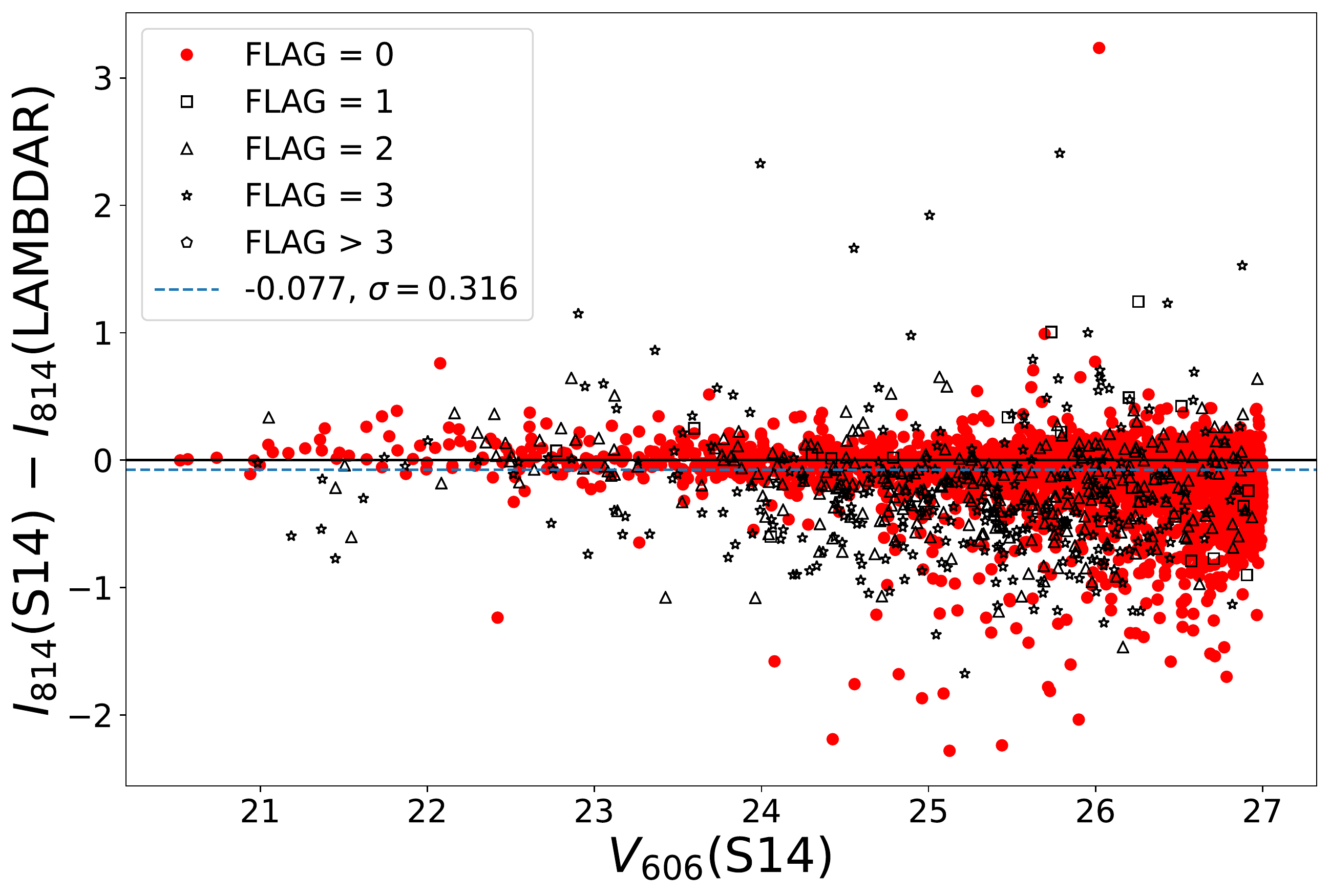}
	\includegraphics[width=0.95\columnwidth,trim={0.0 0.0 0.0 0.0},clip]{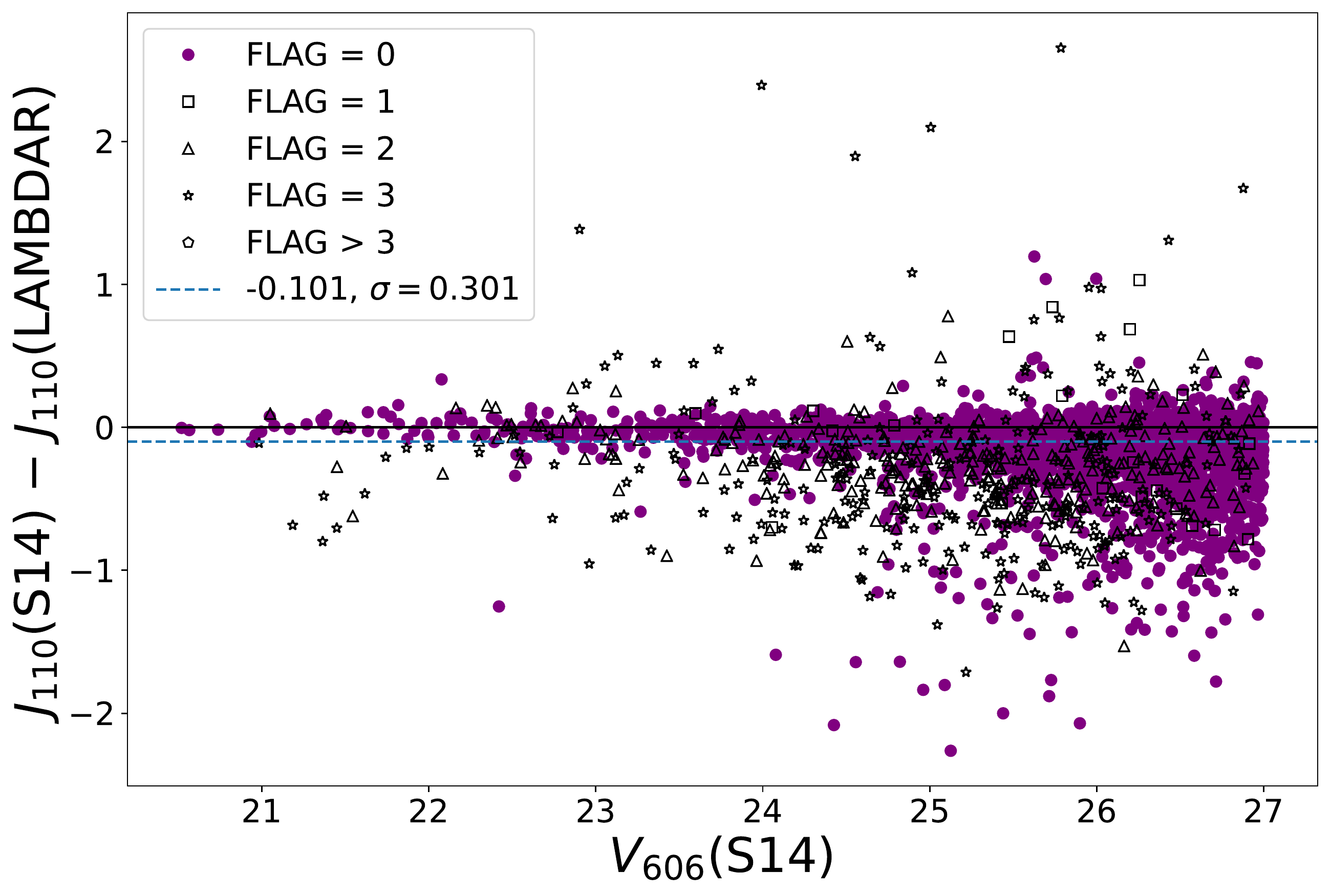}

	\caption{Magnitude differences between \citetalias{Skelton2014} and LAMBDAR photometry for the $U_\mathrm{HIGH}$, $V_{606}$, $I_{814}$, and $J_{110}$ magnitudes. The blue dashed lines represent the median, and we indicate the scatter of the respective bands in the legend label. We show all matched galaxies down to \mbox{$V_{606} < 27.0$}\,mag. We note the different scales on the y-axis for the $U$ magnitudes and the \textit{HST}-based magnitudes.}
    \label{Fig:S14+LAMBDAR photometry comparison - magnitudes}
\end{figure*}

In this appendix, we define all offsets of the magnitudes or colours in terms of \citetalias{Skelton2014} photometry minus LAMBDAR photometry. 
In \mbox{Fig. \ref{Fig:S14+LAMBDAR photometry comparison - magnitudes}}, 
% In \mbox{Fig. F1} (available as supplementary material), 
we show how our magnitude measurements with LAMBDAR compare to the \citetalias{Skelton2014} photometry. We found a negative shift with a median offset of up to $\sim -0.1$\,mag between \citetalias{Skelton2014} and LAMBDAR in all of the \textit{HST} bands with a scatter of $\sim 0.3$\,mag. In part, this negative shift is caused by sources with a \texttt{Source Extractor} detection flag of \mbox{$\mathrm{FLAG} > 0$} (based on our detection in the F606W band). For these sources, \texttt{Source Extractor} recognises, for instance, contamination by nearby sources or blending. We found that the magnitude differences of these sources are predominantly negative in the direct comparison of \citetalias{Skelton2014} and LAMBDAR, meaning that \citetalias{Skelton2014} measurements are systematically brighter than LAMBDAR measurements. This is consistent with the expectation given the measurement techniques. \citetalias{Skelton2014} utilise aperture photometry, where fluxes are measured within apertures of fixed size with a diameter of $0\farcs7$ for \textit{HST} images. In contrast to that, LAMBDAR actively deblends photometry and thus measures fainter magnitudes for blended sources. But also for sources with $\mathrm{FLAG} = 0$, we found a slight asymmetry skewed towards more negative magnitude differences between the \citetalias{Skelton2014} and LAMBDAR photometry. 

For the $U_\mathrm{HIGH}$ band, we found a median offset of $-0.062$\,mag with a scatter of 0.703\,mag, which is a considerably larger scatter than for the \textit{HST} bands. This is likely connected to the difference in depth between the $U_\mathrm{VIMOS}$ stack from \citetalias{Skelton2014} ($5\sigma$ depth 27.4\,mag) and our $U_\mathrm{HIGH}$ stack ($5\sigma$ depth 26.6\,mag) and the difference of the seeing ($0\farcs8$ for $U_\mathrm{VIMOS}$ versus $1\farcs0$ for $U_\mathrm{HIGH}$).
We found that including a conversion from the $U_\mathrm{VIMOS}$ band to the $U_\mathrm{HIGH}$ band based on the respective filter curves does not reduce this scatter. However, Fig. \ref{Fig:S14+LAMBDAR photometry comparison - magnitudes} reveals that the scatter is a strong function of magnitude, suggesting that it is indeed related to the shallower depth of the $U_\mathrm{HIGH}$ data. When limited to bright $V_{606}<25$ galaxies, it reduces to 0.426\,mag.

\begin{figure*}
\centering

	\includegraphics[width=0.95\columnwidth,trim={0.0 0.0 0.0 0.0},clip]{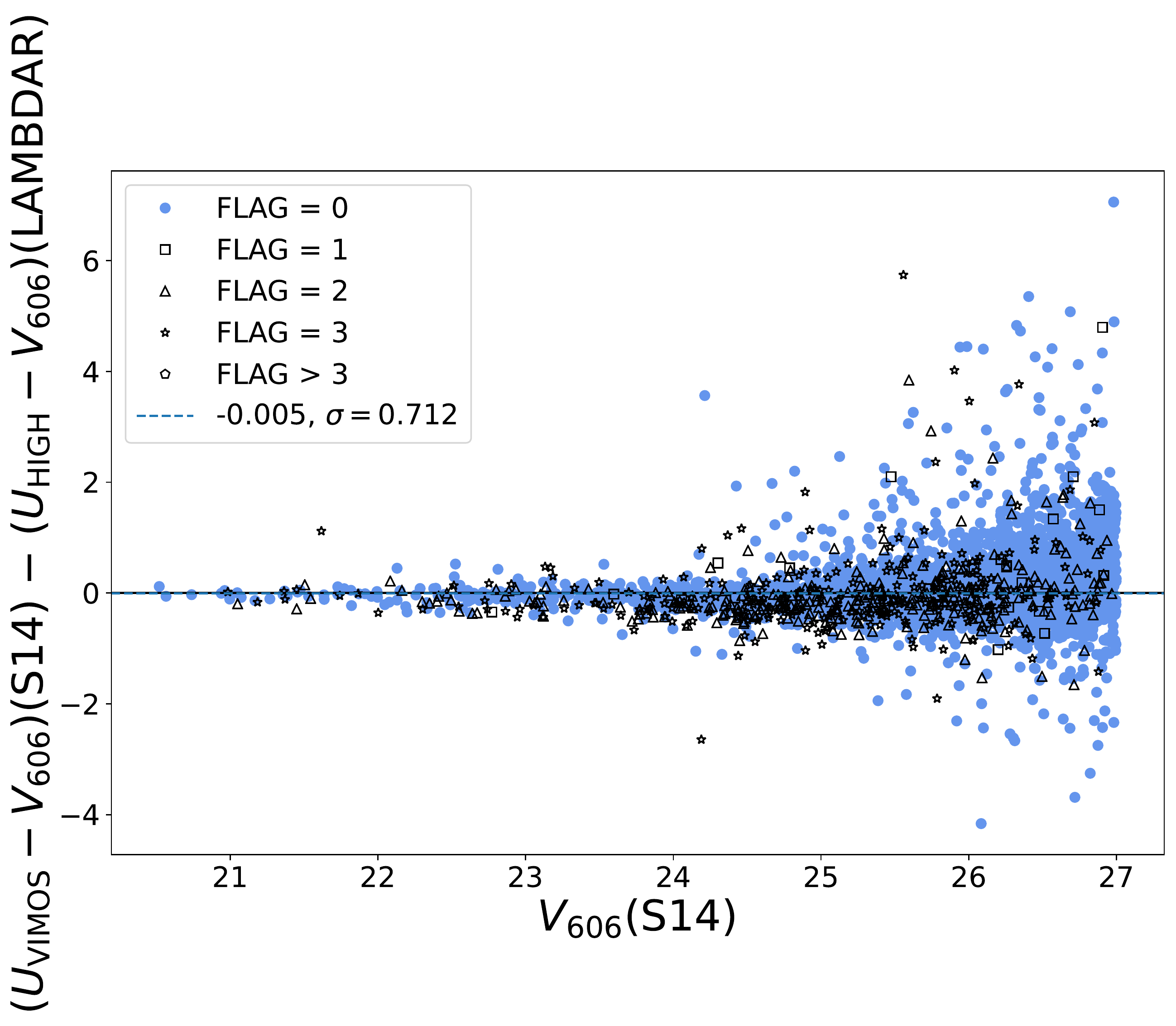}
	\includegraphics[width=0.95\columnwidth,trim={0.0 0.0 0.0 0.0},clip]{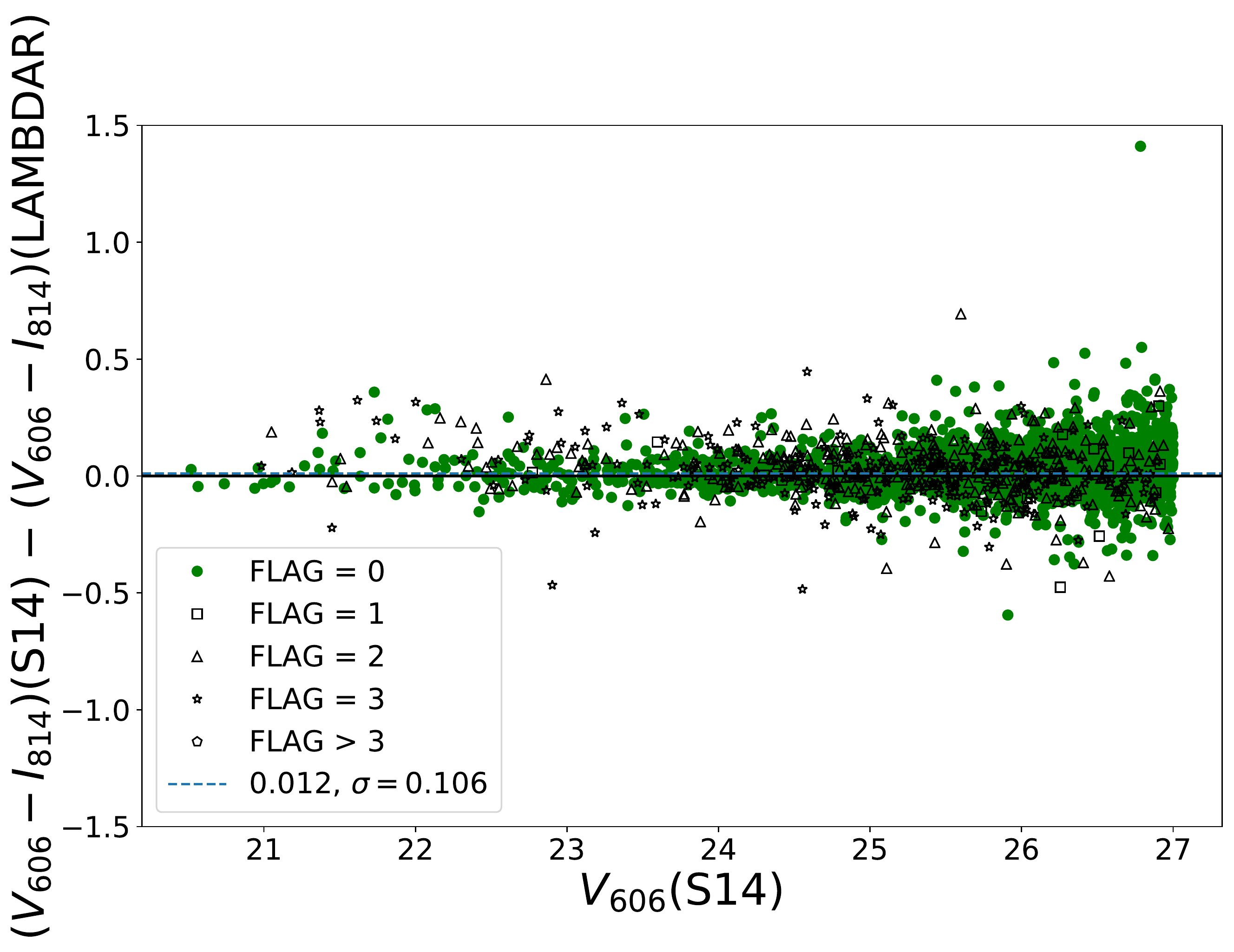}
	\includegraphics[width=0.95\columnwidth,trim={0.0 0.0 0.0 0.0},clip]{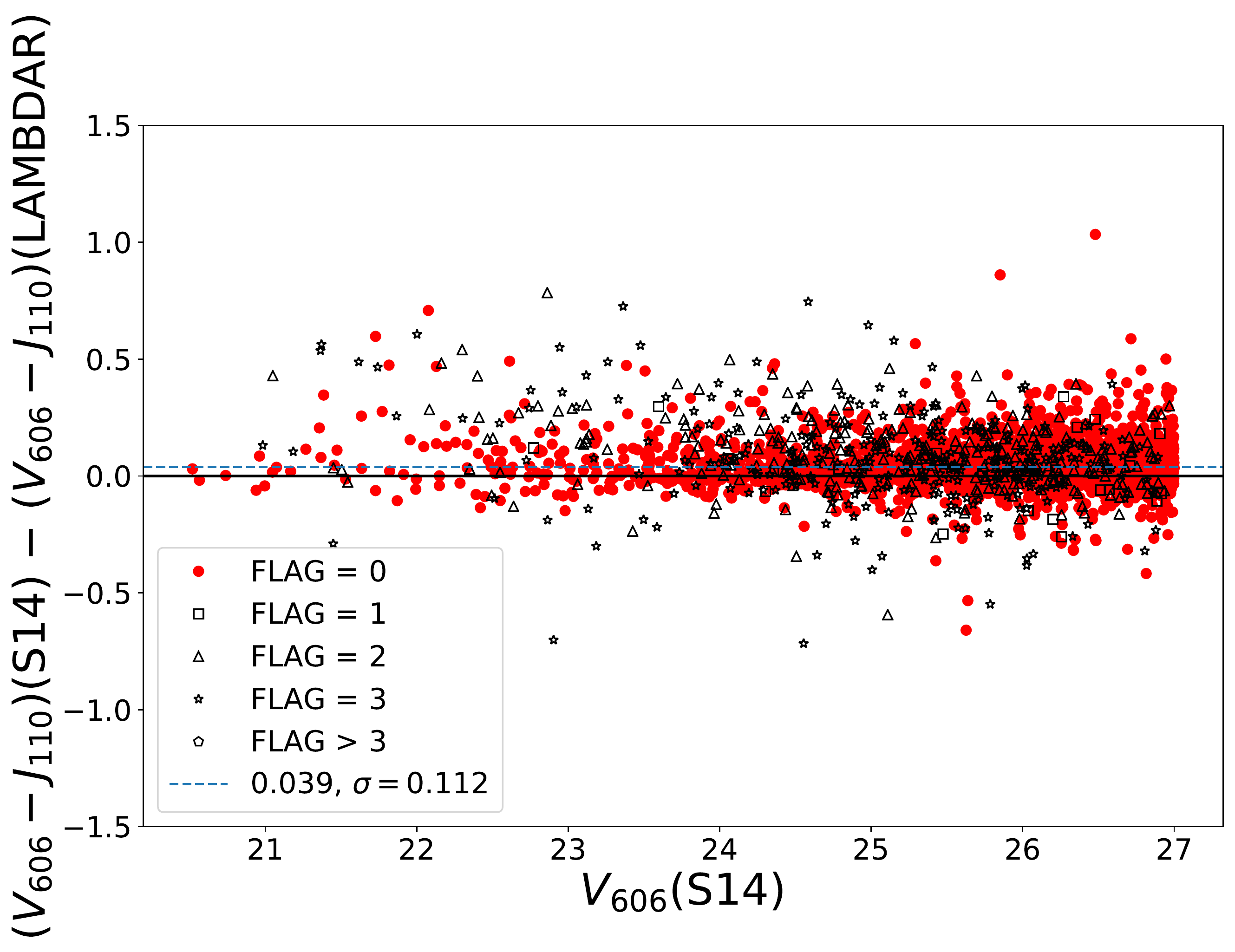}
	\includegraphics[width=0.95\columnwidth,trim={0.0 0.0 0.0 0.0},clip]{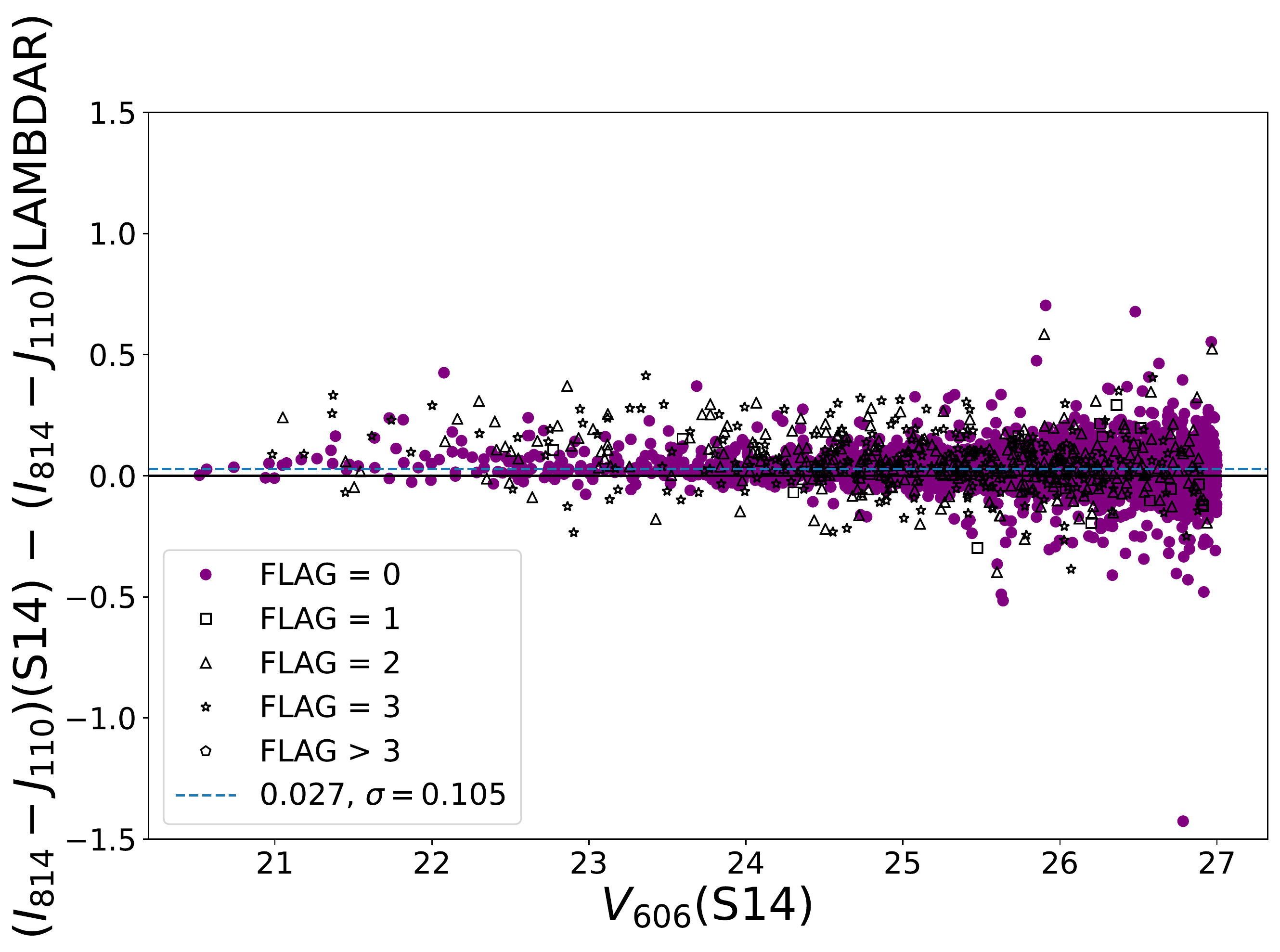}

	\caption{Colour differences between \citetalias{Skelton2014} and LAMBDAR photometry for the colours $U_\mathrm{HIGH} - V_{606}$, $V_{606} - I_{814}$, $V_{606} - J_{110}$, and $I_{814} - J_{110}$. The blue dashed lines represent the median and we indicate the scatter of the respective colours in the legend label. We show all matched galaxies down to \mbox{$V_{606} < 27.0$\,mag}. We note the different scales on the y-axis for the $U-V_{606}$ colour and the \textit{HST}-based colours.}
    \label{Fig:S14+LAMBDAR photometry comparison - colours}
\end{figure*}

Regarding the comparisons of colour measurements (see \mbox{Fig. \ref{Fig:S14+LAMBDAR photometry comparison - colours}}), 
we found slightly positive shifts for all colours based on \textit{HST} bands. In particular, these colours typically exhibited small shifts of up to $\sim 0.04$\,mag with a scatter of up to \mbox{$\sim 0.11$\,mag}. The shift for \mbox{$U_\mathrm{HIGH} - V_{606}$} is $-0.005$\,mag with a scatter of $0.712$\,mag. Systematic shifts of this order will only mildly impact the estimates of the average lensing efficiency $\langle \beta \rangle$, as we show in Appendix \ref{Appendix:Impact of syst. shifts in photom}.
We additionally reduced a data set in the filter F110W (GO programme 14043, PI: F. Bauer) located within the GOODS-South field and compared our F110W photometry with the results from the \citetalias{Skelton2014} photometric catalogues. We found only mild offsets of $-0.010$\,mag and $-0.022$\,mag between the \citetalias{Skelton2014} and our photometry for the colours \mbox{$V_{606}-J_{110}$} and \mbox{$I_{814}-J_{110}$}, respectively.

When we calculated the average lensing efficiency for the cluster fields, we could, in principle, apply the scatter that we measured when comparing the \citetalias{Skelton2014} and LAMBDAR photometry to all CANDELS/3D-HST catalogues to account for the different measurement techniques.  However, we have to keep in mind that the comparison, which we presented here, is limited in some respects: the $U$ bands we compared here have different depths so that we cannot clearly distinguish between effects due to depth and due to the different filter curves of $U_\mathrm{HIGH}$ and $U_\mathrm{VIMOS}$. Additionally, the CANDELS/3D-HST fields employed different $U$ bands, and also each field has different depths in different filters. Therefore, we decided to account for differences in depth in a consistent way for all five CANDELS/3D-HST fields by adding Gaussian noise based on the difference to the depths in our cluster fields (see Table \ref{tab:Cluster sample properties}). However, we did investigate how shifts in the photometry as presented in this section can affect the average lensing efficiency and added the related uncertainties to our error budget (see Table \ref{Tab:Errorbudget of photometry,beta} and Appendix \ref{Appendix:Impact of syst. shifts in photom}).

% \FloatBarrier

\section{Robustness of the photometric zeropoint estimation via the galaxy locus method}
\label{Appendix:ZP robustness with gal locus}

For our $U$ band calibration purposes, we defined the galaxy locus to comprise all galaxies in the magnitude range \mbox{$24.2 < V_{606} < 27.0$}, but excluding galaxies approximately at the cluster redshift \mbox{($1.2 \lesssim z \lesssim 1.7$)} through a cut in the $VIJ$ colour plane (see \mbox{Fig. \ref{Fig:Cuts to remove only cluster gals}}). As described in Sect. \ref{Sec:common photometric system}, we corrected for small shifts in the $U$ band photometry among the five CANDELS/3D-HST fields based on the peak position of highest density in the $UVI$ colour plane. These shifts are listed in \mbox{Table \ref{Tab:Galloc_comparisons}}.

In order to estimate how well the zeropoint calibration of the $U_\mathrm{HIGH}$ band works for the observations of our cluster fields, we tested the zeropoint estimation in the CANDELS/3D-HST fields using only subsets of galaxies that approximately match the number of galaxies available in the cluster fields. Our cluster field observations roughly cover a field of view of 11\,arcmin$^2$. We, therefore, only used galaxies from a region of this size from a random position in the respective CANDELS/3D-HST fields. A number of around 400 to 600 galaxies per subsample belongs to our galaxy locus (as defined by the magnitude and colour cuts in Sect. \ref{Sec: Photometric zeropoints}), which approximately equals the expected number of locus galaxies in our cluster fields. Since we had already applied a shift to the $U$ bands in the CANDELS/3D-HST fields as explained above, this means that we measured the residual zeropoint offset for 100 different (possibly overlapping) subsamples and report the average residual zeropoint offset and scatter in Table \ref{Tab:Galloc_comparisons}.  
Overall, we found that the offsets did not exceed a value of $\sim -0.04$\,mag with a scatter of 0.08\,mag. The impact of such offsets is studied in Appendix \ref{Appendix:Impact of syst. shifts in photom}.

\begin{table}

        \caption{Overview about absolute and residual zeropoint offsets between CANDELS/3D-HST fields. }
	\begin{center}
	\begin{threeparttable}

		\begin{tabular}{ l c c} 
			\hline\hline
			&\multicolumn{2}{c}{Zeropoint offsets}\\
			\cmidrule(lr){2-3}
			Field & full & 100 samples  \\ 
			& [mag] & [mag] \\
			\hline
		    AEGIS & $0.121$ & $-0.013\,,\sigma = 0.053$ \\ 
			COSMOS & $0.121$ & $-0.021\,,\sigma = 0.062$  \\
			UDS & $0.121$ & $-0.037\,,\sigma = 0.076$ \\
			GOODS-North & $-0.040$ & $-0.020\,,\sigma = 0.080$  \\
			GOODS-South & $0.0$ & $-0.027\,,\sigma = 0.055$  \\ 
			\hline
		\end{tabular}
	\end{threeparttable}
		
	\end{center}
		\textbf{Notes.}
		\textit{First column}: Names of the CANDELS/3D-HST fields. \textit{Second column}: Overview about the measured zeropoint offsets in the $U$ band between the galaxy loci from the five CANDELS/3D-HST catalogues from \citetalias{Skelton2014} with respect to the locus in the GOODS-South field, which serves as an anchor. \  \textit{Third column}: Average residual offset computed from 100 subsamples in the CANDELS/3D-HST fields (drawn from areas with a similar field of view as \textit{HST}/ACS)  after applying the `full' correction (second column). The values correspond to the average and scatter.

		\label{Tab:Galloc_comparisons}
		\vspace{-4mm}
\end{table}

% \FloatBarrier

\section{Effect of systematic offsets in the photometry on $\langle \beta \rangle$}
\label{Appendix:Impact of syst. shifts in photom}

\begin{table}

        \caption{
        Impact of expected photometric uncertainties of relevant colours on the average lensing efficiency. 
         }
	\begin{center}
	\begin{threeparttable}

		\begin{tabular}{ l c c c} 
			\hline\hline
			Colour & expected uncert. & $\left( \frac{\Delta \langle \beta \rangle}{\langle \beta \rangle} \right) _\mathrm{HUDF,R20}$ & $\left( \frac{\Delta \langle \beta \rangle}{\langle \beta \rangle} \right) _\mathrm{CAND}$  \\ 
			\hline

		    $U-V_{606}$ & $\pm 0.08$\,mag & 2.7\,\% & 4.1\,\% \\ 
			$V_{606}-I_{814}$ & $\pm 0.02$\,mag & 2.9\,\% & 2.2\,\% \\
			$V_{606}-J_{110}$ & $\pm 0.05$\,mag & 2.7\,\% & 2.2\,\% \\
			$I_{814}-J_{110}$ & $\pm 0.05$\,mag & 0.3\,\% & 0.1\,\% \\
 
			\hline
		\end{tabular}
		
	\end{threeparttable}
	\end{center}
		\textbf{Notes.}
		We quantified this by calculating the difference $\Delta  \langle \beta \rangle$ between the results for $\langle \beta \rangle$ (at reference redshift $z_\mathrm{l} = 1.4$) based on the \citetalias{Skelton2014} photometry shifted by the expected uncertainty in a positive and negative direction. We divide this by the average lensing efficiency $\langle \beta \rangle$ without shift of the photometry.
        \textit{First column:} Colour. \textit{Second column:} Expected uncertainty of the colour. \textit{Third column:} Impact on the average lensing efficiency for matched galaxies in the HUDF region.
        We report the value based on the \citetalias{Raihan2020} photometric redshifts. \textit{Fourth column:} Average impact on the average lensing efficiency for galaxies in the five CANDELS/3D-HST fields using the \citetalias{Raihan2020} photometric redshifts.
		
		\label{Tab:Syst offset in photometry, effect on beta}
		\vspace{-4mm}
\end{table}

In order to estimate how systematic shifts in the photometry affect the average lensing efficiency, we applied different systematic shifts to the colours \mbox{$U-V_{606}$}, \mbox{$V_{606} - I_{814}$}, \mbox{$V_{606} - J_{110}$}, and \mbox{$I_{814} - J_{110}$} from the \citetalias{Skelton2014} photometry. We then calculated $\langle \beta \rangle$ based on the photometric redshifts for the colour-selected galaxies. Since we applied a Gaussian noise to the $U$ band from the GOODS-South field, we evaluated five noise realisations.  
A summary of the uncertainty level of the photometric shifts (based on our results presented in Appendix \ref{Appendix:Comparison of S14 and LAMBDAR photometry} and \ref{Appendix:ZP robustness with gal locus}) and the consequential uncertainties of the average lensing efficiency are presented in Table \ref{Tab:Syst offset in photometry, effect on beta}.

% \FloatBarrier

\section{Alternative colour selection strategies for clusters at $z \sim 1.2$}
\label{Appendix:Coloursel_alternatives}

\begin{figure*}
\centering
	
	\includegraphics[width=0.97\columnwidth]{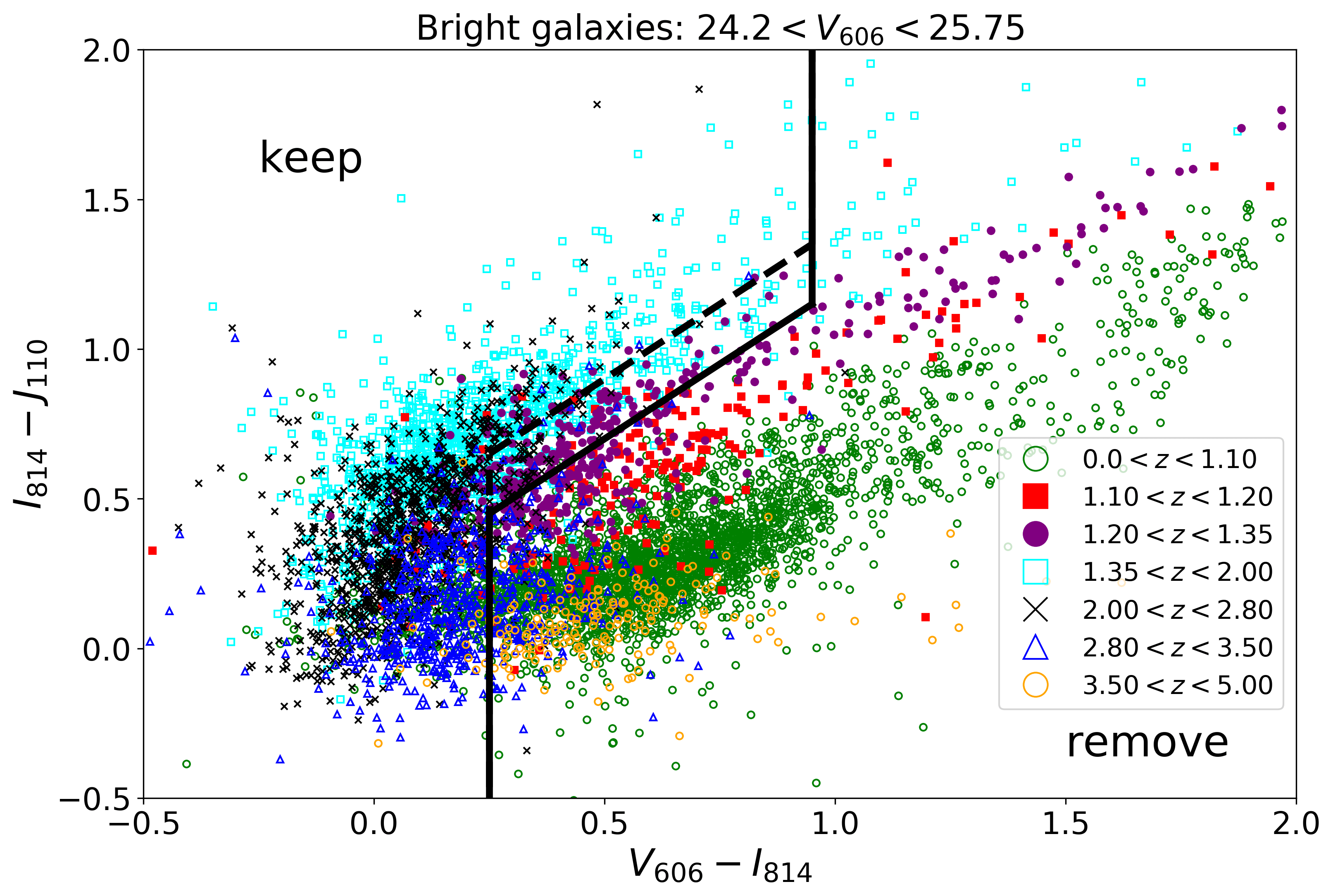}
	\includegraphics[width=0.97\columnwidth]{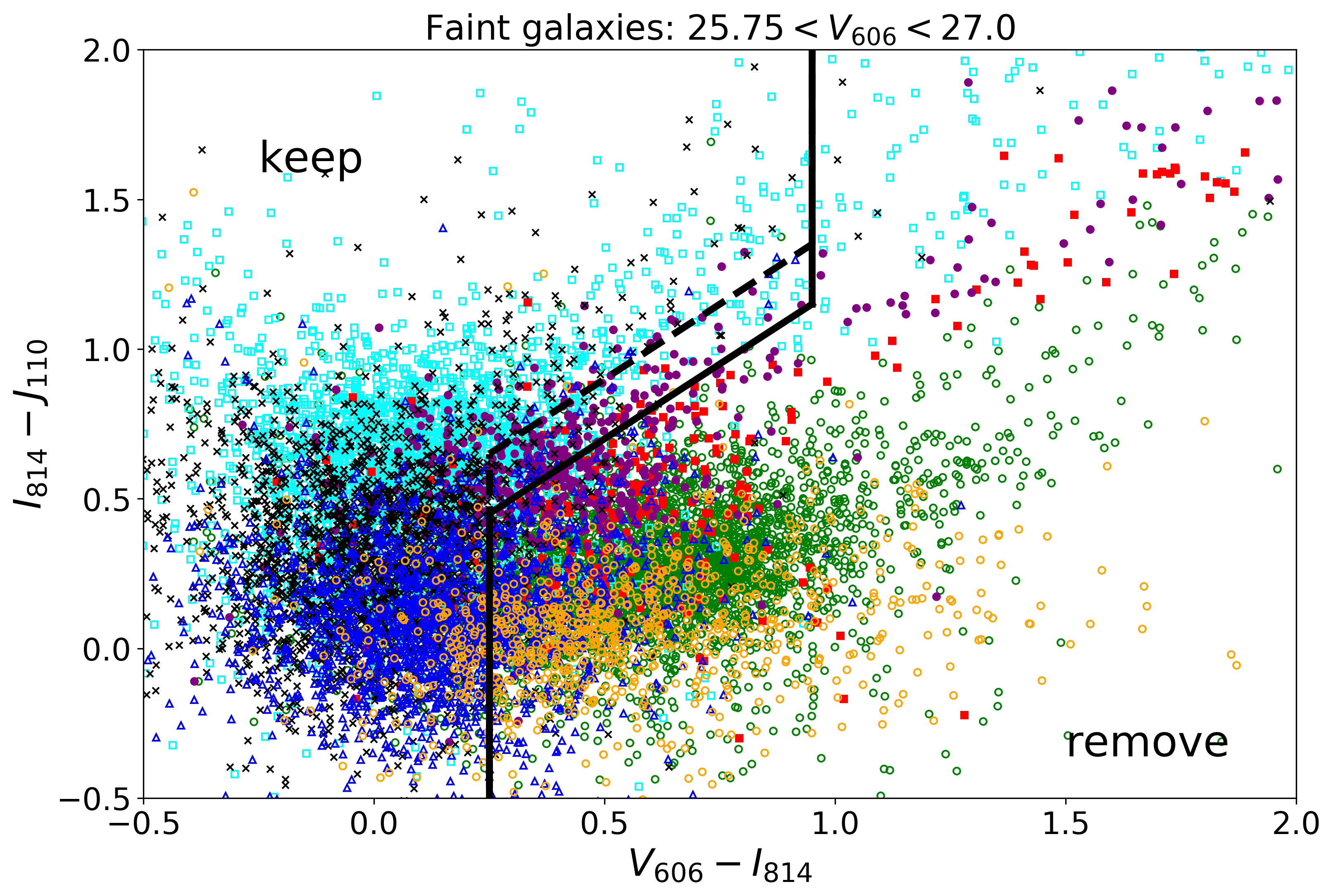}
	\includegraphics[width=0.97\columnwidth]{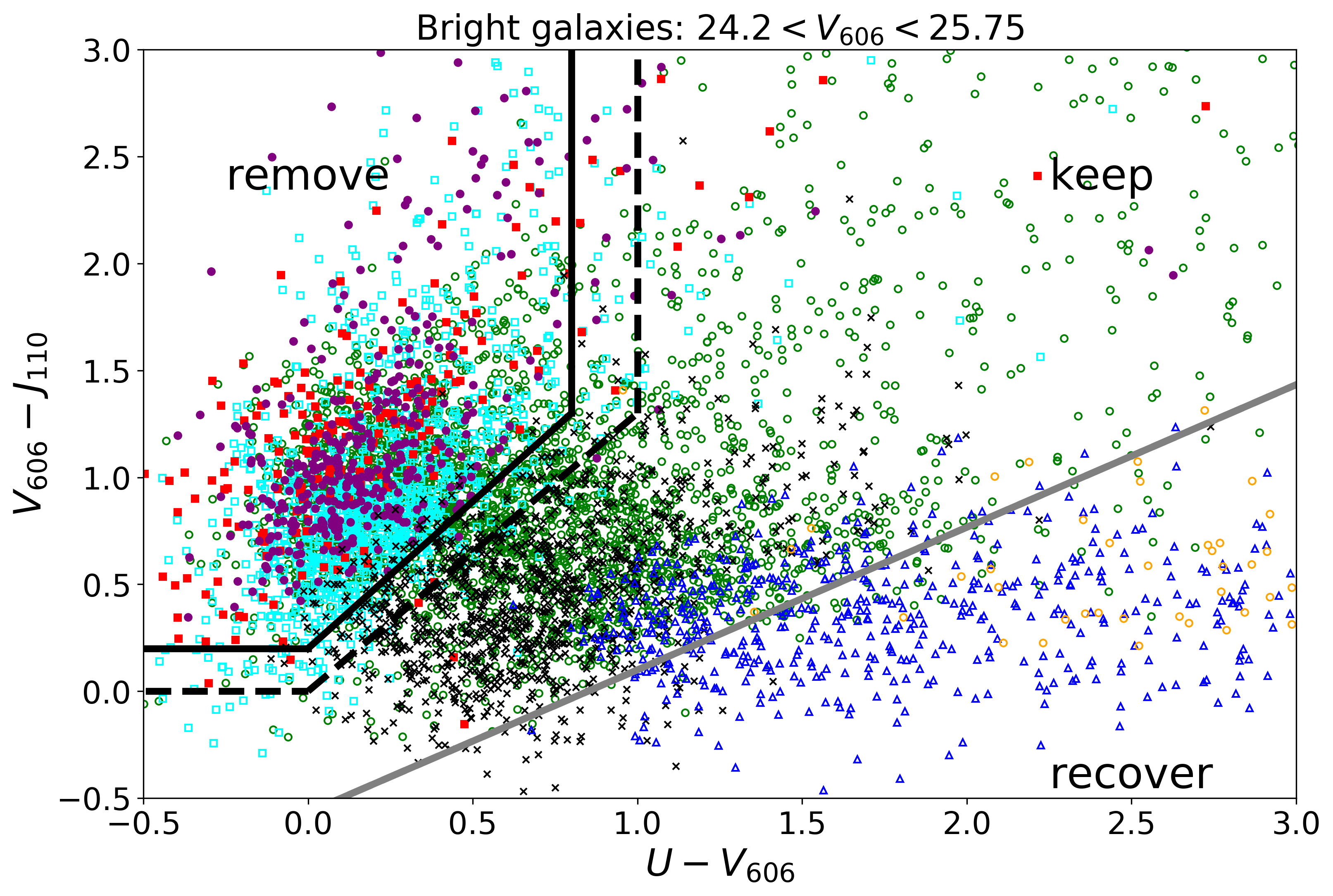}
	\includegraphics[width=0.97\columnwidth]{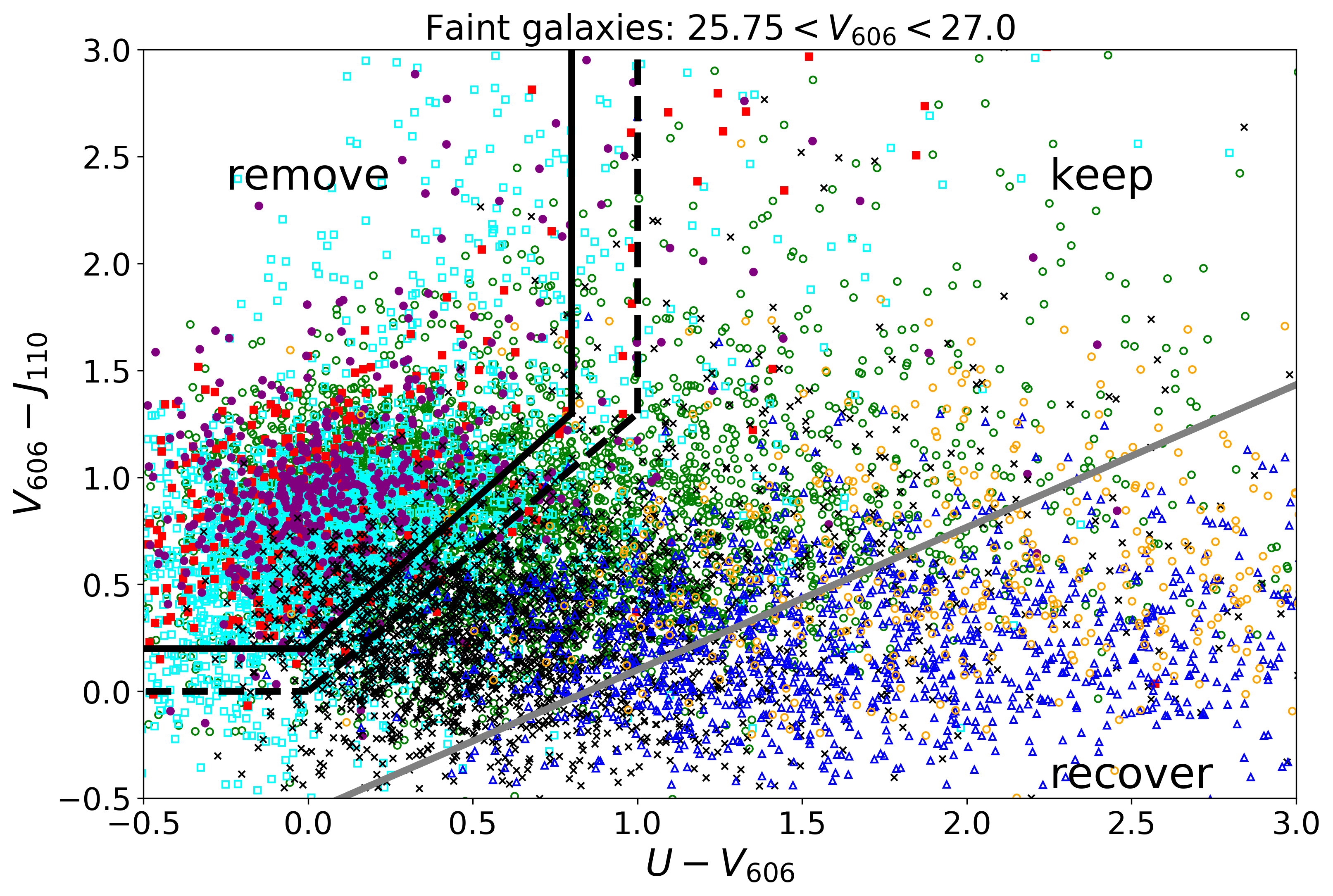}

    \caption{First alternative colour selection for galaxy clusters at redshift \mbox{$ z \sim 1.2$}. The selected source galaxies are at redshift \mbox{$z \gtrsim 1.3$}. We display galaxies based on their photometry from \citetalias{Skelton2014} in the GOODS-South field. \textit{Top:} First selection step in the $VIJ$ plane for bright galaxies on the left and faint galaxies on the right. \textit{Bottom:}  Second selection step in the $UVJ$ plane for bright galaxies on the left and faint galaxies on the right. The solid black lines indicate cuts applied for bright galaxies, the dashed black lines show cuts for faint galaxies. Galaxies below the diagonal grey line are recovered in both the bright and the faint regime.}
    \label{Fig:1.2selection conservative}
\end{figure*}

As mentioned before, galaxies at redshift $1.3 < z < 1.7$ could, in principle, be used for a lensing analysis for a cluster at redshift $z \sim 1.2$, but have to be removed for a cluster at redshift $z \sim 1.7$. We explored two alternative colour selection strategies of background galaxies for a cluster at redshift $z\sim1.2$ aiming to add the galaxies at $1.3 < z < 1.7$ into the selection, which would increase the signal-to-noise ratio of the lensing measurement. In our first alternative, we left the first step of the selection in the $VIJ$ colour plane unchanged, because it serves the removal of (the same) foreground galaxies as in the default selection strategy. However, we noticed that the galaxies at the cluster redshift in the $UVJ$ colour plane occupy a smaller space in the upper left corner. Therefore, we modified the cuts slightly so that fewer background galaxies are cut from this corner (see Fig. \ref{Fig:1.2selection conservative}). At a lens redshift of $z=1.2$ and using the matched sources from the HUDF region as in Sect. \ref{Sec:Colour_selection, defining mag and colour cuts}, the default selection strategy achieved an average lensing efficiency of $\langle \beta \rangle = 0.324$ with a number density of $n=13.4$\,arcmin$^{-2}$, resulting in a weak lensing sensitivity factor of $\tau_\mathrm{WL} = 1.19$. In comparison to that the alternative strategy achieved $\langle \beta \rangle = 0.317$ with a number density of $n=15.3$\,arcmin$^{-2}$, resulting in a weak lensing sensitivity factor of $\tau_\mathrm{WL} = 1.23$. In conclusion, this alternative provides only a negligible improvement of the weak lensing sensitivity factor, which would be even less for clusters at higher redshifts $1.2 < z \lesssim 1.6$. 

\begin{figure*}
\centering

	\includegraphics[width=0.97\columnwidth]{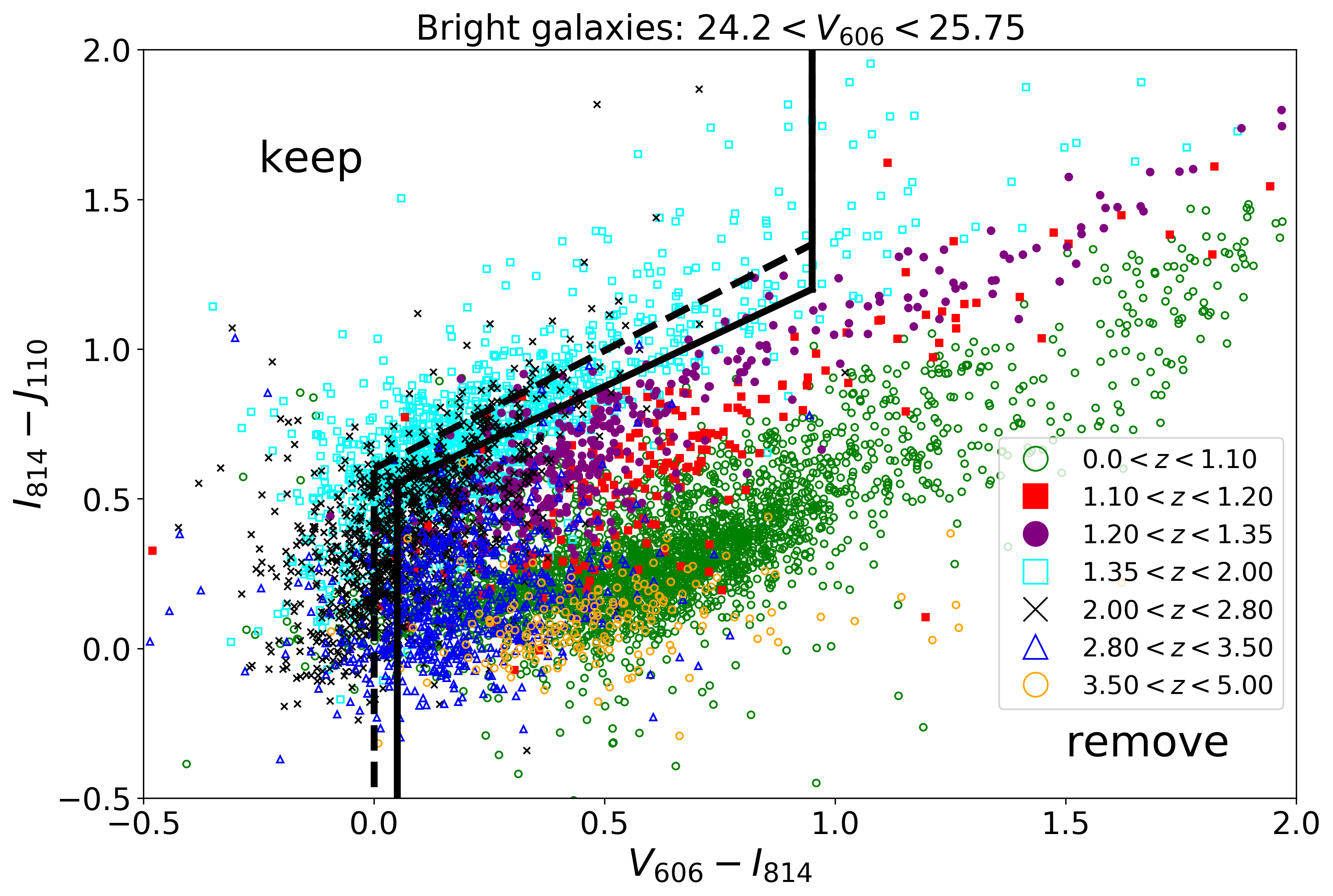}
	\includegraphics[width=0.970\columnwidth]{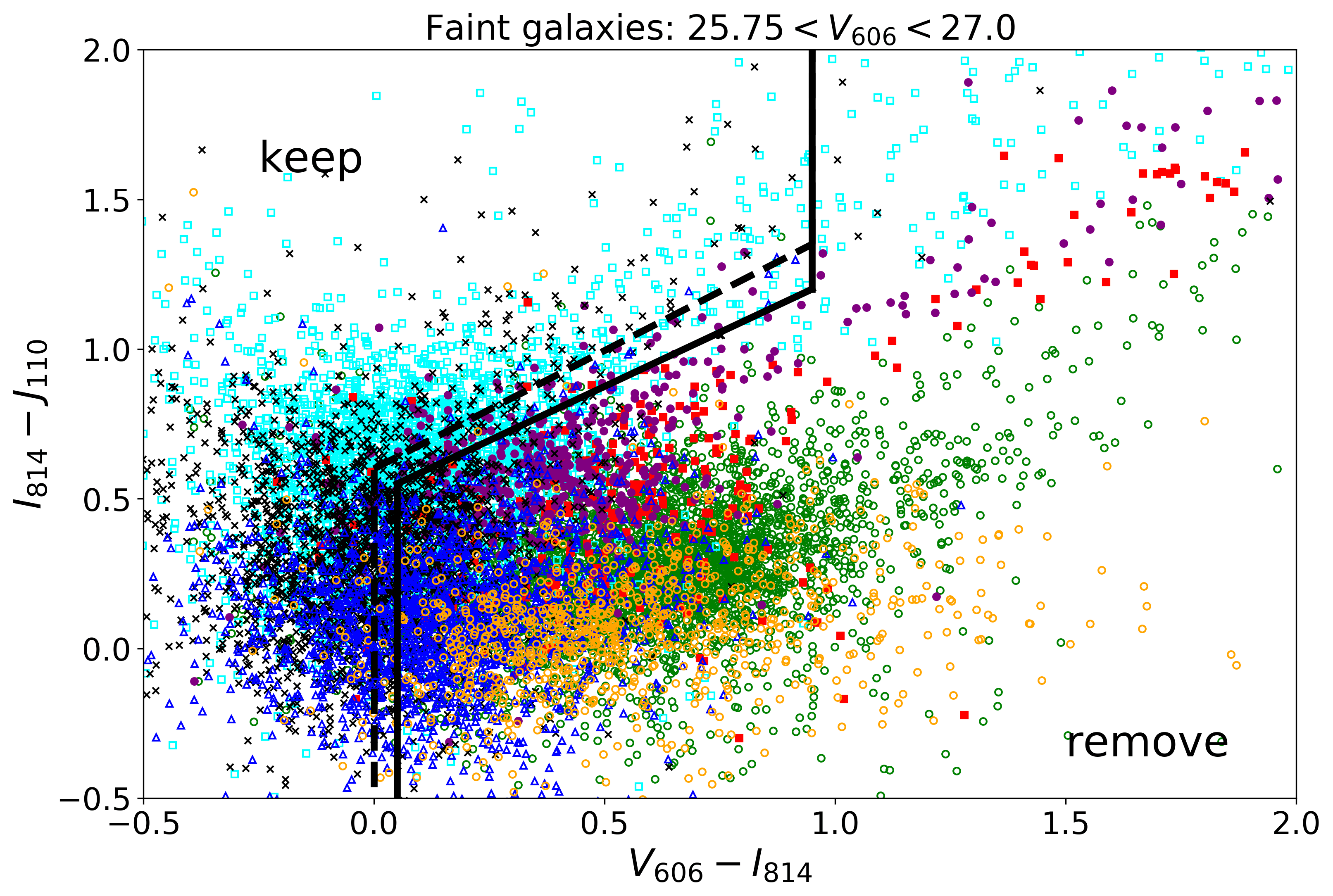}
	\includegraphics[width=0.97\columnwidth]{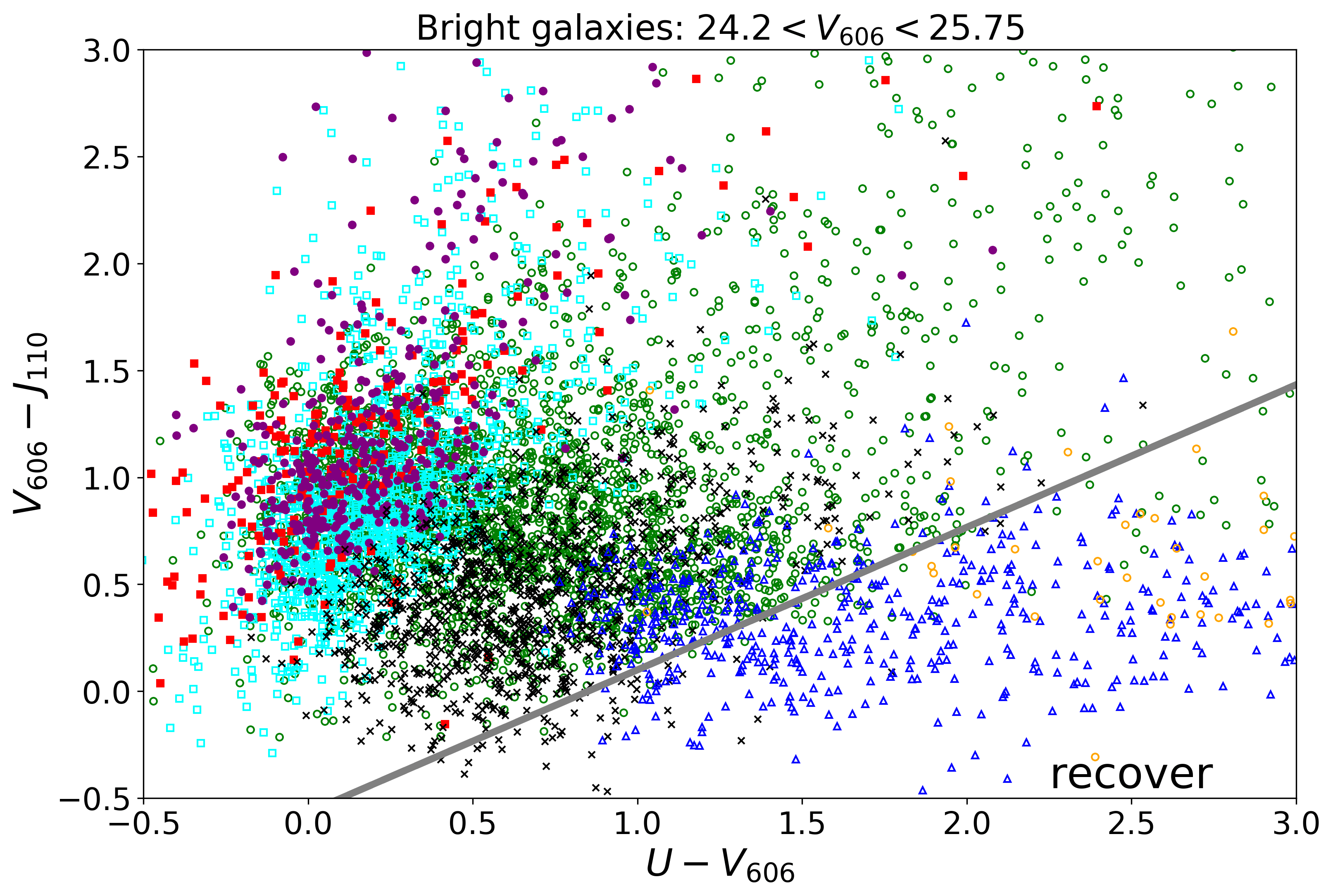}
	\includegraphics[width=0.97\columnwidth]{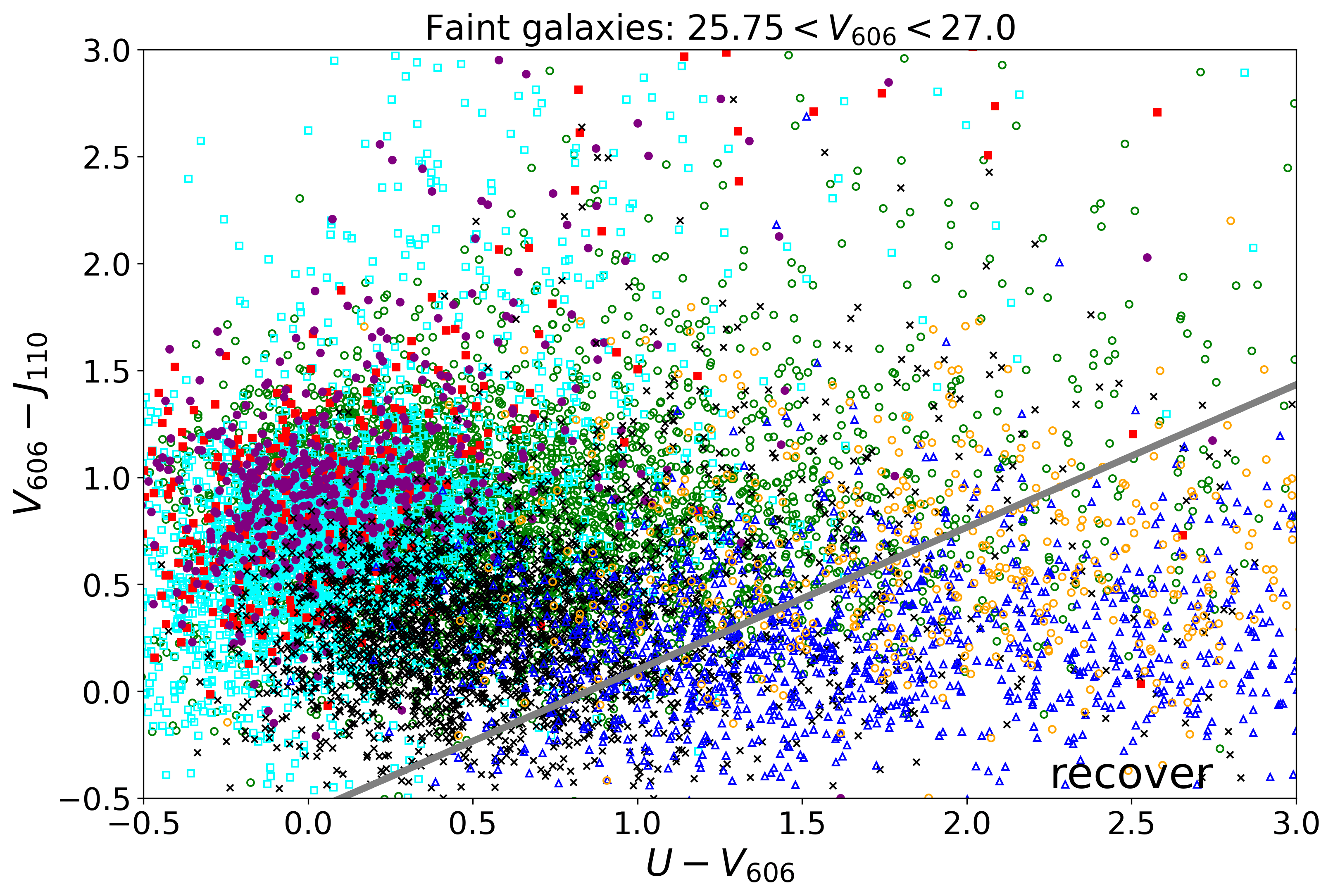}

    \caption{Second alternative colour selection for galaxy clusters at redshift \mbox{$ z \sim 1.2$}. The selected source galaxies are at redshift \mbox{$z \gtrsim 1.3$}. We display galaxies based on their photometry from \citetalias{Skelton2014} in the GOODS-South field. \textit{Top:} First selection step in the $VIJ$ plane for bright galaxies on the left and faint galaxies on the right. The solid black lines indicate cuts applied for bright galaxies, the dashed black lines show cuts for faint galaxies. \textit{Bottom:}  Second selection step in the $UVJ$ plane for bright galaxies on the left and faint galaxies on the right. Galaxies below the diagonal grey line are recovered in both the bright and the faint regime.}
    \label{Fig:1.2selection alternative}
\end{figure*}

As a second alternative selection strategy, we made use of the fact that the galaxies at the cluster redshift for a cluster at $z=1.2$ are concentrated more towards the lower right of the $VIJ$ colour plane than for a cluster, for instance, at $z=1.7$. In this strategy, we used the $VIJ$ plane to cut not only the foreground but also the galaxies at the cluster redshift (see \mbox{Fig. \ref{Fig:1.2selection alternative}}). To cut all galaxies at the cluster redshift this way, the cuts need to be extended further towards bluer $V-I$ colour (to the left in the $VIJ$ plane). Consequently, cutting the galaxies at the cluster redshift in the upper left corner of the $UVJ$ colour plane is not necessary anymore, which allows us to keep more background galaxies (mainly the close background galaxies indicated by cyan symbols in \mbox{Fig. \ref{Fig:1.2selection alternative}}). With this strategy, we found an average lensing efficiency of $\langle \beta \rangle = 0.276$ with a number density of $n=16.9$\,arcmin$^{-2}$, resulting in a weak lensing sensitivity factor of $\tau_\mathrm{WL} = 1.13$. Thus, we found we cannot increase the weak lensing sensitivity factor with this strategy. While the number density did increase, mainly in the regime of near background galaxies, we also lost a notable fraction of the far background galaxies at high redshift due to the more extended cut in the $VIJ$ plane. As a result, the average geometric lensing efficiency decreased strongly and this could not be compensated by the higher source number density.

From exploring these two alternative background source selection strategies, we concluded that it is not beneficial to introduce a selection strategy that is optimised based on the cluster redshift for clusters with redshifts between $1.2 \lesssim z \lesssim 1.7$. We, therefore, applied the selection strategy presented in Sect. \ref{Sec:Colour_selection, defining mag and colour cuts} for all clusters in our sample with $1.2 \lesssim z \lesssim 1.7$. 
However, for the cluster SPT-CL{\thinspace}$J$0646$-$6236, which is located at a lower redshift of $z=0.995$, an alternative selection strategy did increase the weak lensing sensitivity factor noticeably as presented in Appendix \ref{Appendix:Coloursel_alternatives0995}.

% \FloatBarrier

\section{Colour selection strategy for the cluster SPT-CL{\thinspace}$J$0646$-$6236 at $z=0.995$}
\label{Appendix:Coloursel_alternatives0995}

\begin{figure*}
\centering
	
	\includegraphics[width=0.97\columnwidth]{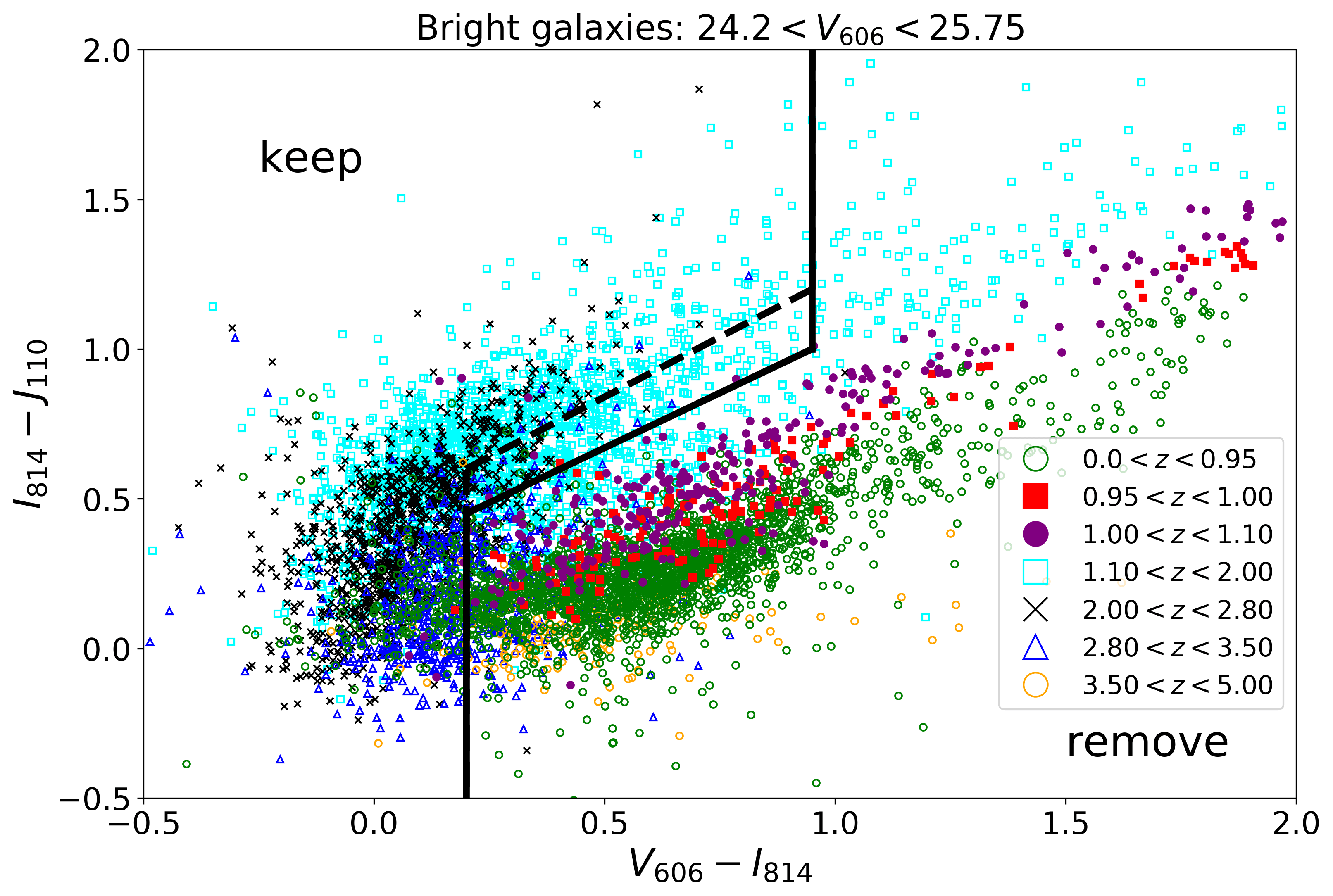}
	\includegraphics[width=0.97\columnwidth]{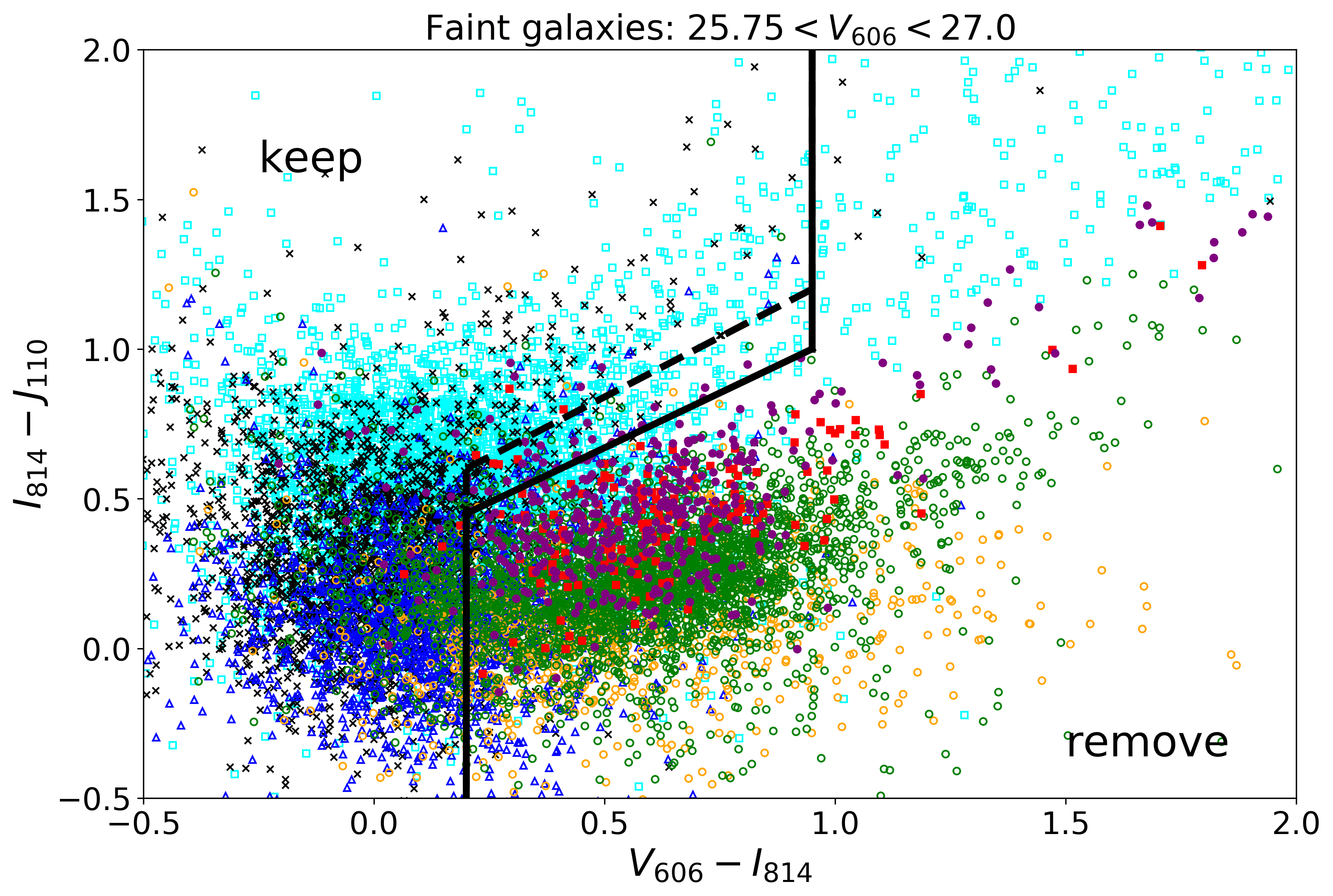}
	\includegraphics[width=0.97\columnwidth]{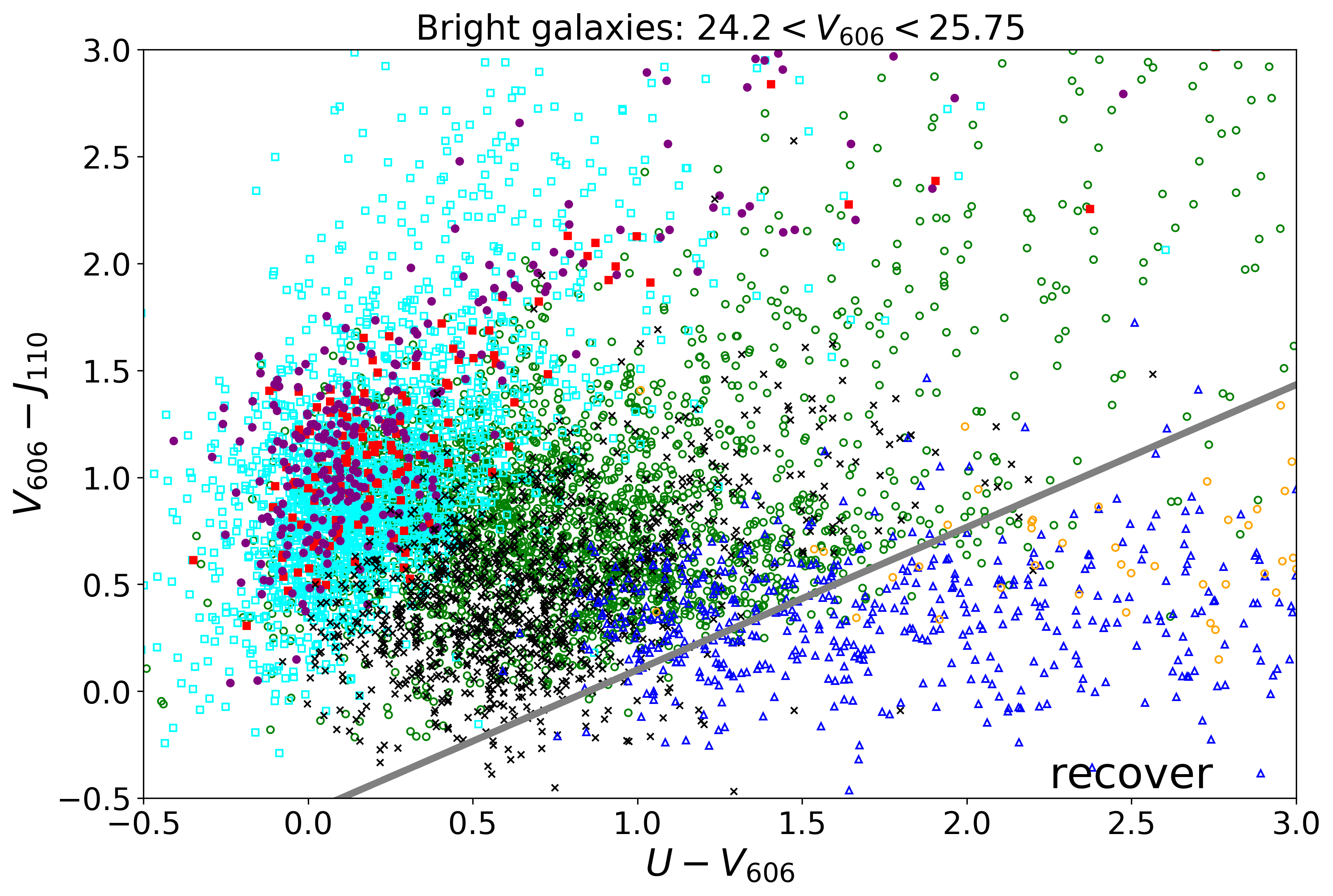}
	\includegraphics[width=0.97\columnwidth]{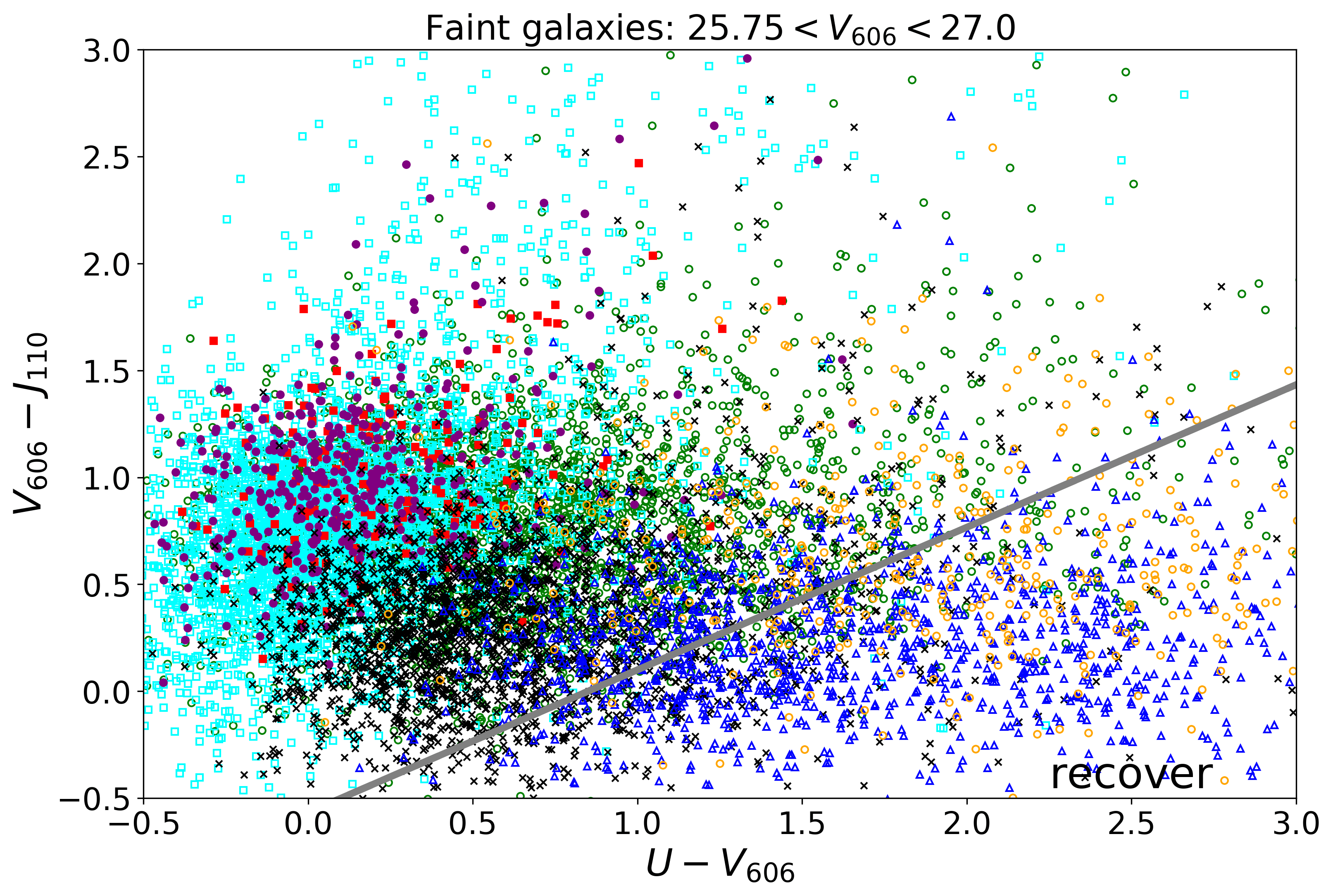}
	
    \caption{Colour selection for galaxy clusters at redshift \mbox{$ z \sim 1.0$}. The selected source galaxies are at redshift \mbox{$z \gtrsim 1.1$}. We display galaxies based on their photometry from \citetalias{Skelton2014} in the GOODS-South field. \textit{Top:} First selection step in the $VIJ$ plane for bright galaxies on the left and faint galaxies on the right. The solid black lines indicate cuts applied for bright galaxies, the dashed black lines show cuts for faint galaxies. \textit{Bottom:}  Second selection step in the $UVJ$ plane for bright galaxies on the left and faint galaxies on the right. Galaxies below the diagonal grey line are recovered in both the bright and the faint regime.}
    \label{Fig:0995selection}
\end{figure*}

The cluster SPT-CL{\thinspace}$J$0646$-$6236 has the lowest redshift in our sample with $z=0.995$. With the default background source selection strategy presented in Sect. \ref{Sec:Colour_selection, defining mag and colour cuts}, we do miss the galaxies in the redshift regime $1.1 \lesssim z \lesssim 1.7$, which we could incorporate for the lensing analysis of this cluster. In contrast to the alternative background source selection strategies presented in Appendix \ref{Appendix:Coloursel_alternatives}, we found that it is possible to achieve a significantly higher weak lensing sensitivity factor with a modification of the default selection strategy for this cluster. The original cut in the $VIJ$ plane already removed the majority of the galaxies at the cluster redshift $z\sim1$, so that we could omit the cut of sources in the upper left corner of the $UVJ$ plane (see Fig. \ref{Fig:0995selection}). As a result, we achieved a number density of selected background source galaxies, which was two times higher (27.4\,arcmin$^{-2}$)
%, 13.4 old vs 27.4 new) 
than for the default selection while the average geometric lensing efficiency only mildly decreased. At a lens redshift of $z=0.995$, we found $\langle \beta \rangle = 0.392$ for the default selection and $\langle \beta \rangle = 0.336$ for the optimised selection. As a consequence, the weak lensing sensitivity factor increased by about 23 per cent from $\tau_\mathrm{WL} = 1.43$ for the default selection strategy to $\tau_\mathrm{WL} = 1.76$ for the optimised strategy. Therefore, we used this optimised strategy in the lensing analysis of the cluster SPT-CL{\thinspace}$J$0646$-$6236 at $z=0.995$. 

\section{Consistency of weak lensing mass results with SZ or X-ray masses}
\label{Appendix:Consistency of WL with SZ + X-ray}

Some of the clusters in our sample have a lensing mass that has scattered high or low with respect to the reference mass measured from SZ or X-ray data (see Table \ref{tab:Cluster sample properties} for SZ masses, \citet{McDonald2017} for X-ray masses). This concerns in particular the clusters SPT-CL{\thinspace}$J$2040$-$4451  and SPT-CL{\thinspace}$J$0205$-$5829. To quantify the tension between weak lensing mass and SZ or X-ray mass for individual targets, we employed a simple model to test for the level at which the mass ratios are consistent with unity. To this end, we randomly drew 10,000 weak lensing masses $M_{\mathrm{WL,rand,}i}$ from a Normal distribution $\mathcal{N}(M_{500\mathrm{c}}^\mathrm{biased,ML}, \sigma_\mathrm{stat}(M_{500\mathrm{c}}^\mathrm{biased,ML}))$ given the best-fit weak lensing mass estimates and statistical uncertainties (see Sect. \ref{Sec: fits_to_tangential_reduced_shear_profiles} and Table \ref{tab:mass-Xray}).
We divided these by correction factors randomly drawn from the corresponding log-normal mass bias distributions (described in Sect. \ref{Sec:corr_for_mass_modelling_bias}). 
Similarly, we drew 10,000 SZ (or X-ray) masses $M_{\mathrm{SZ,rand,}i}$ (or $M_{\mathrm{X,rand,}i}$) from the best-fit values in conjunction with their uncertainties (Table \ref{tab:Cluster sample properties} for SZ masses, \citet{McDonald2017} for X-ray masses), using a Normal distribution. In case of asymmetric uncertainties, a two-piece Normal distribution \citep[e.g. ][]{John1982} was employed. We proceeded to take ratios of the weak lensing and SZ (or X-ray) mass distributions $M_{\mathrm{WL,rand,}i}/M_{\mathrm{SZ,rand,}i}$ (or $M_{\mathrm{WL,rand,}i}/M_{\mathrm{X,rand,}i}$). For a given target, the resulting ratio distribution was analysed for its consistency with unity. In particular, we constructed confidence intervals based on the shortest possible interval containing a given fraction (the confidence level) of the distribution. In this way, we found the lowest level of confidence making the mass ratio consistent with one. For SPT-CL{\thinspace}$J$2040$-$4451, which has a best-fit weak lensing mass noticeably higher than the SZ mass (X-ray mass), we found this confidence level to be 70 per cent (75 per cent), corresponding to a probability of 0.3 (0.25) of seeing an outlier with this degree or more of discrepancy (for an individual cluster). Similarly, for SPT-CL{\thinspace}$J$0205$-$5829, the probability of an outlier with or exceeding the observed degree of discrepancy is 0.09 for the SZ mass (0.21 for the X-ray mass). We conclude that the observed scatter between lensing masses and SZ or X-ray masses is well within the expectation given the large statistical uncertainties of our study, and given that these two clusters are the most extreme outliers within our sample of nine clusters.
 
% \FloatBarrier
 
\section{Weak lensing results: mass maps and tangential reduced shear profiles}
\label{Appendix:WeakLensingResults}

\begin{figure}
\centering
 \includegraphics[clip, trim=0.5cm 5.3cm 0.5cm 3.4cm,width=0.82\columnwidth]{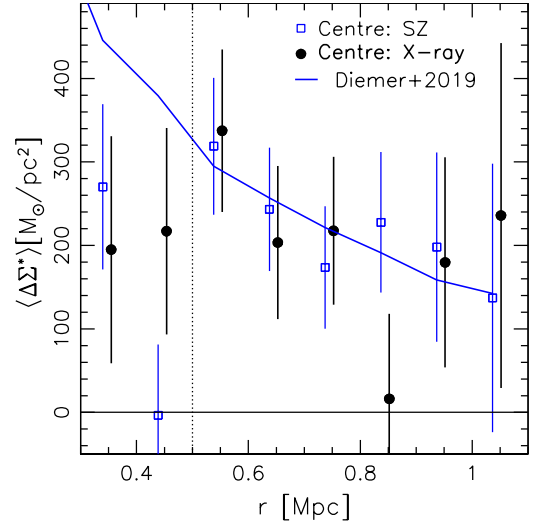}
 \caption{Weighted average of the rescaled differential surface mass density profiles for the clusters in our sample. The black points and blue squares refer to measurements using the X-ray (all clusters except SPT-CL{\thinspace}$J$0646$-$6236 for which we do not have X-ray measurements) and SZ centres, respectively. The blue line shows the average weighted model NFW function assuming a fixed concentration--mass relation following \citet{Diemer2015} with updated parameters from \citet{Diemer2019} for measurements from the SZ centres. The vertical dotted line indicates the lower limit of our fit range.
          \label{Fig:Stacked-profile}}
\end{figure}

\begin{figure*}
% \centering
  \includegraphics[width=0.98\columnwidth]{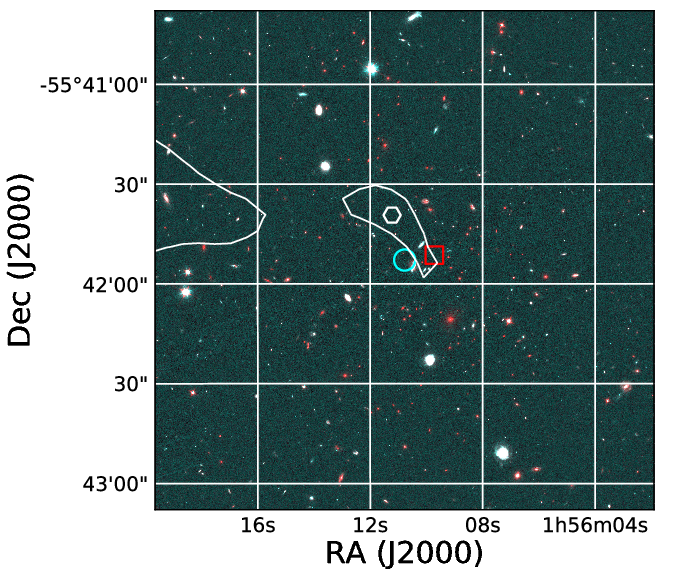}
 \hspace{2cm}
 \includegraphics[clip, trim=0.5cm 5cm 0.5cm 3cm,width=0.83\columnwidth]{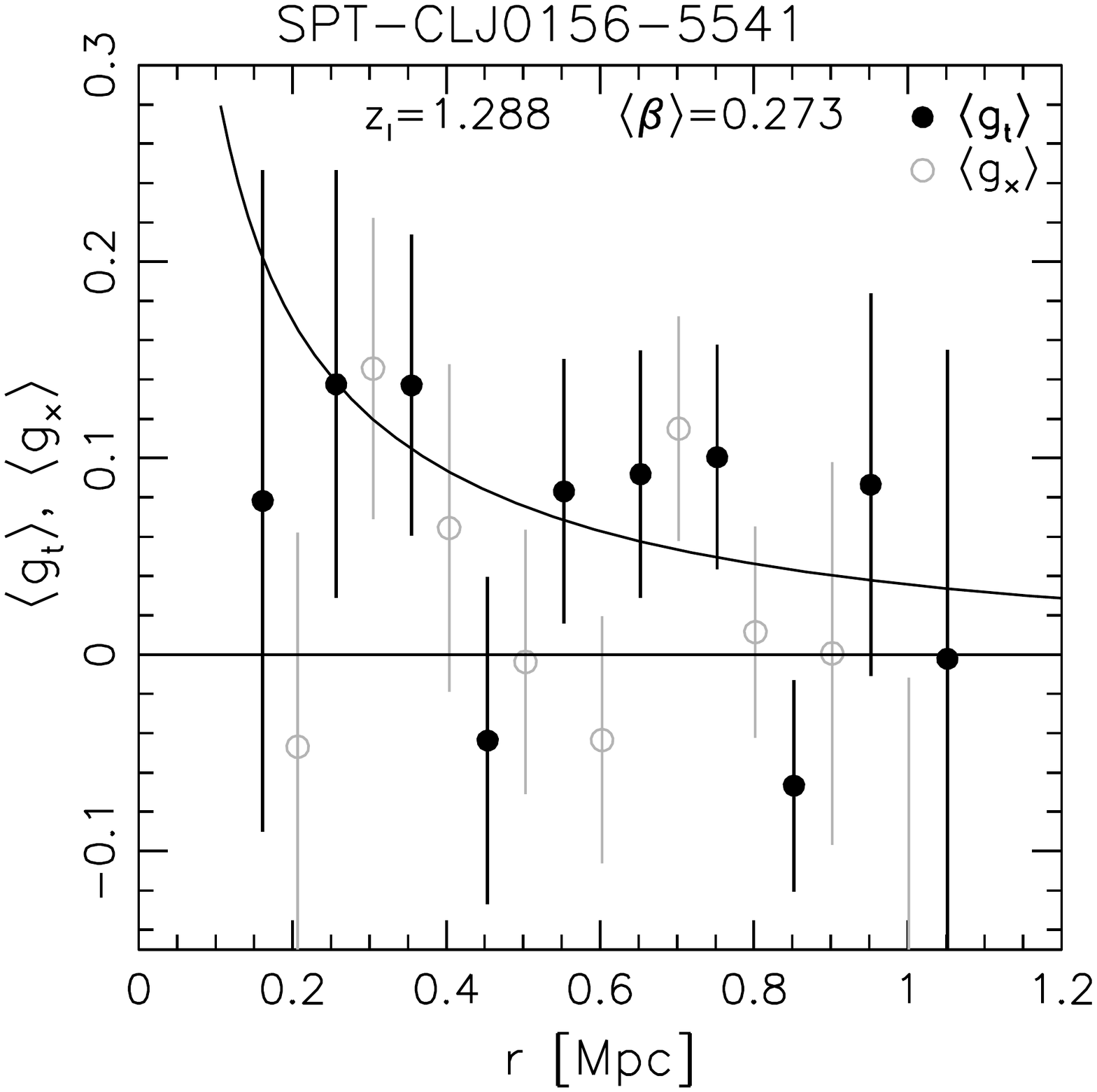}
  \includegraphics[width=0.98\columnwidth]{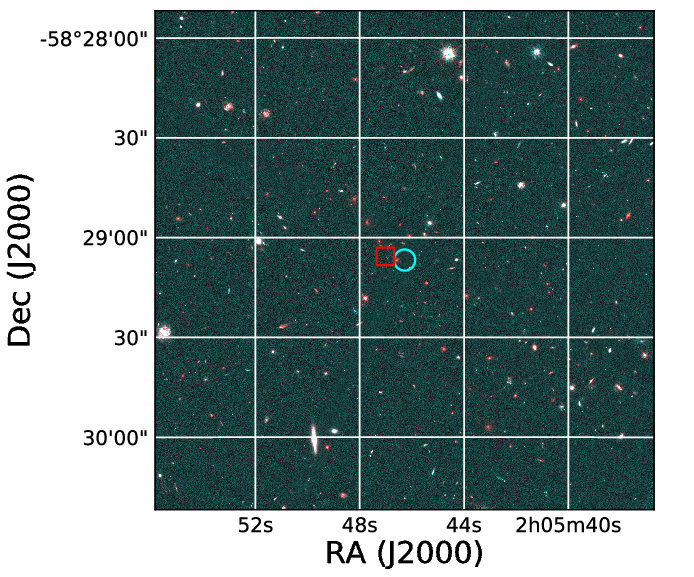}
 \hspace{2cm}
 \includegraphics[clip, trim=0.5cm 5cm 0.5cm 3cm,width=0.83\columnwidth]{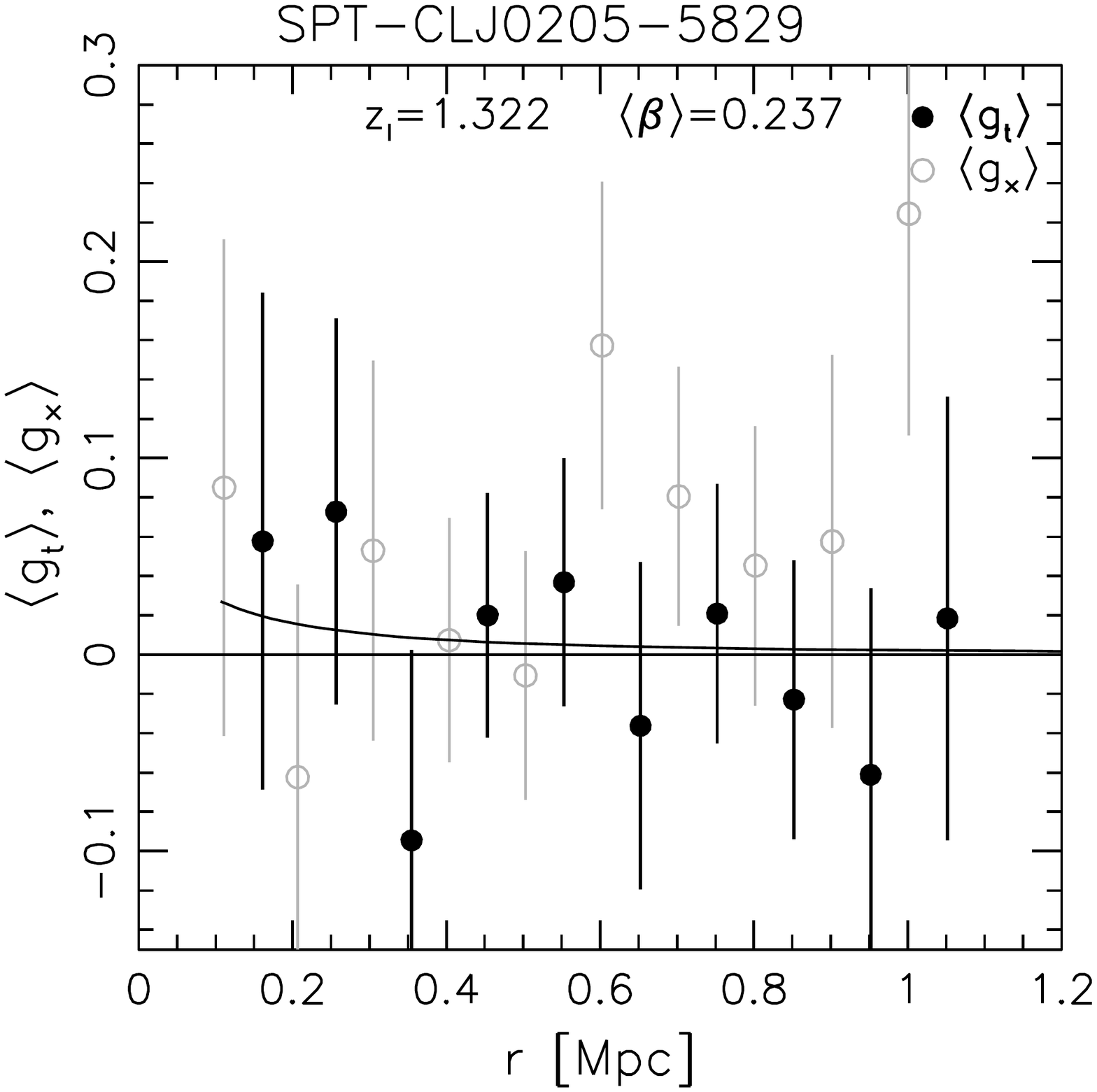}
    \includegraphics[width=0.98\columnwidth]{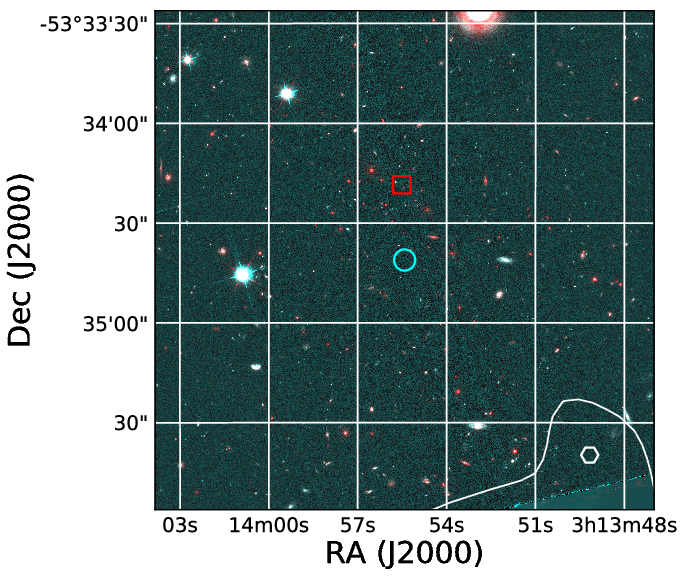}
  \hspace{2cm}
  \includegraphics[clip, trim=0.5cm 5cm 0.5cm 3cm,width=0.83\columnwidth]{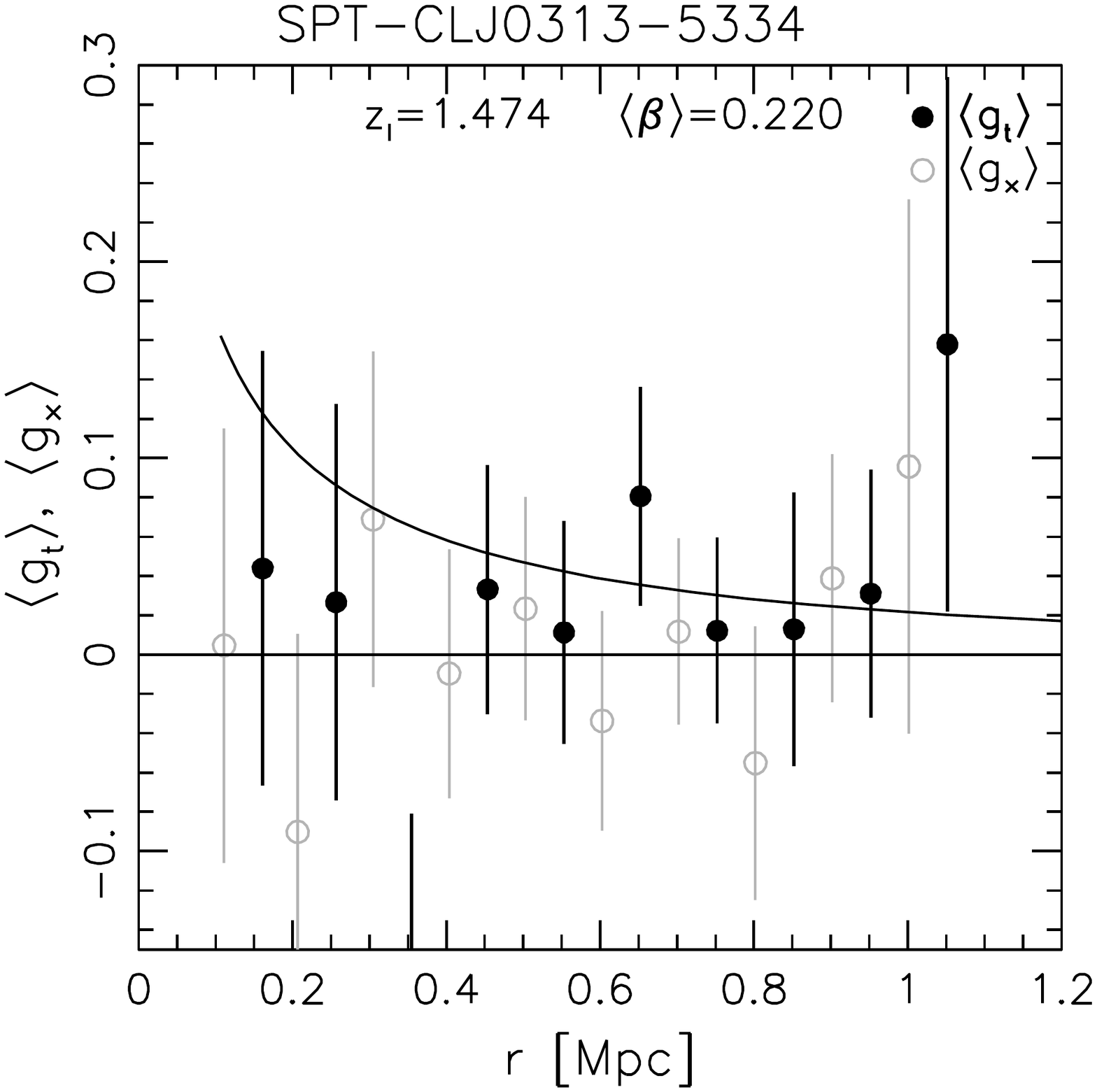}
  
 \caption{Weak lensing results for the clusters in our sample (see the caption of \mbox{Fig. \ref{fi:wl_results_1}} for details).
          \label{fi:wl_results_2}}
\end{figure*}

\begin{figure*}
%  \centering
   \includegraphics[width=0.98\columnwidth]{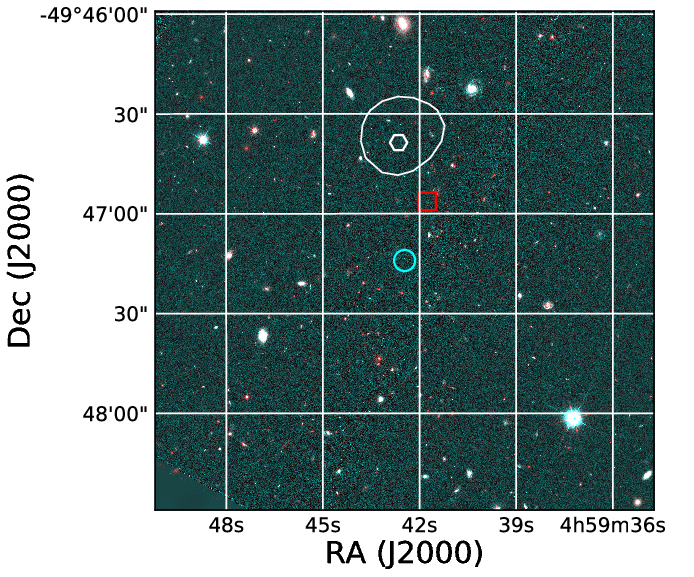}
  \hspace{2cm}
  \includegraphics[clip, trim=0.5cm 5cm 0.5cm 3cm,width=0.83\columnwidth]{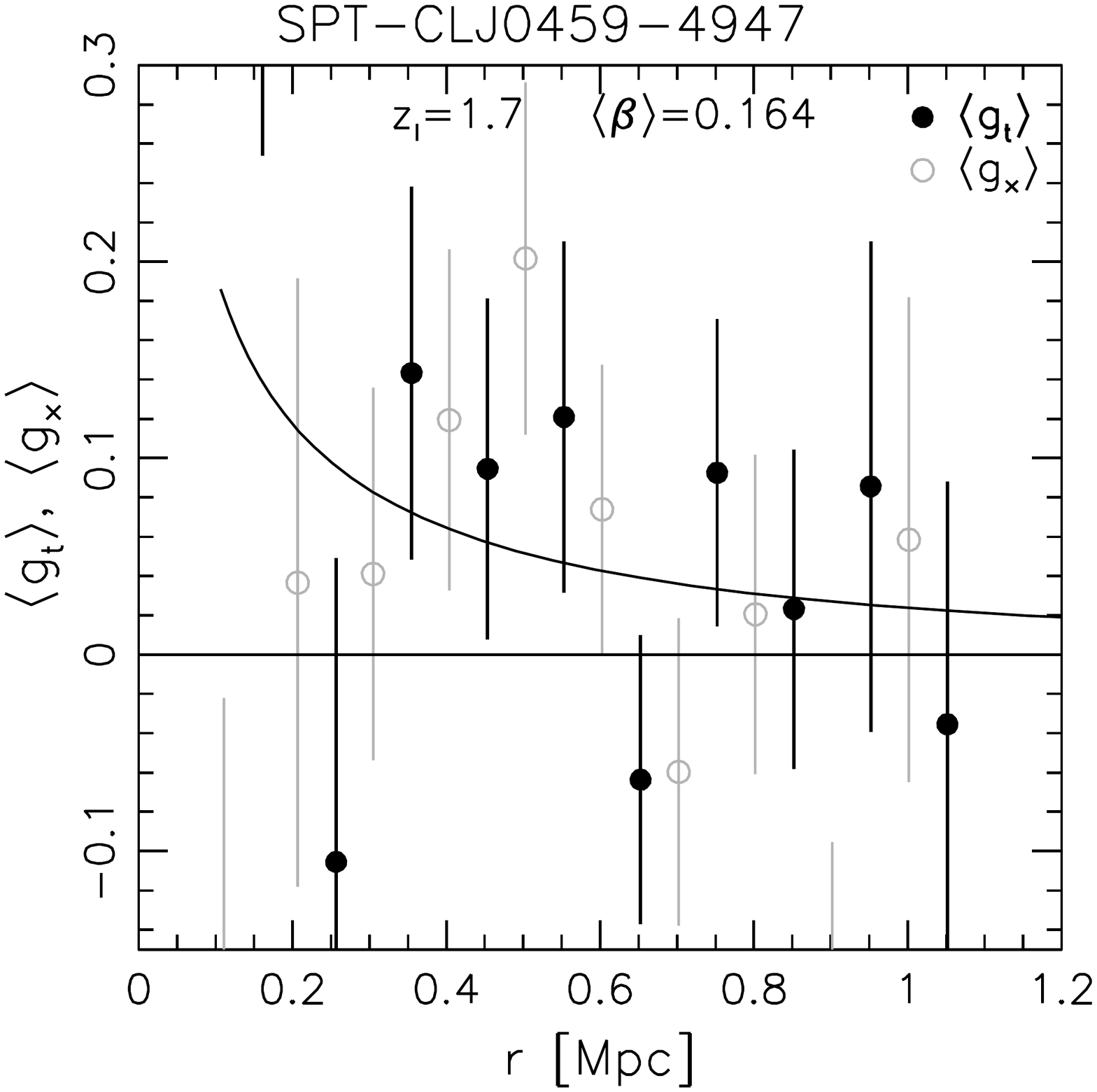}
    \includegraphics[width=0.98\columnwidth]{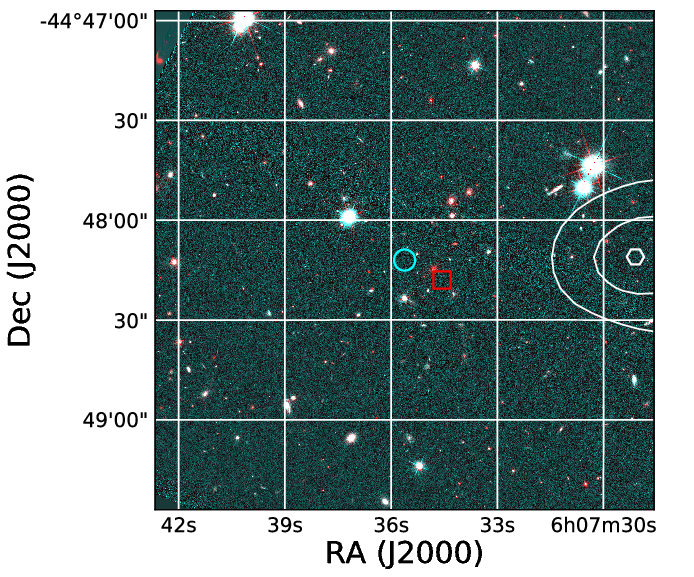}
  \hspace{2cm}
  \includegraphics[clip, trim=0.5cm 5cm 0.5cm 3cm,width=0.83\columnwidth]{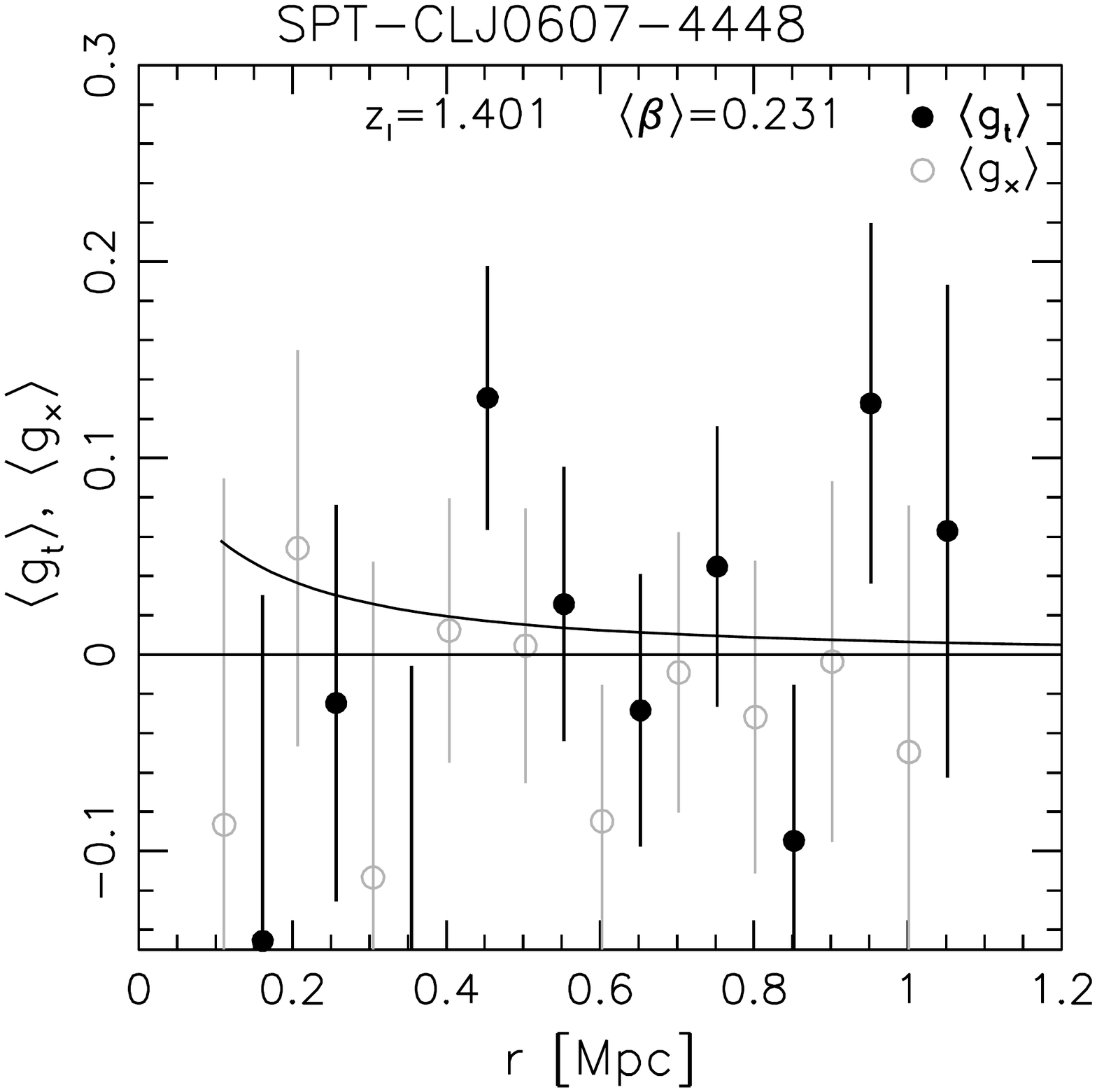}
   \includegraphics[width=0.98\columnwidth]{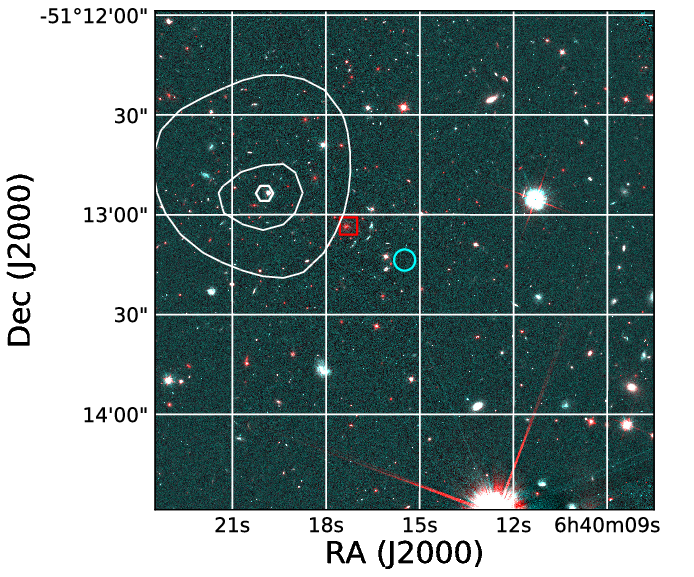}
  \hspace{2cm}
  \includegraphics[clip, trim=0.5cm 5cm 0.5cm 3cm,width=0.83\columnwidth]{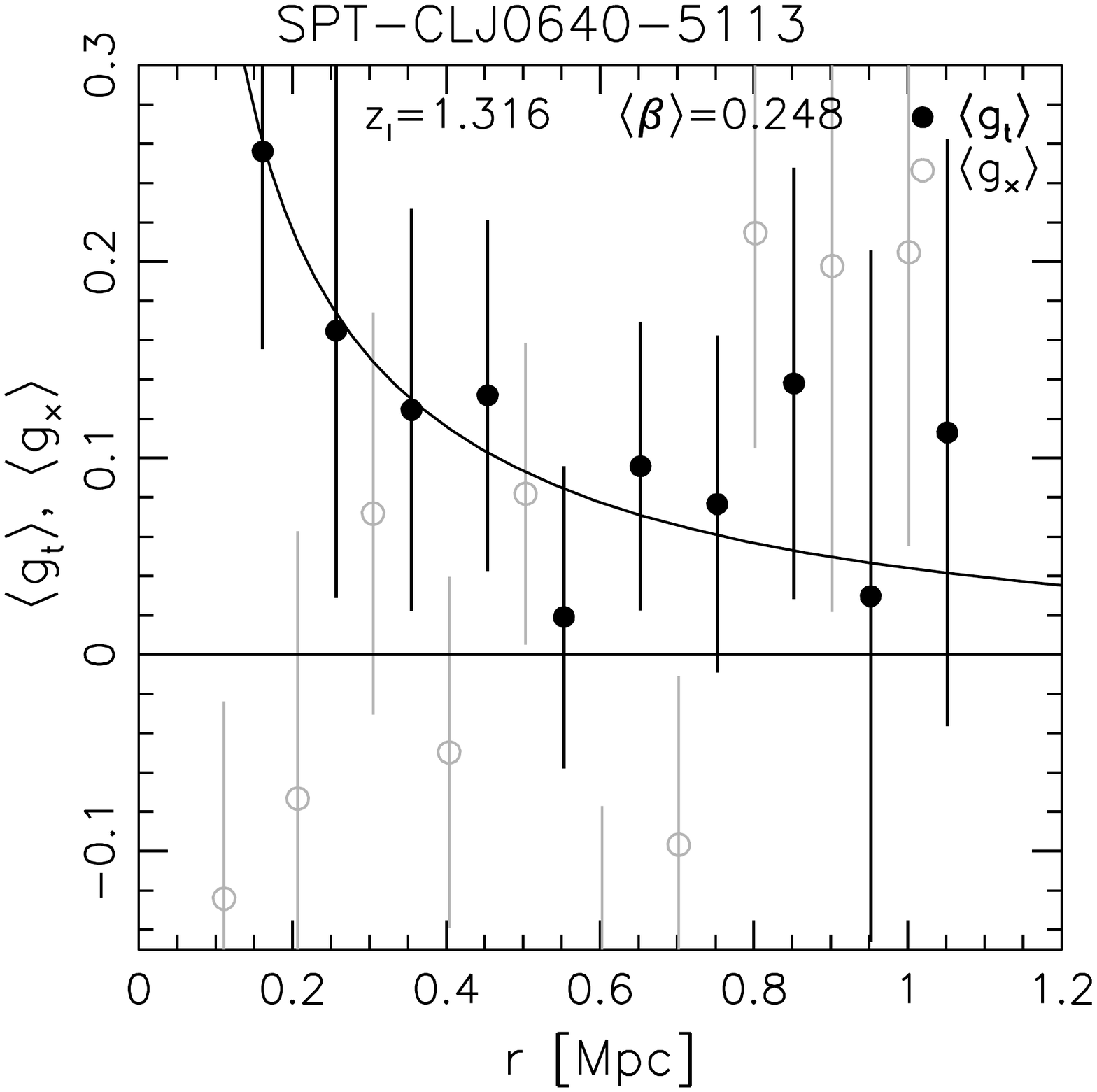}
  
  \caption{Weak lensing results for the clusters in our sample (continued, see the caption of \mbox{Fig. \ref{fi:wl_results_1}} for details). \label{fi:wl_results_3}}
 \end{figure*}

\begin{figure*}
%  \centering
    \includegraphics[width=0.98\columnwidth]{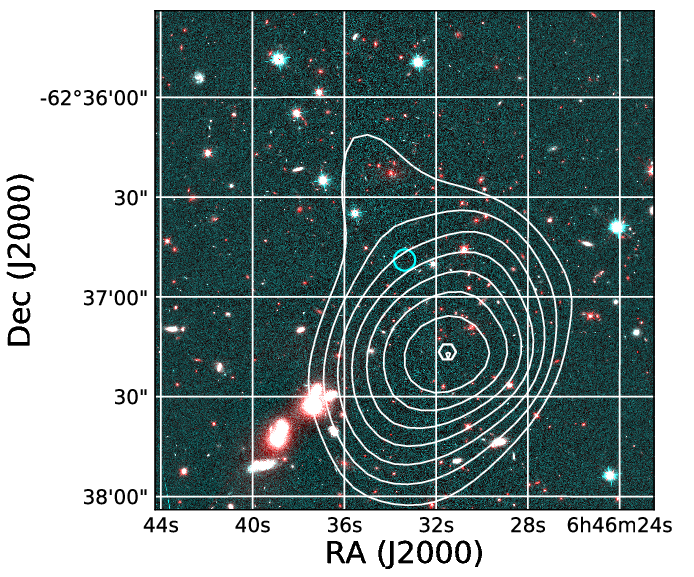}
  \hspace{2cm}
  \includegraphics[clip, trim=0.5cm 5cm 0.5cm 3cm,width=0.83\columnwidth]{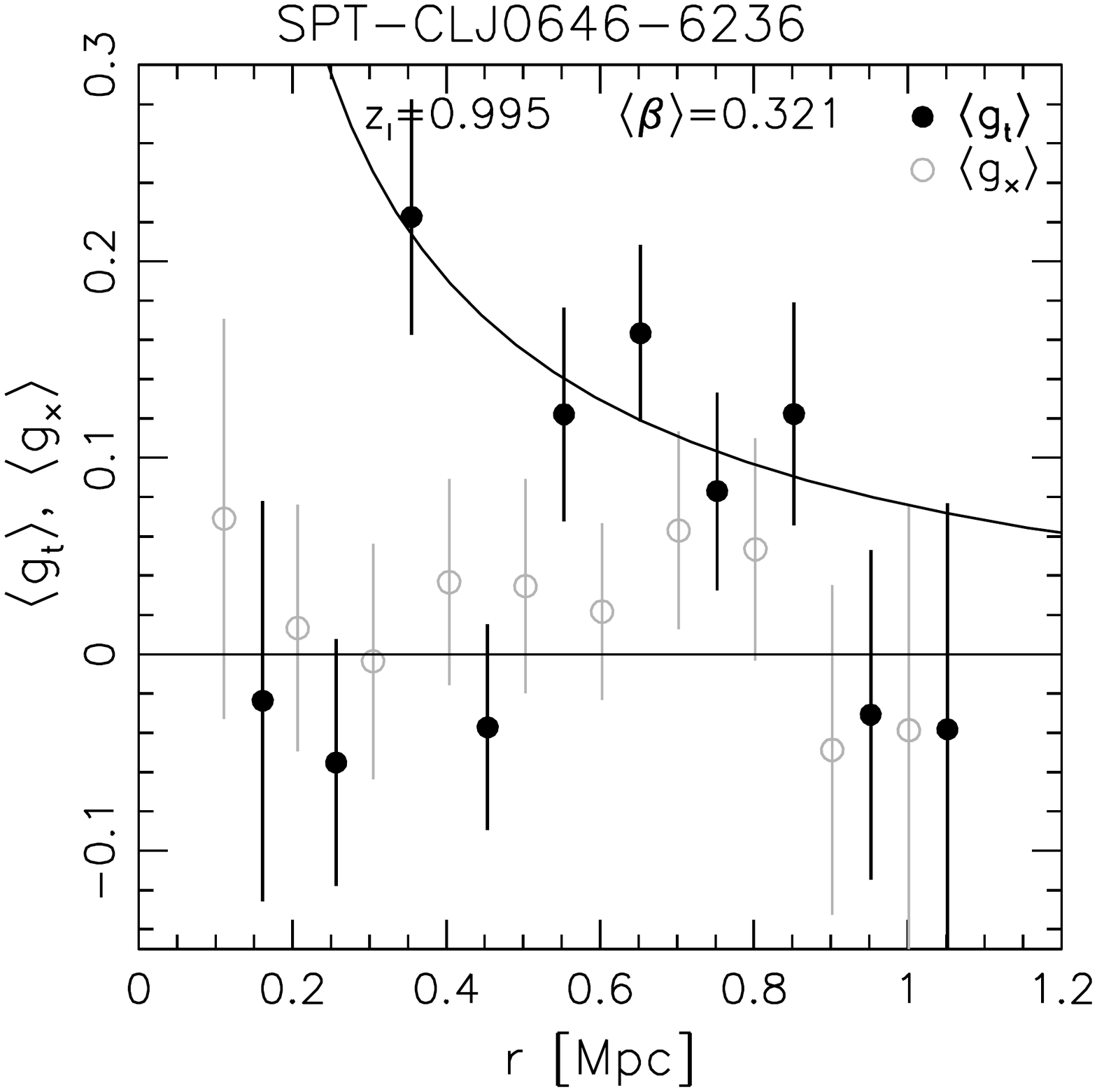}
    \includegraphics[width=0.98\columnwidth]{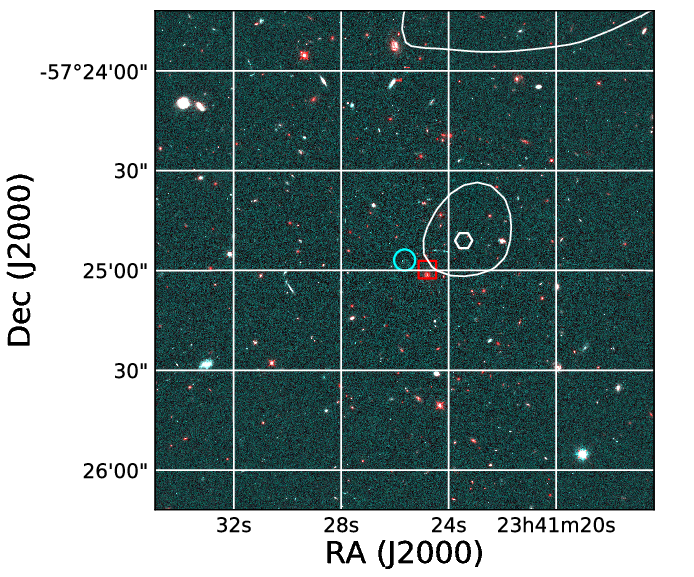}
  \hspace{2cm}
  \includegraphics[clip, trim=0.5cm 5cm 0.5cm 3cm,width=0.83\columnwidth]{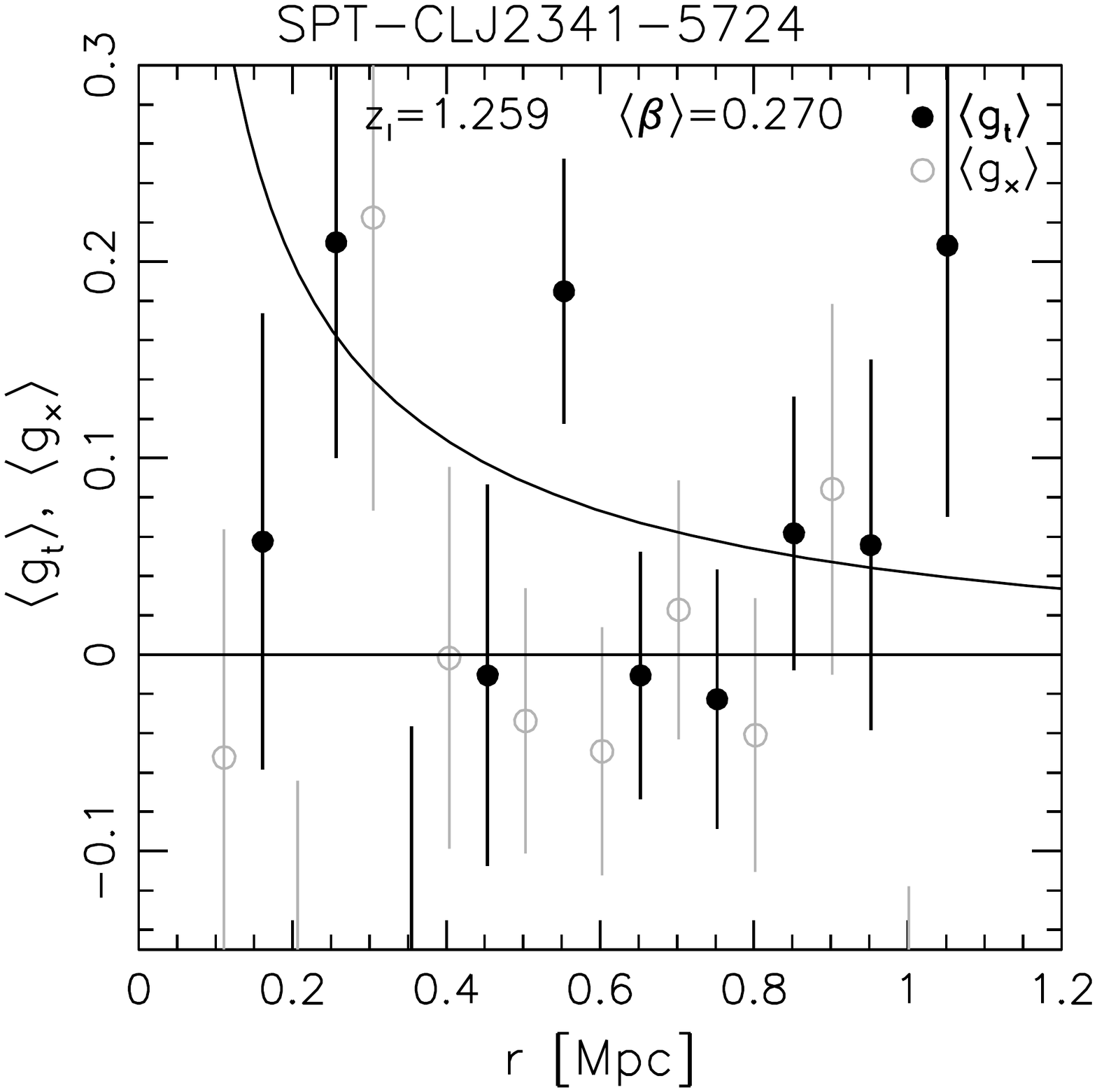}
  
  \caption{Weak lensing results for the clusters in our sample (continued, see the caption of \mbox{Fig. \ref{fi:wl_results_1}} for details). For SPT-CL{\thinspace}$J$0646$-$6236 the reduced shear profile was computed with respect to the SZ centre.
    \label{fi:wl_results_4}}
 \end{figure*}

We show the weak lensing results, including the mass maps and tangential reduced shear profiles for the studied cluster sample in Figs. \ref{fi:wl_results_2} to \ref{fi:wl_results_4}. In addition, we display the stacked profile of the cluster sample in Fig. \ref{Fig:Stacked-profile}. 
Following \citetalias{Schrabback2018} (their sect. 7.3), we stacked the lensing signal of the clusters in terms of the differential surface mass density $\Delta \Sigma(r)$, where we computed $\Sigma_\mathrm{crit}$ based on the average lensing efficiency $\langle \beta \rangle$ from the individual clusters, respectively. Since the clusters vary in mass, we rescaled them to an approximately similar signal amplitude with the help of the SZ masses listed in Table \ref{tab:Cluster sample properties}. Based on this mass and assuming the concentration--mass relation by \citet{Diemer2015} with updated parameters from \citet{Diemer2019}, we computed a theoretical NFW model for the differential surface mass density $\Delta \Sigma_\mathrm{model}$. We then rescaled the cluster lensing signal by a factor $s$ according to 
\begin{equation}
 \Delta \Sigma^{\ast} (r) = s\Delta \Sigma(r) \equiv \frac{\langle \Delta \Sigma_\mathrm{model}(800\,\mathrm{kpc})\rangle}{\Delta \Sigma_\mathrm{model}(800\,\mathrm{kpc})} \Delta\Sigma(r) \,,
\end{equation}
where we used $r=800\,\mathrm{kpc}$ as the reference scale to evaluate the theoretical model. The weighted average then reads
\begin{equation}
 \langle \Delta \Sigma^{\ast} \rangle (r_j) = \sum_{i \in \mathrm{clusters}} \Delta \Sigma^{\ast}_i (r_j) \hat{W}_{ij} / \sum_{i \in \mathrm{clusters}} \hat{W}_{ij} \,,
\end{equation}
with $\hat{W}_{ij} = \left[ s\sigma(\Delta \Sigma(r_j))\right]^{-2}$ and $\sigma(\Delta \Sigma(r_j))$ as the $1\sigma$ uncertainty of $\Delta \Sigma(r_j)$.

\end{appendix}

\end{document}